%% file: jnl_ws_computation.tex
\documentclass[journal, romanappendices]{IEEEtran}

\usepackage{mathtools}
\usepackage{filecontents}
\usepackage{pgfplots}
 \usepackage{tikz} 
 \usepackage{hyperref}

\ifCLASSINFOpdf
 
\else
 
\fi

\input{myDefinitions}

\begin{document}
%
\title{Wigner-Smith Time Delay Matrix for Electromagnetics: Computational Aspects for Radiation and Scattering Analysis}
%
%
%

\author{Utkarsh~R. Patel        and Eric~Michielssen
\thanks{U. R. Patel and E. Michielssen are with the Department
of Electrical  Engineering and Computer Science, University of Michigan, Ann Arbor,
MI, 48109 USA. e-mail: urpatel@umich.edu and emichiel@umich.edu.}
}

%
%

\markboth{}%
{}
%



\maketitle

\begin{abstract}
The WS time delay matrix relates a lossless and reciprocal system's scattering matrix to its frequency derivative, and enables the synthesis of modes that experience well-defined group delays when interacting with the system. The elements of the WS time delay matrix for surface scatterers and antennas comprise renormalized energy-like volume integrals involving electric and magnetic fields that arise when exciting the system via its ports. Here, direct and indirect methods for computing the WS time delay matrix are presented. The direct method evaluates the energy-like volume integrals using surface integral operators that act on the incident electric fields and current densities for all excitations characterizing the scattering matrix. The indirect method accomplishes the same task by computing scattering parameters and their frequency derivatives. Both methods are computationally efficient and readily integrated into existing surface integral equation codes. The proposed techniques facilitate the evaluation of frequency derivatives of antenna impedances, antenna patterns, and scatterer radar cross sections in terms of renormalized field energies derived from a single frequency characterization of the system.
\end{abstract}


%
\IEEEpeerreviewmaketitle

\input{introduction}

\input{theory}

\input{results}

\section{Conclusion}
\label{sec:conclusions}

Two methods for computing the WS time delay matrix for systems composed of lossless PEC radiators and/or scatterers were presented. The direct method casts entries of $\matr{Q}$, viz. volume integrals of energy-like quantities, in terms of surface integral operators acting on incident fields and the current densities they induce. The indirect method computes $\matr{Q}$ using its defining equation~\eqref{eq:WS_Main_Relation}, by evaluating $\matr{S}$ and $\matr{S'}$ in terms of the same quantities. Computation and diagonalization of $\matr{Q}$ enables the synthesis of WS modes that experience well-defined group delays upon interacting with the system. It furthermore allows $\matr{S}'$ and frequency sensitivities of port observables to be expressed in terms of renormalized energies of the system's WS modes. The proposed methods for computing $\matr{Q}$ shed light on the spatial origin's role in the evaluation of group delays (renormalized energies) of electromagnetic fields. More specifically, they show that the sum of all group delays experienced by all modes is independent of the basis that expresses incident fields and translations of the system w.r.t. the spatial origin.

The proposed integral equation techniques for computing the WS time delay matrix of PEC structures can be expanded along several dimensions. First, they can be extended to allow for the characterization of penetrable and potentially inhomogeneous antennas and scatterers. Second, they can be used to construct fast frequency-sweep computational methods for scattering problems. Work on the above topics is in progress and will be reported in future papers.

\input{Appendix}


%


\ifCLASSOPTIONcaptionsoff
  \newpage
\fi



%
\bibliographystyle{IEEEtran}
\bibliography{IEEEabrv,biblio}

%




\end{document}

%% file: myDefinitions.tex
\usepackage{tikz}
\usepackage{subfig}
\usetikzlibrary{calc}
\usetikzlibrary{arrows}
\usetikzlibrary{spy}
\usepackage{relsize}
\usepackage{wrapfig}
\usepackage{eurosym}
\usepackage[siunitx]{circuitikz}
\usepackage{amsmath}
\usepackage{amssymb}
\usepackage{amsfonts}
\usepackage{esint}
\usepackage{mathrsfs}
\usepackage{amsthm}
\usepackage{amscd}
\usepackage{listings}
\usepackage{booktabs,longtable}
\usepackage{graphicx}
\usepackage{multirow}
\usepackage{textcomp}
\usetikzlibrary{patterns}
\usepackage{soul}

\newcommand{\vect}[1]{\boldsymbol{#1}}

\newcommand{\matr}[1]{\mathbf{#1}}
\newcommand{\abs}[1]{\left \lvert #1 \right\rvert }

\renewcommand{\Re}[1]{\mathbb{R}\mathrm{e}\left \{#1\right\} }
\renewcommand{\Im}[1]{\mathbb{I}\mathrm{m}\left \{#1\right\} }

\newcommand{\vrq}[0]{\vect{r}''}
\newcommand{\vrp}[0]{\vect{r}'}
\newcommand{\vr}[0]{\vect{r}}

 \newcommand{\unit}[1]{\vect{\hat{#1}}}

\newcommand{\junk}[1] {}

\def\XXint#1#2#3{{\setbox0=\hbox{$#1{#2#3}{\int}$}
\vcenter{\hbox{$#2#3$}}\kern-.5\wd0}}

\newcommand*\widebar[1]{%
  \hbox{%
    \vbox{%
      \hrule height 0.4pt 
      \kern0.3ex
      \hbox{%
        \kern-0.07em
        \ensuremath{#1}%
        \kern-0.07em
      }%
    }%
  }%
}

\usepackage[siunitx]{circuitikz}
\usepackage{etoolbox}
\makeatletter

\pgfcircdeclarebipole{}
    {\ctikzvalof{bipoles/tline/height}}{tline}
    {\ctikzvalof{bipoles/tline/height}}
    {\ctikzvalof{bipoles/tline/width}}
{   
    \pgfpointdiff{\pgfpointanchor{\ctikzvalof{bipole/name}start}{center}}
                 {\pgfpointanchor{\ctikzvalof{bipole/name}end}{center}}
    \pgfmathparse{veclen(\the\pgf@x,\the\pgf@y)}
    \pgftransformxshift{\pgfmathresult/2-30pt}
    \pgf@circ@res@left=\dimexpr-\pgfmathresult pt+40pt\relax
    \pgf@circ@res@step=.2\pgf@circ@res@right 
    \pgfsetlinewidth{\pgfkeysvalueof{/tikz/circuitikz/bipoles/thickness}\pgfstartlinewidth}
    \pgfpathellipse{\pgfpoint{\pgf@circ@res@right-\pgf@circ@res@step}{0}}
                   {\pgfpoint{\pgf@circ@res@step}{0}}
                   {\pgfpoint{0}{-\pgf@circ@res@up}}
    \pgfpathmoveto{\pgfpoint{\pgf@circ@res@right-\pgf@circ@res@step}{\pgf@circ@res@up}}
    \pgfpathlineto{\pgfpoint{\pgf@circ@res@left+\pgf@circ@res@step}{\pgf@circ@res@up}}
    \pgfpatharc{-90}{90}{-\pgf@circ@res@step and -\pgf@circ@res@up}
    \pgfpathlineto{\pgfpoint{\pgf@circ@res@right-\pgf@circ@res@step}{\pgf@circ@res@down}}
    \pgfsetfillcolor{white}
    \pgfusepath{draw,fill}
    \pgfpathmoveto{\pgfpoint{\pgf@circ@res@right-2*\pgf@circ@res@step}{0pt}}
    \pgfpathlineto{\pgfpoint{\pgf@circ@res@right}{0pt}}
    \pgfusepath{draw}
}

\newcommand{\myparallel}{{\mkern3mu\vphantom{\perp}\vrule depth 0pt\mkern2mu\vrule depth 0pt\mkern3mu}}

\allowdisplaybreaks

\usetikzlibrary{shapes.arrows}   

\tikzset{arrowstyle/.style={draw=FireBrick,arrowfill, single arrow,minimum height=#1, single arrow,
single arrow head extend=.2cm,}}

\DeclareMathOperator{\Tr}{Tr}

%% file: introduction.tex
\section{Introduction}


Wigner-Smith (WS) techniques, developed 60 years ago to characterize time delays experienced by interacting fields and particles~\cite{Smith_1960}~\cite{Texier_2016}, increasingly are being applied to the study of optical and microwave phenomena.  Illustrative applications of WS concepts include the characterization of wave propagation in multimode fibers~\cite{Carpenter_2015}, the optimization of light storage in highly scattering environments~\cite{Durand_2019}, and the focusing of light in disordered media~\cite{Ambichl_2017}. Experimental observations of WS fields coupling into microwave resonators and micro-manipulating targets recently have been reported as well~\cite{Hougne2020}--\cite{Horodynski2020optimal}.

This paper is a follow-on to a study by the same authors that  analyzed the ``WS time delay matrix"
\begin{equation}
\matr{Q} = j \matr{S}^\dag \dfrac{d\matr{S}}{d\omega}\,,
\label{eq:WS_Main_Relation}
\end{equation}
of guiding, scattering, and radiating electromagnetic systems~\cite{TAP_1} with scattering matrix $\matr{S}$. The diagonal elements of $\matr{Q}$
were interpreted as average group delays experienced by incoming waves as they interact with the system prior to exiting via its ports. For guiding systems excited by Transverse Electromagnetic (TEM) waves, the entries of the WS time delay matrix $\matr{Q}$ were shown to be volume integrals of energy-like densities involving the electric and magnetic fields that arise upon excitation of the system's ports. For guiding systems with non-TEM excitations, scattering or radiating systems, correction terms and renormalization procedures were required to restore the energy interpretation of \eqref{eq:WS_Main_Relation}. Reference \cite{TAP_1} also elucidated the use of WS modes, viz. eigenvectors of $\matr{Q}$, to synthesize excitations that experience well-defined group delays when interacting with microwave networks, to untangle resonant, corner/edge, and ballistic scattering phenomena, and to estimate the frequency sensitivity of antenna impedances.

\input{graphics/radiating_system_graphics2}

This paper extends the methodology of~\cite{TAP_1} by enabling its application in the integral equation-based analysis of perfect electrically conducting (PEC) radiators and/or scatterers (Fig. \ref{fig:mixed_scatterer}).  Its contributions are threefold.
\begin{itemize}
\item It introduces a direct technique for computing the WS time delay matrix $\matr{Q}$ that casts formulas derived in \cite{TAP_1} for elements of $\matr{Q}$ involving volume integrals of renormalized energy-like quantities in terms of surface integrals of operators acting on incident electric fields and associated current densities. The method realizes significant computational savings over those proposed in \cite{TAP_1}, and extends  techniques by Vandenbosch \cite{VDB_2010} and Gustafsson et. al. \cite{Gustafsson_2015} for evaluating the energy stored by single port antennas to multiport systems subject to external excitations. 
\item It introduces an indirect technique for computing $\matr{Q}$ that evaluates $\matr{S}$ and $\matr{S}'$ from knowledge of the current densities excited by the incident fields that define $\matr{S}$, and then computes their product as in~\eqref{eq:WS_Main_Relation}. Both the direct and indirect methods for computing $\matr{Q}$ are interpreted in a method of moments context, providing the reader with an easy to implement recipe for evaluating group delays, renormalized energies, and frequency sensitivities of scattering parameters w.r.t. frequency.  A connection between the direct and indirect methods for computing $\matr{Q}$ is established, yielding an alternative proof of WS relationship~\eqref{eq:WS_Main_Relation}. 
\item It leverages the direct and indirect methods for evaluating $\matr{Q}$ to show that average group delays incurred by fields that interact with antennas and/or scatterers are independent of both the basis for expressing  incoming waves as well as the location of the spatial origin. It furthermore introduces a scheme for evaluating the frequency derivative of scattering parameters in terms of group delays of WS modes. Applied to single-port antennas, the scheme yields new insights into the conditions under which the celebrated Yaghian-Best-Vandenbosch equation for the magnitude of the frequency derivative of an antenna's input impedance yields accurate results \cite{Best_2005}\cite{VDB_2010}.

\end{itemize}

This paper is organized as follows. Section \ref{sec:computational_framework} describes the radiating and scattering systems under consideration and defines their scattering matrices. Sections \ref{sec:IE_Form} through \ref{sec:WSModes} detail the paper's principal contributions summarized above. Section \ref{sec:numerical_results} applies the proposed methods to the computation of frequency sensitivities of antenna input impedances and radiation patterns, as well as scatterer radar cross sections. Section \ref{sec:conclusions} presents conclusions and avenues for future research.


\vskip 10pt

\noindent{\underline{Notation}}. This paper borrows most notation from its precursor \cite{TAP_1}.  Specifically:
\begin{itemize}
\item $\,^\dag$, $\,^T$, $\,^*$, and $'$ represent adjoint, transpose, complex conjugate, and angular frequency derivative $(d/d\omega)$ operations, respectively. The trace of a square matrix $\matr{X}$, i.e. the sum of $\matr{X}$'s diagonal elements, is denoted by $\Tr \left( \matr{X} \right)$.
\item A time dependence $e^{j\omega t}$ with $\omega = 2 \pi f$ is assumed and suppressed. Additionally, the
free-space permeability, permittivity, impedance, wavelength, and wavenumber are denoted by
$\mu$, $\varepsilon$, $Z= \sqrt{\mu/\varepsilon}$, $\lambda$, and $k =\omega \sqrt{\mu \varepsilon}$, respectively.
\item Free-space electric fields often are expanded in terms of Vector Spherical Wave (VSW) functions $\vect{\cal B}_{\tau l m}(\vr)$. $\vect{\cal B}_{\tau l m}(\vr) = \vect{\cal I}_{\tau l m}(\vr)$, $\vect{\cal O}_{\tau l m}(\vr)$, and $\vect{\cal W}_{\tau l m}(\vr) = \vect{\cal I}_{\tau l m}(\vr) + \vect{\cal O}_{\tau l m}(\vr)$ when modeling \emph{incoming}, \emph{outgoing}, and \emph{standing} waves; here $\tau = 1,2$ denotes polarization, and $l$ and $m$ are modal indices.
While VSWs $\vect{\cal I}_{\tau l m}(\vr)$ and $\vect{\cal O}_{\tau l m}(\vr)$ are singular at the origin, $\vect{\cal W}_{\tau l m}(\vr)$ is regular throughout space. Associated magnetic fields are expanded in terms of the VSWs $\tilde{\vect{\cal B}}_{\tau l m} (\vr) = \frac{1}{k} \nabla \times \vect{\cal B}_{\tau l m}(\vr) = j (-1)^{\tau + 1} \vect{\cal B}_{\bar{\tau} l m}$ where $\bar{1} = 2$ and $\bar{2} = 1$.
The large argument approximations (i.e. ``far-fields'') of $\vect{\cal B}_{\tau l m}(\vr)$ and $\tilde{\vect{\cal B}}_{\tau l m}(\vr)$ are denoted by $\vect{\cal B}_{\tau l m, \myparallel}(\vr)$ and $\tilde{\vect{\cal B}}_{\tau l m, \myparallel}(\vr)$, respectively.
Explicit expressions for the VSWs, their large argument approximations, and the related Vector Spherical Harmonics (VSHs) $\vect{\cal X}_{\tau lm}(\theta,\phi)$ are provided in Appendix~\ref{appendix:VSH_VSW}.
\end{itemize}

%% file: graphics/radiating_system_graphics2.tex
\usetikzlibrary{decorations.pathmorphing}
\begin{figure}
\centering
\begin{circuitikz}[scale=1.2, every node/.style={scale=1.2}]

\draw (0,0) circle (3);
\draw[dashed] (0,0) circle (2.2);
\draw[dashed, ->] (0,0) -- (315:2.2);
\node at (315:1.6) [below] {$a$};
\draw[fill=black] (0,0) circle (0.02);

\draw[dashed, ->] (0,0) -- (45:3);
\node at (47:2.8) [left] {\small $R$};

\begin{scope}[shift = {(-1,0)}]

\begin{scope}[shift = {(0,0.9)},scale=1.5]
\draw [fill=brown!40] (0,0.2) -- (0.4,0.2) -- (1.0,0.6) -- (1.0,-0.6) -- (0.4,-0.2) -- (0,-0.2) -- (0,0.2);

\draw[fill=yellow!20] (0,0.2) -- (-0.4,0.2) -- (-0.4,-0.2) -- (0,-0.2);
\draw[red,very thick] (0,-0.2) -- (0,0.2);
\draw [->,thick] (0,0.12) -- (0.25,0.12);
\node at (0.38, 0.2) [above] {\small $\vect{\cal E}_{1,\myparallel}^{i}$};
\draw [<-,thick] (-0.25,0.08) -- (0.0,0.08);
\node at (-0.09, 0.2) [above] {\small $\vect{\cal E}_{1,\myparallel}^{o}$};
\draw[->] (-0.08,0.3) -- (-0.08, 0.1);
\draw[->] (0.27,0.31) -- (0.14, 0.17);

\draw [<-,thick] (-0.1,-0.1) -- (0.0,-0.1);
\node at (-0.1,-0.1) [left] {$\unit{w}$};

\draw [ultra thick] (-0.2,0) -- (0.4,0);
\node at (0.6,0) {\footnotesize \begin{tabular}{c} \small antenna \\ \small \# 1  \end{tabular}};

\node at (0.28,-0.45) {\small $d\Omega_{g,1}$};
\draw [->] (0.15,-0.35) -- (0,-0.22);
\end{scope}

\begin{scope}[shift = {(0,-1.2)},scale=1.5]
\draw [fill=brown!40] (0,0.2) -- (0.4,0.2) -- (1.0,0.6) -- (1.0,-0.6) -- (0.4,-0.2) -- (0,-0.2) -- (0,0.2);

\draw[fill=yellow!20] (0,0.2) -- (-0.4,0.2) -- (-0.4,-0.2) -- (0,-0.2);
\draw[red,very thick] (0,-0.2) -- (0,0.2);
\draw [->,thick] (0,0.12) -- (0.25,0.12);
\node at (0.38, 0.2) [above] {\small $\vect{\cal E}_{2,\myparallel}^{i}$};
\draw [<-,thick] (-0.25,0.08) -- (0.0,0.08);
\node at (-0.09, 0.2) [above] {\small $\vect{\cal E}_{2,\myparallel}^{o}$};
\draw[->] (-0.08,0.3) -- (-0.08, 0.1);
\draw[->] (0.27,0.31) -- (0.14, 0.17);

\draw [<-,thick] (-0.1,-0.1) -- (0.0,-0.1);
\node at (-0.1,-0.1) [left] {$\unit{w}$};

\draw [ultra thick] (-0.2,0) -- (0.4,0);
\node at (0.6,0) {\footnotesize \begin{tabular}{c} \small antenna \\ \small \# 2  \end{tabular}};

\node at (0.28,-0.45) {\small $d\Omega_{g,2}$};
\draw [->] (0.15,-0.35) -- (0,-0.22);
\end{scope}

\end{scope}

\node at (1.1,1.6) {\small $d\Omega_s$};
\draw[->] (1.1,1.35) -- (1.3, 1);
\draw[->] (1.1,1.35) -- (0.5, 0.7);

\begin{scope}[scale = 0.9, shift={(1.4,0)}, xscale = 0.5]
\draw [black, fill = brown!40] plot [smooth cycle] coordinates {(1.9,0)  (1.6, 0.9) (1.0, 1.2) (0.5, 1.25) (0,1) (-0.5,0.8) (-0.7, 0.65) (-1,0) (-0.8, -0.6) (-0.5,-0.8) (0,-0.9) (0.5, -1)  (1.5, -0.8)};
\end{scope}
\node at (1.4,0) {\small \begin{tabular}{c} \small PEC \\ \small scatterer \end{tabular}};

\foreach \th in {60, 120, ..., 360}
{
\draw [->] (\th:3.2) -- (\th:3);
\draw [->] (\th-10:3) -- (\th-10:3.2);
}
\node at (52:3.25) [right] {\small $\vect{\cal E}_{p,\myparallel}^o$};
\node at (67:3.25) [right] {\small $\vect{\cal E}_{p,\myparallel}^i$};

\node at (58:3.6) [right] {\small $p>2$};

\node at (-2.4,1.55) [right] {\small $d\Omega_f$};
\draw[->] (-2.3,1.5) -- (-2.5,1.5);


\end{circuitikz}
\caption{Example setup with two antennas and a PEC scatterer. The system is excited by guided and free-space TEM waves that originate at the antenna terminals (here the coaxial apertures $d\Omega_{g,1}$ and $d\Omega_{g,2}$) and free-space ports ($d\Omega_f$), respectively.}
\label{fig:mixed_scatterer}
\end{figure}
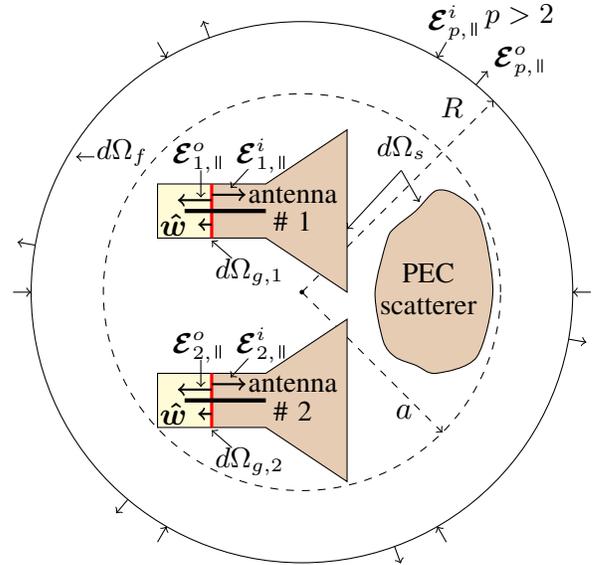

%% file: theory.tex
\section{Computational Framework: Incoming and Outgoing Fields, Incident and Scattered Fields, and their Integral Equation-based Computation}
\label{sec:computational_framework}

This section describes the electromagnetic systems under consideration. It also defines incoming and outgoing fields that feature in the definition of their scattering matrices, as well as closely related incident and scattered fields that permit these matrices' integral equation-based characterization. 

\subsection{Setup and Incoming Fields}

Consider a system composed of PEC antennas and/or scatterers with combined surface $d\Omega_s$ that resides in free space (Fig.~\ref{fig:mixed_scatterer}).  The system is excited by guided and free-space waves described in terms of incoming fields defined on port surfaces.

\begin{enumerate}
    \item \underline{Guided waves}. The antennas are excited by $M_g$  guided waves indexed $p = 1,\hdots,M_g$.   Wave $p$  propagates on an air-filled, closed and lossless, two-conductor TEM transmission line with planar port surface $d\Omega_{g,p}$.  For $\vr$  near $d\Omega_{g,p}$, let $\{\vect{\cal E}_{p,\myparallel}^i(\vr), \vect{\cal H}_{p,\myparallel}^i(\vr)\}$ denote the incoming TEM guided wave  
\begin{subequations}
\begin{align}
\vect{\cal E}_{p,\myparallel}^i(\vr) &= \sqrt{Z} e^{jkw} \vect{\cal X}_{p}(u,v) \label{eq:Ei_antenna} \\
\vect{\cal H}_{p,\myparallel}^i(\vr) &= \frac{1}{\sqrt{Z}} e^{jkw} \left(-\unit{w} \times \vect{\cal X}_{p}(u,v) \right)\,.
\label{eq:Hi_antenna}
\end{align}
\end{subequations}
Here $(u,v,w = 0)$ parametrizes a surface containing $d\Omega_g = \cup_{p=1}^{M_g} d\Omega_{g,p}$.
The $(u,v,w)$ coordinate system is locally Cartesian near each $d\Omega_{g,p}$, and $w$ increases in the direction of the unit vector $\unit{w} \perp d\Omega_{g,p}$, which points away from the antenna. The mode profiles $\vect{\cal X}_{p}(u,v)$ are real and assumed normalized, i.e.
\begin{align}
    \int_{d\Omega_{g,p}} \abs{\vect{\cal X}_{p}(u,v)}^2 du dv = 1\,,
\end{align}
meaning that $\{\vect{\cal E}_{p,\myparallel}^i(\vr), \vect{\cal H}_{p,\myparallel}^i(\vr) \}$ carries unit power. In what follows, the port surfaces $d\Omega_{g,p}$ are assumed electrically small and sufficiently removed from the physical antenna terminals, implying their fields do not contain any higher order modes.
The mode profile for a given $d\Omega_{g,p}$ can be constructed using the procedure detailed in \cite[Appendix A]{TAP_1}. Note: while this paper devotes significant attention to guided excitations of $d\Omega_s$, the proposed methods also apply to pure scatterers, i.e. when $M_g = 0$.

\item \underline{Free-space waves}. The antennas and/or scatterers also are excited by $M_f$ free-space waves indexed $p = M_g + 1, \hdots, M$ with $M= M_g + M_f$. Let $d\Omega_f$ denote the surface of an origin-centered sphere of radius $R \gg \mathrm{max} \{ a, \lambda \}$, where $a$ is the radius of a sphere circumscribing $d\Omega_s$.
For $\vr$ near $d\Omega_f$, let $\{\vect{\cal E}_{p,\myparallel}^i(\vr), \vect{\cal H}_{p,\myparallel}^i(\vr) \}$ denote the incoming free-space wave
\begin{subequations}
\begin{align}
    \vect{\cal E}_{p,\myparallel}^i(\vr) &= \sqrt{Z} \vect{\cal I}_{p,\myparallel}(\vr) \label{eq:Ei_freespace_1} \\
    &= \sqrt{Z} \frac{e^{jkr}}{r} \vect{\cal X}_{p}(\theta,\phi) \nonumber \\
    \vect{\cal H}_{p,\myparallel}^i(\vr) &= \frac{-\unit{r} \times \vect{\cal E}_{p,\myparallel}^i(\vr)}{Z} \label{eq:Hi_freespace_1}\\
    &= \frac{j}{\sqrt{Z}} \tilde{\vect{\cal I}}_{p,\myparallel}(\vr) \nonumber  \\
    &= \frac{1}{\sqrt{Z}} \frac{e^{jkr}}{r} \left( -\unit{r} \times \vect{\cal X}_{p}(\theta,\phi)\right)\,.  \nonumber
\end{align}
\end{subequations}
Here and in what follows, $p$ maps to the triplet $(\tau, l, m)$ where $\tau$ denotes polarization ($\tau = 1 \rightarrow \text{TE}$ to $r$, $\tau = 2 \rightarrow \text{TM to } r$), $l = 1,\hdots,l_{\text{max}}$, $m = -l,\hdots,l$, and $l_{\text{max}} = ka + c (ka)^{1/3}$ where $2 < c<4$~\cite{Wiscombe1980}. This choice for $l_{\text{max}}$ is warranted by the observation that incoming fields with $l > l_{\text{max}}$ do not appreciably couple to the antennas and/or scatterers and implies $M_f = {\cal O}((ka)^2)$. 
Irrespective of $\tau$, the fields $\{ \vect{\cal E}_{p,\myparallel}^i(\vr), \vect{\cal H}_{p,\myparallel}^i(\vr)\}$ are TEM near $d\Omega_f$. 
Note that the VSHs $\vect{\cal X}_{p}(\theta,\phi)$ obey the orthonormality relation
\begin{align}
     \int_{0}^{2\pi} \int_{0}^{\pi} \vect{\cal X}_{p}(\theta,\phi) \cdot \vect{\cal X}_{p'}^*(\theta,\phi) \sin \theta d\theta d\phi = \delta_{pp'} \,,
\end{align}
implying free-space wave $\{\vect{\cal E}_{p,\myparallel}^i(\vr), \vect{\cal H}_{p,\myparallel}^i(\vr) \}$ carries unit power across $d\Omega_f$.
\end{enumerate}

In what follows, let $d\Omega_{g,f} = d\Omega_g \cup d\Omega_f$  (union of all port surfaces), $d\Omega_{g,s} = d\Omega_g \cup d\Omega_s$  (union of antenna port and PEC surfaces), and  $d\Omega = d\Omega_s \cup d\Omega_g \cup d\Omega_f$ (union of all surfaces).  Also, let $\Omega$  denote the volume bounded by  $d\Omega$.

\subsection{Scattering Matrix and Outgoing Fields}

Total fields for $\vr$ near port surfaces $d\Omega_{g,f}$  consist of incoming and outgoing TEM waves.  Outgoing waves $\{\vect{\cal E}_{p,\myparallel}^o(\vr), \vect{\cal H}_{p,\myparallel}^o(\vr) \}$  generated in response to  
$\{\vect{\cal E}_{p,\myparallel}^i(\vr), \vect{\cal H}_{p,\myparallel}^i(\vr) \}$
typically involve all modes, with modal contributions weighed by scattering coefficients.  Near the antenna ports,
\begin{subequations}
\begin{align}
    \vect{\cal E}_{p,\myparallel}^o(\vr) &= \sum_{t=1}^{M_g} \matr{S}_{tp} \sqrt{Z} e^{-jkw} \vect{\cal X}_t(u,v) \label{eq:Eo_antenna}\\
    \vect{\cal H}_{p,\myparallel}^o(\vr) &=  \sum_{t=1}^{M_g} \matr{S}_{tp} \frac{1}{\sqrt{Z}} e^{-jkw} \left(\unit{w} \times \vect{\cal X}_t(u,v) \right)\,. \label{eq:Ho_antenna}
\end{align}
\end{subequations}
Likewise, near the free-space port $d\Omega_f$,
\begin{subequations}
\begin{align}
    \vect{\cal E}_{p,\myparallel}^o(\vr) &= \sum_{t=M_g+1}^{M} \matr{S}_{tp} \sqrt{Z} \vect{\cal I}_{t,\myparallel}^*(\vr) \label{eq:Eo_freespace} \\
    \vect{\cal H}_{p,\myparallel}^o(\vr) &= \frac{\unit{r} \times \vect{\cal E}_{p,\myparallel}^o(\vr)}{Z} \nonumber \\
    &= \sum_{t=M_g+1}^{M} \matr{S}_{tp} \frac{j}{\sqrt{Z}} \tilde{\vect{\cal I}}_{t,\myparallel}^*(\vr)\,. \label{eq:Ho_freespace}
\end{align}
\end{subequations}
Here, $\vect{\cal I}^*_{\tau l m}(\vr) = (-1)^{l + \tau - m} \vect{\cal O}_{\tau l (-m)}$ is an outgoing VSW. Note that the sums on the RHSs of \eqref{eq:Eo_freespace}--\eqref{eq:Ho_freespace} include contributions from all $\vect{\cal O}_{t,\myparallel}(\vr)$ for $t = M_g+1, \hdots, M$. The above construction, including the ``pairing'' of $\matr{S}_{tp}$ with outgoing wave $\vect{\cal I}_{t,\myparallel}^*(\vr)$, however guarantees that the $M \times M$ scattering matrix $\matr{S} = \begin{bmatrix} \matr{S}_{tp}; t,p = 1,\hdots, M \end{bmatrix}$ not only is unitary ($ \matr{S}^\dag \matr{S} = \matr{I}$) but also symmetric ($\matr{S}=\matr{S}^T$)~\cite{Pozar_2005}.

The total fields associated with incoming wave $1 \le p \le M$ for $\vr$ near the ports $d\Omega_{g,f}$ are
\begin{subequations}
\begin{align}
    \vect{\cal E}_{p,\myparallel}(\vr) &= \vect{\cal E}_{p,\myparallel}^i(\vr) + \vect{\cal E}_{p,\myparallel}^o(\vr) \label{eq:Eio_total}\\
    \vect{\cal H}_{p,\myparallel}(\vr) &= \vect{\cal H}_{p,\myparallel}^i(\vr) + \vect{\cal H}_{p,\myparallel}^o(\vr)\,. \label{eq:Hio_total}
\end{align}
\end{subequations}
Let $\{ \vect{\cal E}_p(\vr), \vect{\cal H}_p(\vr) \}$ denote the fields that exist throughout $\Omega$ when the unit-power incoming field $\{\vect{\cal E}_{p,\myparallel}^i(\vr), \vect{\cal H}_{p,\myparallel}^i(\vr) \}$ enters $\Omega$ via $d\Omega_{g,f}$ while all ports are matched. These fields are unique extensions of 
$\{ \vect{\cal E}_{p,\myparallel}(\vr), \vect{\cal H}_{p,\myparallel}(\vr) \}$ on $d\Omega_{g,f}$ inside $\Omega$ in the presence of a vanishingly small loss \cite{Harrington_2001}. Note that decompositions \eqref{eq:Eio_total}--\eqref{eq:Hio_total} of $\{\vect{\cal E}_{p}(\vr), \vect{\cal H}_{p}(\vr) \}$ into incoming and outgoing waves generally speaking do not apply away from $d\Omega_{g,f}$.

\input{graphics/radiating_antenna_feed_model}

\subsection{Integral Equation-based Characterization of Total Fields $\{ \vect{\cal E}_p(\vr), \vect{\cal H}_p(\vr) \}$}

\label{sec:IE_Form_1}

The above decomposition of $\{ \vect{\cal E}_p(\vr), \vect{\cal H}_p(\vr) \}$ into \emph{incoming} and \emph{outgoing} waves, while useful to define the scattering matrix $\matr{S}$, does not lend itself well to computation. 

To compute $\{\vect{\cal E}_p(\vr), \vect{\cal H}_p(\vr) \}$, 
consider their decomposition into \emph{incident} and \emph{scattered} fields,
\begin{subequations}
\begin{align}
    \vect{\cal E}_p(\vr) &= \vect{\cal E}_p^{\text{inc}}(\vr) + \vect{\cal E}_p^{\text{sca}}(\vr) \label{eq:Eis_total} \\
    \vect{\cal H}_p(\vr) &= \vect{\cal H}_p^{\text{inc}}(\vr) + \vect{\cal H}_p^{\text{sca}}(\vr) \label{eq:His_total}
\end{align}
\end{subequations}
where the scattered fields $\{\vect{\cal E}_p^{\text{sca}}(\vr), \vect{\cal H}_p^{\text{sca}}(\vr) \}$ are generated by the surface current density
\begin{align}
    \vect{\cal J}_p(\vr) = \unit{n} \times \vect{\cal H}_p(\vr) \quad \quad \quad \vr \in d\Omega_s
\end{align}
where $\unit{n}$ is the outward pointing normal to  $d\Omega_s$, and
\begin{subequations}
\begin{align}
    \vect{\cal E}_p^{\text{sca}}(\vr) 
    &=  \vect{\cal L} \left[\vect{\cal J}_p\right](\vr) \label{eq:Es_1} \\
    \vect{\cal H}_p^{\text{sca}}(\vr) &= \vect{\cal K} \left[\vect{\cal J}_p\right](\vr)\,. \label{eq:Hs_1} 
\end{align}
\end{subequations}
Here, the operators $\vect{\cal L} \left[\vect{\cal J}\right](\vr)$ and $\vect{\cal K} \left[\vect{\cal J}\right](\vr)$ are 
\begin{subequations}
\begin{align}
    \vect{\cal L} \left[\vect{\cal J}\right](\vr) &= -j\omega \mu \int_{d\Omega_{g,s}} \bigg[ G(\vr,\vrp) \vect{\cal J}(\vrp) \nonumber \\
    &\quad   + \frac{1}{k^2} \nabla G(\vr,\vrp) \nabla' \cdot \vect{\cal J}(\vrp)\bigg] d\vrp \label{eq:LOperator} \\
    \vect{\cal K} \left[\vect{\cal J}\right](\vr)  &=  \nabla \times \int_{d\Omega_{g,s}}  G(\vr,\vrp)  \vect{\cal J}(\vrp) d\vrp
\end{align}
\end{subequations}
where
\begin{align}
    G(\vr,\vrp) = \frac{e^{-jk\abs{\vr-\vrp}}}{4\pi\abs{\vr - \vrp}} \,.
\end{align}
The above definition of $\vect{\cal J}_p(\vr)$ assumes that $d\Omega_s$ is closed, which always can be achieved by assigning open antenna or scatterer surfaces a finite thickness.

The incident fields $\{\vect{\cal E}_p^{\text{inc}}(\vr) = \vect{\cal E}_p(\vr) - \vect{\cal E}_p^{\text{sca}}(\vr), \vect{\cal H}_p^{\text{inc}}(\vr) = \vect{\cal H}_p(\vr) - \vect{\cal H}_p^{\text{sca}}(\vr) \}$ arise when the incoming fields $\{\vect{\cal E}_p^{i}(\vr), \vect{\cal H}_p^i(\vr) \}$ are injected into $d\Omega_{g,f}$ in the absence of the antenna and scatterer surfaces $d\Omega_s$, i.e. when $\vect{\cal J}_p(\vr) = 0$.

\begin{enumerate}
    \item \underline{Guided waves.} In the absence of antennas and scatterers, the incoming traveling waves $\{\vect{\cal E}_p^i(\vr), \vect{\cal H}_p^i(\vr) \}$ for $p \le M_g$ reflect upon reaching the open antenna terminal, producing standing waves that for $\vr \in d\Omega_{g,p}$ equate to
\begin{subequations}
\begin{align}
    \vect{\cal E}_p^{\text{inc}}(\vr) &= 2 \vect{\cal E}_p^i(\vr) \label{eq:Einc_antenna} \\
    \vect{\cal H}_p^{\text{inc}}(\vr) &= 0\,. \label{eq:Hinc_antenna} 
\end{align}
\end{subequations}
\item \underline{Free-space waves}. In the absence of antennas and scatterers, the incoming traveling waves $\{\vect{\cal E}_p^i(\vr), \vect{\cal H}_p^i(\vr) \}$ for $p > M_g$ ``reflect'' upon reaching the origin, producing standing waves that for $\vr \in \Omega$ equate to
\begin{subequations}
\begin{align}
    \vect{\cal E}_p^{\text{inc}}(\vr) &= \sqrt{Z} \vect{\cal W}_p(\vr) \label{eq:Einc_freespace}\\
    &= \sqrt{Z} \left(\vect{\cal I}_p(\vr) + \vect{\cal O}_p(\vr) \right) \nonumber \\
    \vect{\cal H}_p^{\text{inc}}(\vr) &= \frac{j}{\sqrt{Z}} \tilde{\vect{\cal W}}_p(\vr) \label{eq:Hinc_freespace1}\\
    &= \frac{j}{\sqrt{Z}} \left( \tilde{\vect{\cal I}}_p(\vr) + \tilde{\vect{\cal O}}_p(\vr) \right)\,. \nonumber 
\end{align}
\end{subequations}
Mathematically, the singular incoming VSWs $\vect{\cal I}_p(\vr)$ and $\tilde{\vect{\cal I}}_p(\vr)$ in \eqref{eq:Ei_freespace_1}--\eqref{eq:Hi_freespace_1} are paired with outgoing VSWs $\vect{\cal O}_p(\vr)$ and $\tilde{\vect{\cal O}}_p(\vr)$ so that their sums $\vect{\cal W}_p(\vr)$ and $\tilde{\vect{\cal W}}_p(\vr)$ carry zero net power across $d\Omega_f$ and are regular (i.e. source-free) at the origin.
\end{enumerate}

The incident and scattered electric fields $\vect{\cal E}_p^{\text{inc}}(\vr)$ and  $\vect{\cal E}_p^{\text{sca}}(\vr)$, and current density $\vect{\cal J}_p(\vr)$ satisfy the following electric field integral equation (EFIE) on the PEC surface and waveguide apertures:
\begin{align}
     \left[ \vect{\cal L}\left[\vect{\cal J}_p\right](\vr) - Z_{\text{gap}}(\vr) \vect{\cal J}_p(\vr) \right]_{\text{tan}}  = -  \vect{\cal E}_p^{\text{inc}}(\vr) \big|_{\text{tan}}\,. \label{eq:MoM_IE_operator}
\end{align}
Here $\vect{\cal F (\vr)}\big |_{\text{tan}} = -\unit{n} \times \unit{n} \times \vect{\cal{F}}(\vr)$ and
\begin{align}
    Z_{\text{gap}}(\vr) = \begin{cases}
    Z & \vr \in d\Omega_g \\
    0 & \vr \in d\Omega_s
    \end{cases}\,. \label{eq:IE_BC}
\end{align}
On the PEC antenna and scatterer surface $d\Omega_s$, \eqref{eq:MoM_IE_operator}--\eqref{eq:IE_BC} impose $\vect{\cal E}_p(\vr)\big|_{\text{tan}} = 0$.
In the waveguide aperture $d\Omega_g$, \eqref{eq:MoM_IE_operator}--\eqref{eq:IE_BC} impose $\vect{\cal E}_p(\vr)\big|_{\text{tan}} = Z\vect{\cal J}_p(\vr) = Z \left( \unit{w} \times \vect{\cal H}_p(\vr) \right)$, implying the presence of a matched load.

\subsection{Method of Moments Implementation}

EFIE \eqref{eq:MoM_IE_operator} can be solved via standard moment method algorithms using Rao-Wilton-Glisson (RWG) basis functions \cite{Chew2001fast}--\cite{RWG}. 
The practical implementation of the moment method algorithm oftentimes uses circuit analogs to model loads (second term on the LHS of \eqref{eq:MoM_IE_operator}) and the incident field for guided waves (eq. \eqref{eq:Einc_antenna}--\eqref{eq:Hinc_antenna}) (Fig. 2).  
Assume that transmission line $t$, $t = 1,\hdots, M_g$, has characteristic impedance $Z_t$ and is terminated in a matched load $R=Z_t$.  When $p \le M_{g}$, line $p$ is excited by a voltage source $V_p^{\text{inc}} = 2 \sqrt{Z_p}$, which launches a unit-power wave $\vect{\cal E}_p^i(\vr)$ that travels towards, and couples into, the attached antenna. The voltage source can be modeled as a delta-gap or magnetic frill \cite{Butler1973alternate}. In what follows, $V_p(t)$ and $I_p(t)$ denote the voltage and current in port $t\le M_g$ under excitation $p$; $I_p(t)$ can be computed as 
\begin{align}
    I_p(t) = \oint_{C_t} \vect{\cal J}_p(\vr) \cdot \unit{w} dl
\end{align}
where $C_t$ is an appropriately chosen boundary of $d\Omega_{g,t}$. Clearly, $V_p(t) = \delta_{pt} V_p^{\text{inc}} - Z_t I_p(t)$.

To solve EFIE \eqref{eq:MoM_IE_operator} with the method of moments, $\vect{\cal J}_p(\vr)$ is expanded as
\begin{align}
    \vect{\cal J}_{p}(\vr) = \sum_{n=1}^{N} J_{np} \vect{f}_n(\vr)\,, \label{eq:Jexpand_Basis1}
\end{align}
where $N$ is the total number of basis functions, i.e. edges on $d\Omega_{s}$, and $\vect{f}_n(\vr)$ is the $n$-th RWG basis function~\cite{RWG}. 
Substituting \eqref{eq:Jexpand_Basis1} into \eqref{eq:MoM_IE_operator} and applying the Galerkin testing procedure yields
\begin{align}
    \matr{V}_p = \matr{Z} \matr{J}_p \label{eq:MoM}
\end{align}
where $\matr{J}_p = \begin{bmatrix} J_{1p} &\hdots & J_{Np} \end{bmatrix}^T$ is the vector containing unknown expansion coefficients, $\matr{V}_p$ is the excitation vector, and $\matr{Z}$ is the $N\times N$ impedance matrix.
The $n$-th entry of $\matr{V}_p$ reads
\begin{align}
    \matr{V}_{np} &= -\int_{d\Omega_{g,s}} \vect{f}_n(\vr) \cdot \vect{\cal E}_{p}^{\text{inc}}(\vr) d\vr  \,,
\end{align}
which, using the circuit analog of the antenna excitation and \eqref{eq:Einc_freespace}, can be expressed as
\begin{align}
\matr{V}_{np} = \begin{cases} - 2 \sqrt{Z_p} \delta_{n \in \{\text{port}\,\, p\}} & p \le M_g \\  -\sqrt{Z} \int_{d\Omega_{s}} \vect{f}_n(\vr) \cdot \vect{\cal W}_{p}(\vr) d\vr & p > M_g \end{cases} \,,   
\label{eq:VMoM_Entries}
\end{align}
where it is assumed that RWG basis functions are normalized to carry a unit current across their defining edge. In \eqref{eq:VMoM_Entries}, $\delta_{C} = 1(0)$ if $C$ is true (false)  and set $\{\text{port } p \}$ contains all edges that define port $p$.
The $(m,n)$-th entry  of $\matr{Z}$ is given by
\begin{align}
    \matr{Z}_{mn} &= \int_{d\Omega_{g,s}} \vect{f}_m(\vr)  \cdot \big[ \vect{\cal L} \left[\vect{f}_n \right](\vr) - Z_{\text{gap}}(\vr) \vect{f}_n(\vr) \big] d\vr. 
\end{align}
Knowledge of $\vect{\cal J}_p(\vr)$ permits the computation of $\{ \vect{\cal E}_p^{\text{sca}}(\vr), \vect{\cal H}_p^{\text{sca}}(\vr) \}$   throughout $\Omega$  via \eqref{eq:Es_1}--\eqref{eq:Hs_1}, which together with the specification of the incident fields in \eqref{eq:Einc_antenna}--\eqref{eq:Hinc_freespace1} leads to a complete characterization of $\{\vect{\cal E}_p(\vr), \vect{\cal H}_p(\vr) \}$  throughout $\Omega$. 

Equation \eqref{eq:MoM} can be compactly expressed as
\begin{align}
    \matr{V} = \matr{Z} \matr{J}
    \label{eq:MoM2}
\end{align} 
where $\matr{V} = \begin{bmatrix} \matr{V}_1 & \hdots & \matr{V}_M \end{bmatrix}$ and $\matr{J} = \begin{bmatrix} \matr{J}_1 & \hdots & \matr{J}_{M} \end{bmatrix}$.

\section{Direct Computation of $\matr{Q}$ Via Integration of Energy-like Quantities}
\label{sec:IE_Form}

As alluded to in the introduction and demonstrated in \cite{TAP_1}, $\matr{Q}$'s defining equation~\eqref{eq:WS_Main_Relation} implies that its diagonal elements $\matr{Q}_{pp}$ represent average group delays experienced by incoming waves $\vect{\cal E}_p^i(\vr)$ as they interact with the system prior to exiting via its ports. Reference \cite{TAP_1} however also showed that $\matr{Q}$'s elements can be cast as volume integrals of renormalized energy-like quantities. This section presents a computationally efficient method for directly evaluating these integrals using surface integral operators that act on the incident electric fields and current densities for all excitations characterizing the scattering matrix.

\subsection{Volume Integral Expressions of the WS Time delay Matrix}
It was shown in \cite[Sec. III.C]{TAP_1} that elements of the WS time delay matrix can be expressed as renormalized energy-like overlap integrals involving the electric and magnetic fields that arise upon excitation of the system's ports\footnote{Equations  \eqref{eq:WS_Main_Relation} and \eqref{eq:Qeqn} are slight simplifications of those in \cite[Sec. III.C]{TAP_1} because waveguides here only support TEM waves whereas those in \cite{TAP_1} also supported non-TEM fields.}
\begin{align}
{\matr{Q}}_{qp} =& \frac{\varepsilon}{2}   \int_{\mathbb{R}^3} \left[   \vect{\cal E}_q^*(\vr) \cdot \vect{\cal E}_p(\vr) - \widehat{\vect{\cal E}}_{q,\myparallel}^*(\vr) \cdot \widehat{\vect{\cal E}}_{p,\myparallel}(\vr)
  \right] d\vr  \label{eq:Qeqn}  \\
  & + \frac{\mu}{2}  \int_{\mathbb{R}^3} \left[  \vect{\cal H}_q^*(\vr) \cdot \vect{\cal H}_p(\vr)-   \widehat{\vect{\cal H}}_{q,\myparallel}^*(\vr) \cdot \widehat{\vect{\cal H}}_{p, \myparallel}(\vr) \right] d\vr. \nonumber 
\end{align}
In the above equation,
\begin{subequations}
\begin{align}
    \widehat{\vect{\cal E}}_{p,\myparallel}(\vr) &= \delta_{p>M_g} \sqrt{Z} \vect{\cal I}_{p,\myparallel}(\vr) \nonumber \\
    &\quad \quad \quad +  \sum_{t=M_g+1}^{M} \matr{S}_{tp} \sqrt{Z} \vect{\cal I}_{t,\myparallel}^*(\vr) \label{eq:Ehat_1} \\
    &= \delta_{p>M_g} \sqrt{Z} \frac{e^{jkr}}{r} \vect{\cal X}_p(\theta,\phi) \nonumber \\
    &\quad \quad \quad + \sum_{t=M_g+1}^{M} \matr{S}_{tp} \sqrt{Z} \frac{e^{-jkr}}{r} \vect{\cal X}_{t}^*(\theta,\phi) \nonumber  \\
    \widehat{\vect{\cal H}}_{p,\myparallel}(\vr) &= \delta_{p>M_g}\frac{j}{\sqrt{Z}} \tilde{\vect{\cal I}}_{p,\myparallel}(\vr) \nonumber \\
    &\quad \quad \quad +  \sum_{t=M_g+1}^{M} \matr{S}_{tp} \frac{j}{\sqrt{Z}}  \tilde{\vect{\cal I}}_{t,\myparallel}^*(\vr) \label{eq:Hhat_1} \\
    &= \delta_{p>M_g} \frac{1}{\sqrt{Z}} \frac{e^{jkr}}{r} \left( - \unit{r} \times \vect{\cal X}_{p}(\theta,\phi) \right) \nonumber  \\
    &\quad  + \sum_{t=M_g+1}^{M} \matr{S}_{tp} \frac{1}{\sqrt{Z}} \frac{e^{-jkr}}{r} \left( \unit{r} \times \vect{\cal X}_{t}^*(\theta,\phi) \right) \,. \nonumber
\end{align}
\end{subequations}
Note that the quantities $\{ \widehat{\vect{\cal E}}_{p,\myparallel}(\vr), \widehat{\vect{\cal H}}_{p,\myparallel}(\vr) \}$
do not obey Maxwell's equations away from $d\Omega_f$. Rather, they extend the fields $\{\vect{\cal E}_{p,\myparallel}(\vr), \vect{\cal H}_{p,\myparallel}(\vr) \}$ 
that exist near the free-space port $d\Omega_f$ to all $\vr \in \Omega$. The first and second terms on the RHS of \eqref{eq:Ehat_1}--\eqref{eq:Hhat_1} represent incoming and outgoing waves near $d\Omega_f$; the former only exist if $p > M_g$. The presence of $\{ \widehat{\vect{\cal E}}_{p,\myparallel}(\vr), \widehat{\vect{\cal H}}_{p,\myparallel}(\vr) \}$ in \eqref{eq:Qeqn} renders the $\matr{Q}_{qp}$ integrals convergent.

Unfortunately, the direct computation of $\matr{Q}$ is no sinecure. While $\matr{Q}$'s entries in principle can be evaluated using the volume integral in \eqref{eq:Qeqn}, the computational cost of doing so would be exorbitant. 
Indeed, the computation of the integrand of \eqref{eq:Qeqn} calls for the evaluation of $\{\vect{\cal E}^{\text{inc}}_{p/q}(\vr) + \vect{\cal E}^{\text{sca}}_{p/q}(\vr), \vect{\cal H}^{\text{inc}}_{p/q}(\vr) + \vect{\cal H}^{\text{sca}}_{p/q}(\vr)\}$, which requires the evaluation of surface integrals \eqref{eq:Es_1}--\eqref{eq:Hs_1} for both $p$ and $q$, rendering integral \eqref{eq:Qeqn} effectively seven dimensional.

\subsection{Direct Surface Integral Computation of the WS Time Delay Matrix}

This section presents expressions for the entries of $\matr{Q}$ composed of surface integral operators acting on the incident electric fields and associated currents densities $\vect{\cal E}^{\text{inc}}_{p/q}(\vr)$ and $\vect{\cal J}_{p/q}(\vr)$, reducing the dimensionality of the integral to four.

Substituting the incident-scattered field decomposition \eqref{eq:Eis_total}--\eqref{eq:His_total} of $\{\vect{\cal E}_p(\vr), \vect{\cal H}_p(\vr)\}$ and $\{\widehat{\vect{\cal E}}_{p,\myparallel}(\vr), \widehat{\vect{\cal H}}_{p,\myparallel}(\vr)\}$ into \eqref{eq:Qeqn} yields
\begin{align}
\matr{Q}_{qp} = 
\matr{Q}_{qp}^{\text{inc,inc}} + \matr{Q}_{qp}^{\text{inc,sca}} + \matr{Q}_{qp}^{\text{sca,inc}} + \matr{Q}_{qp}^{\text{sca,sca}}
\label{eq:Qeqn_decompose_1}
\end{align}
where
\begin{align}
\matr{Q}_{qp}^{\alpha, \beta} =& \frac{\varepsilon}{2}   \int_{\mathbb{R}^3} \left[   \vect{\cal E}_q^{\alpha*}(\vr) \cdot \vect{\cal E}_p^\beta(\vr) - \widehat{\vect{\cal E}}_{q,\myparallel}^{\alpha*}(\vr)  \cdot \widehat{\vect{\cal E}}_{p,\myparallel}^\beta(\vr)
  \right] d\vr \label{eq:Qqp_s}  \\
 + \frac{\mu}{2}&  \int_{\mathbb{R}^3} \left[  \vect{\cal H}_q^{\alpha*}(\vr) \cdot \vect{\cal H}_p^\beta(\vr) -   \widehat{\vect{\cal H}}_{q,\myparallel}^{\alpha*}(\vr) \cdot \widehat{\vect{\cal H}}_{p, \myparallel}^\beta(\vr) \right] d\vr   \nonumber
\end{align}
and the superscripts $\alpha$ and $\beta$ are $\text{inc}$ or $\text{sca}$.
A lengthy and technical derivation presented in Appendix~\ref{App:QComputation} shows that the four terms in \eqref{eq:Qeqn_decompose_1} can be evaluated as 
\begin{align}
    \matr{Q}_{qp}^{\text{inc,inc}} &= 0 \label{eq:Qincinc_final}\\
\matr{Q}_{qp}^{\text{sca,inc}} &= \delta_{p> M_g } {\cal Q}^{\text{sca,inc}}\left(\vect{\cal E}_p^{\text{inc}}, \vect{\cal J}_q \right) \label{eq:Qscainc_final} \\
\matr{Q}_{qp}^{\text{inc,sca}} &= \left[\matr{Q}_{pq}^{\text{sca,inc}}\right]^*  \label{eq:Qincsca_final}\\
\matr{Q}_{qp}^{\text{sca,sca}}  
 &= \matr{Q}_{i,qp}^{\text{sca,sca}}  +  \matr{Q}_{d,qp}^{\text{sca,sca}}  \label{eq:Qscasca_final} \\
 \matr{Q}&_{i,qp}^{\text{sca,sca}} = {\cal Q}^{\text{sca,sca}}_{i}\left(\vect{\cal J}_q, \vect{\cal J}_p \right) \\
 \matr{Q}&_{d,qp}^{\text{sca,sca}} = {\cal Q}^{\text{sca,sca}}_{d}\left(\vect{\cal J}_q, \vect{\cal J}_p \right)
\end{align}
where
\begin{align}
    &{\cal Q}^{\text{sca,inc}}(\vect{\cal E}, \vect{\cal J}) = \frac{j}{2} \int_{d\Omega_{g,s}}  \vect{\cal E}'(\vr) \cdot \vect{\cal J}^*(\vr) d\vr \quad \quad \quad \quad \quad \label{eq:Operator_Qscainc_final}
\end{align}
\begin{subequations}
\begin{align}
    &{\cal Q}^{\text{sca,sca}}_i(\vect{\cal J}_a, \vect{\cal J}_b ) \nonumber \\
    &= \frac{Z k }{\omega} \int_{\Omega_{g,s}} \int_{\Omega_{g,s}} \bigg \{ \vect{\cal J}_b(\vrp) \cdot \vect{\cal J}_a^*(\vrq)\bigg[ \frac{\cos(k D)}{8 \pi D}    \nonumber \\
    &   \quad \quad - \frac{k \sin (k D)}{8 \pi} \bigg] +  \frac{1}{k^2} \nabla' \cdot \vect{\cal J}_b(\vrp) \nabla'' \cdot \vect{\cal J}_a^*(\vrq)    \nonumber \\
    & \quad \quad \quad \quad \left[ \frac{\cos(k D)}{8 \pi D} + \frac{k \sin (kD) }{8 \pi}  \right]  \bigg \} d\vrp d\vrq \label{eq:Operator_Qscasca_i_final} \\
    &= \frac{j}{4} \int_{d\Omega_{g,s}} \vect{\cal J}_a^*(\vr) \cdot  \big \{    \vect{\cal L}' \left[\vect{\cal J}_b\right](\vr) - \vect{\cal L}'^* \left[\vect{\cal J}_b\right](\vr)\} d\vr \label{eq:Operator_Qscasca_i_final_2}
\end{align}
\end{subequations}
\begin{subequations}
\begin{align}
&{\cal Q}^{\text{sca,sca}}_d(\vect{\cal J}_a, \vect{\cal J}_b ) \nonumber \\
&=  \frac{j Z k^3 }{8 \pi \omega} \int_{\Omega_{g,s}} \int_{\Omega_{g,s}} \big \{  \vect{\cal J}_b(\vrp) \cdot \vect{\cal J}_a^*(\vr) - \frac{1}{k^2}  \nabla' \cdot \vect{\cal J}_b(\vrp)  \nonumber \\
 & \quad    \nabla \cdot \vect{\cal J}_a^*(\vr) \big \} \left(\vrp + \vr \right) \cdot \unit{d} \left( \frac{\sin(kD)}{(kD)^2} - \frac{\cos (kD)}{kD} \right)  d\vrp d\vr\,. \label{eq:Operator_Qscasca_d_final}  \\
 &= \frac{j}{8} \sum_{t=M_g+1}^{M} \bigg \{ \int_{d\Omega_{g,s}} \vect{\cal E}_{t}^{\text{inc}}{'}(\vrp) \cdot \vect{\cal J}_b(\vrp) d\vrp  \nonumber \\
 & \quad \quad  \int_{d\Omega_{g,s}} \vect{\cal E}_t^{\text{inc}*}(\vrq) \cdot \vect{\cal J}_a^*(\vrq) d\vrq -  \int_{d\Omega_{g,s}} \vect{\cal E}_{t}^{\text{inc}}(\vrp)  \nonumber \\
 & \quad \quad \cdot \vect{\cal J}_b(\vrp) d\vrp    \int_{d\Omega_{g,s}} \vect{\cal E}_t^{\text{inc}*}{'}(\vrq) \cdot \vect{\cal J}_a^*(\vrq) d\vrq
 \bigg \}\,. \label{eq:Operator_Qscasca_d_final_2}
\end{align}
\end{subequations}
In the above equations, $D = \abs{\vr - \vrp}$, $\unit{d} = (\vr - \vrp)/D$,
\begin{align}
    &\vect{\cal L}'\left[\vect{\cal J}\right](\vr) \nonumber \\
    &= -j\omega \mu \int_{d\Omega_{g,s}} \bigg \{ \left( \frac{G(\vr,\vrp)}{\omega} + G'(\vr,\vrp) \right) \vect{\cal J}(\vrp) \label{eq:L_1prime} \\
    &\quad + \frac{1}{k^2} \left(\frac{-\nabla G(\vr,\vrp)}{\omega} + \nabla G'(\vr,\vrp) \right)  \nabla'\cdot \vect{\cal J}(\vrp) \bigg\} d\vrp\nonumber 
\end{align}
is the frequency derivative of $\vect{\cal L}\left[ \vect{\cal J} \right](\vr)$ with current density $\vect{\cal J}(\vr)$ kept constant, and $
        G'(\vr,\vrp) = \frac{-j}{4\pi} \sqrt{\mu \varepsilon} e^{-jk\abs{\vr-\vrp}}
$
is the frequency derivative of the free-space Green's function.

Equations \eqref{eq:Qeqn_decompose_1} through \eqref{eq:L_1prime} allow for the surface integral evaluation of matrix $\matr{Q}$ in \eqref{eq:WS_Main_Relation}. 
Specifically,  they illustrate that the volume integrals $\matr{Q}_{qp}$ of renormalized energy-like quantities involving \emph{total} electric and magnetic fields in \eqref{eq:Qeqn} can be expressed as the sum of  three distinct contributions: $\matr{Q}_{qp}^{\text{inc,sca}} + \matr{Q}_{qp}^{\text{sca,inc}} + \matr{Q}_{qp}^{\text{sca,sca}}$    ($\matr{Q}_{qp}^{\text{inc,inc}}$  always vanishes identically).   Each of these volume integrals can be cast as a surface integral.
\begin{itemize}
    \item $\matr{Q}_{qp}^{\text{sca,inc}}$, the renormalized volume integral in \eqref{eq:Qqp_s} with $\alpha = \text{sca}$  and $\beta = \text{inc}$  involving incident field $p$  and scattered field $q$  can be expressed as the surface integral \eqref{eq:Operator_Qscainc_final} of the frequency derivative of $\vect{\cal E}_p^{\text{inc}}(\vr)$  and current density $\vect{\cal J}_q^*(\vr)$.
    \item $\matr{Q}_{qp}^{\text{inc,sca}}$, the renormalized volume integral in \eqref{eq:Qqp_s} with $\alpha = \text{inc}$  and $\beta = \text{sca}$ involving scattered field $p$  and incident field $q$ can be expressed as the surface integral \eqref{eq:Operator_Qscainc_final} of the frequency derivative of $\vect{\cal E}_q^{\text{inc}*}(\vr)$  and current density $\vect{\cal J}_p(\vr)$.
    \item $\matr{Q}_{qp}^{\text{sca,sca}}$, the renormalized volume integral in \eqref{eq:Qqp_s} with $\alpha = \text{sca}$  and $\beta = \text{sca}$ involving scattered fields $p$  and $q$ can be expressed as the sum of two terms:
    \begin{itemize}
        \item $\matr{Q}_{i,qp}^{\text{sca,sca}}$, which can be expressed as the surface integral \eqref{eq:Operator_Qscasca_i_final} involving current densities  $\vect{\cal J}_p(\vr)$ and $\vect{\cal J}_q^*(\vr)$  with \emph{real and origin-independent ($i$) kernel}.
        \item $\matr{Q}_{d,qp}^{\text{sca,sca}}$, which can be expressed as the surface integral \eqref{eq:Operator_Qscasca_d_final} involving current densities  $\vect{\cal J}_p(\vr)$ and $\vect{\cal J}_q^*(\vr)$  with \emph{imaginary \emph{and} origin-dependent ($d$) kernel}.
    \end{itemize}
\end{itemize}
Importantly, the surface integral expressions for $\matr{Q}_{qp}^{\text{inc,sca}}$, $\matr{Q}_{qp}^{\text{sca,inc}}$, $\matr{Q}_{i,qp}^{\text{sca,sca}}$ and $\matr{Q}_{d,qp}^{\text{sca,sca}}$ only require knowledge of currents $\vect{\cal J}_p(\vr)$  and $\vect{\cal J}_q(\vr)$, not their frequency derivatives.   In other words, they permit the evaluation of $\matr{Q}(\omega)$  from the solution of the radiation/scattering problem at frequency $\omega$. 

Expressions \eqref{eq:Operator_Qscasca_i_final} and \eqref{eq:Operator_Qscasca_d_final} for $\matr{Q}_{i,qp}^{\text{sca,sca}}$  and $\matr{Q}_{d,qp}^{\text{sca,sca}}$ generalize formulas previously introduced for energy stored by single port antennas, to multiport systems subject to external excitations.
Specifically, when $\vect{\cal J}_a = \vect{\cal J}_b$, ${\cal Q}_i^{\text{sca,sca}}(\vect{\cal J}_a, \vect{\cal J}_b)$ reduces to the origin-independent expression for stored energy developed by Vandenbosch \cite{VDB_2010}, while ${\cal Q}_d^{\text{sca,sca}}(\vect{\cal J}_a, \vect{\cal J}_b)$ agrees with the correction term introduced by Gustafsson \cite{Gustafsson_2015} to account for the renormalized energy’s dependence on the antenna’s position w.r.t. the origin.
Finally, note that the expression for $\matr{Q} = \matr{Q}_{qp}^{\text{inc,sca}} + \matr{Q}_{qp}^{\text{sca,inc}} + \matr{Q}_{qp}^{\text{sca,sca}}$ cast in terms of surface integrals \eqref{eq:Operator_Qscainc_final} through \eqref{eq:Operator_Qscasca_d_final} is manifestly self-adjoint.

\subsection{Method of Moments Implementation}

The above equations can be trivially implemented in the method of moments. Substituting \eqref{eq:Jexpand_Basis1} into \eqref{eq:Qscainc_final} yields
\begin{align}
    {\matr{Q}}^{\text{sca,inc}} &= \matr{J}^\dag \widetilde{\matr{Q}}^{\text{sca,inc}}
    \label{eq:Qscainc_MoM_Final}
\end{align}    
where
\begin{align}
 \widetilde{\matr{Q}}^{\text{sca,inc}}_{np} &= \delta_{p> M_g } {\cal Q}^{\text{sca,inc}} (\vect{\cal E}_p^{\text{inc}}, \vect{f}_n) \nonumber \\
 &= \frac{-j}{2}\matr{V}_{np}' \,. 
\end{align}
Likewise, substituting \eqref{eq:Jexpand_Basis1} into \eqref{eq:Qscasca_final} yields
\begin{align}
    \matr{Q}^{\text{sca,sca}}  &= \matr{J}^\dag \left(\widetilde{\matr{Q}}_{i}^{\text{sca,sca}} + \widetilde{\matr{Q}}_{d}^{\text{sca,sca}}  \right) \matr{J} \label{eq:Qscasca_MoM_Final}
\end{align}
where the entries of the purely real and origin-independent
$\widetilde{\matr{Q}}_{i}^{\text{\text{sca,sca}}}$ and purely imaginary and origin dependent $\widetilde{\matr{Q}}_{d}^{\text{sca,sca}}$
are
\begin{subequations}
\begin{align}
    \widetilde{\matr{Q}}_{i,mn}^{\text{\text{sca,sca}}} &= {\cal Q}^{\text{sca,sca}}_i(\vect{f}_m, \vect{f}_n) \nonumber \\
    &= \frac{j}{4}\left(\matr{Z}_{mn}'-\matr{Z}'^*_{mn} \right) \nonumber \\
    &= -\frac{1}{2} \Im{\matr{Z}_{mn}'}  \\
    \widetilde{\matr{Q}}_{d,mn}^{\text{sca,sca}} &= {\cal Q}^{\text{sca,sca}}_d(\vect{f}_m, \vect{f}_n) \nonumber \\
    &= \frac{j}{8} \sum_{t=M_g+1}^{M} \left( \matr{V}_{mt}^* \matr{V}'_{nt} - \left(\matr{V}_{mt}^*\right)' \matr{V}_{nt} \right)\,. 
\end{align}
\end{subequations}
In the above expressions, matrices $\matr{Z}'$ and $\matr{V}'$ are the frequency derivatives of $\matr{Z}$ and $\matr{V}$ with entries
\begin{subequations}
\begin{align}
\matr{Z}_{mn}' &=  \int_{d\Omega_{g,s}} \vect{f}_m(\vr) \cdot \vect{\cal L}'\left[\vect{f}_n\right](\vr) d\vr \\
\matr{V}_{mn}' &= -\int_{d\Omega_{g,s}} \vect{f}_m(\vr) \cdot \vect{\cal E}_{n}^{\text{inc}}{'}(\vr) d\vr\,.
\end{align}
\end{subequations}
Inserting \eqref{eq:Qscainc_MoM_Final} and \eqref{eq:Qscasca_MoM_Final} into \eqref{eq:Qeqn_decompose_1} yields the following discrete energy-based expression for $\matr{Q}$:
\begin{align}
    \matr{Q} &= -\frac{j}{2} \matr{J}^\dag \matr{V}' +  \frac{j}{2}\matr{V}'^\dag \matr{J}  \label{eq:Q_MoM_Final} \\
    & \quad  - \frac{1}{2} \matr{J}^\dag \left(  \Im{\matr{Z}'} - \frac{j}{4} \left[ \matr{V}^* \matr{V}^T{'} - \matr{V}^*{'} \matr{V}^T \right] \right) \matr{J}\,.\nonumber 
\end{align}
The first, second, and third terms in \eqref{eq:Q_MoM_Final} represent $\matr{Q}^{\text{sca,inc}}$, $\matr{Q}^{\text{inc,sca}}$, and $\matr{Q}^{\text{sca,sca}}$ respectively. 
The first and second contribution to the third term are $\matr{Q}_i^{\text{sca,sca}}$ and $\matr{Q}_d^{\text{sca,sca}}$.

\section{Indirect Computation of $\matr{Q}$ Via the Scattering Matrix and its Frequency Derivative}
\label{sec:IE_Form_2}

This section introduces an indirect technique for computing $\matr{Q}$ by explicitly computing the product $j\matr{S}^\dag \matr{S}'$ from knowledge of the current densities excited by the incident fields that define $\matr{S}$. A connection between the direct and indirect methods for computing $\matr{Q}$ is established, providing an alternative proof of the WS relationship.

\subsection{Computation of the Scattering Matrix}

Solving integral equation \eqref{eq:MoM_IE_operator} for $\vect{\cal J}_p(\vr)$, $p=1,\hdots,M$ not only allows for the computation of $\vect{\cal E}_p(\vr)$, $p=1,\hdots,M$ throughout $\Omega$, but also of the $M \times M$ scattering matrix $\matr{S}$.  Specifically, knowledge of $\vect{\cal J}_p(\vr)$ permits the evaluation of $\matr{S}$'s $p$-th column.  The method for computing the $(t,p)$-th entry of $\matr{S}$ depends on whether $t$ corresponds to a guided or free-space port.
\begin{enumerate}
    \item \underline{Guided waves}. If $t\le M_g$, then $\matr{S}_{tp}$ reads
    \begin{align}
    \matr{S}_{tp} &=   \delta_{tp} + \matr{P}_{tp}\,,
    \label{eq:Smatrix_guided}
    \end{align}
    where 
    \begin{align}
    \matr{P}_{tp} = - \sqrt{Z_t} I_p(t) = -\frac{1}{2} \int_{d\Omega_{g,s}} \vect{\cal E}_t^{\text{inc}}(\vr) \cdot \vect{\cal J}_p(\vr) d\vr.
    \end{align}
    The second term on the RHS of \eqref{eq:Smatrix_guided} is the power amplitude of the outgoing wave on line $t$ due to $\vect{\cal J}_p(\vr)$. The first term represents a correction when the line carries an incoming wave. 
    \item \underline{Free-space waves}. If $t = (\tau, l, m) > M_g$, then $\matr{S}_{tp}$ can be evaluated once $\vect{\cal E}_{p,\myparallel}^{\text{sca}}(\vr)$ and $\vect{\cal H}_{p,\myparallel}^{\text{sca}}(\vr)$, the scattered (far) electric and magnetic fields near $d\Omega_f$, have been computed.  These fields are
    \begin{subequations}
    \begin{align}
    \vect{\cal E}_{p,\myparallel}^{\text{sca}}(\vr) &= -j \omega \mu \int_{d\Omega_{g,s}} \bigg[ G_\infty(\vr,\vrp) \vect{\cal J}_p( \vrp) \nonumber \\
    &\quad + \frac{1}{k^2}\nabla  G_{\infty}(\vr,\vrp) \nabla' \cdot \vect{\cal J}_p(\vrp) \bigg] d\vrp \nonumber \\
    &= \sum_{t=M_g+1}^{M} \matr{P}_{tp} \sqrt{Z} \vect{\cal I}_{t,\myparallel}^*(\vr) \label{eq:E_scatter} \\
    \vect{\cal H}_{p,\myparallel}^{\text{sca}}(\vr) &= \frac{1}{Z} \unit{r} \times \vect{\cal E}_{p,\myparallel}^{\text{sca}}(\vr) \nonumber \\
    &=  \sum_{t=M_g+1}^{M} \matr{P}_{tp} \frac{j}{\sqrt{Z}} \tilde{\vect{\cal I}}_{t,\myparallel}^*(\vr)  \,, \label{eq:H_scatter}
    \end{align}
    \end{subequations}    
where 
\begin{align}
 G_{\infty}(\vr,\vrp) &= \frac{e^{-jkr}}{4 \pi r} e^{jk \unit{r} \cdot \vrp} \label{eq:Ginf}
\end{align}
and 
\begin{align}
    \matr{P}_{tp} &= - \frac{\sqrt{Z}}{2} \int_{d\Omega_s} \vect{\cal W}_t(\vr) \cdot \vect{\cal J}_p(\vr) d\vr \nonumber \\
    &= -\frac{1}{2} \int_{d\Omega_{g,s}} \vect{\cal E}_t^{\text{inc}}(\vr) \cdot \vect{\cal J}_p(\vr) d\vr\,. \label{eq:Ptp}
\end{align}
Equations \eqref{eq:E_scatter}--\eqref{eq:H_scatter} follow from the VSW expansion of the $\vect{\cal L}$ operator in \eqref{eq:VSW_expansion2}~\cite{colton2013integral}--\cite{kristensson}.
Using \eqref{eq:Einc_freespace}--\eqref{eq:Hinc_freespace1} and \eqref{eq:E_scatter}--\eqref{eq:H_scatter}, the total electric and magnetic fields on $d\Omega_f$ read
\begin{subequations}
\begin{align}
    \vect{\cal E}_{p,\myparallel}(\vr) &= \vect{\cal E}_{p,\myparallel}^{\text{inc}}(\vr) + \vect{\cal E}_{p,\myparallel}^{\text{sca}}(\vr)  \nonumber \\
    &= \sqrt{Z} \vect{\cal W}_{p,\myparallel}(\vr) \delta_{p>M_g} + \sum_{t=M_g+1}^{M} \matr{P}_{tp} \sqrt{Z} \vect{\cal I}_{t,\myparallel}^*(\vr) \nonumber \\
    &= \sqrt{Z} \vect{\cal I}_{p,\myparallel}(\vr) {\delta}_{p>M_g}  + \sum_{t=M_g+1}^{M} \bigg((-1)^{m+l+\tau}   \nonumber \\
    &\quad \quad \quad \quad \quad  \delta_{\hat{t}p} {\delta}_{p>M_g} + \matr{P}_{tp}  \bigg) \sqrt{Z} \vect{\cal I}_{t,\myparallel}^*(\vr)  \label{eq:Ep_parallel_4} \\
    \vect{\cal H}_{p,\myparallel}(\vr) &= \vect{\cal H}_{p,\myparallel}^{\text{inc}}(\vr) + \vect{\cal H}_{p,\myparallel}^{\text{sca}}(\vr) \nonumber \\
    &= \frac{j}{\sqrt{Z}} \tilde{\vect{\cal W}}_{p,\myparallel}(\vr) {\delta}_{p>M_g} +  \sum_{t=M_g+1}^{M} \matr{P}_{tp} \frac{j}{\sqrt{Z}} \tilde{\vect{\cal I}}_{t,\myparallel}^*(\vr) \nonumber \\
    &= \frac{j}{\sqrt{Z}}\tilde{\vect{\cal I}}_{p,\myparallel}(\vr) {\delta}_{p>M_g}    +  \sum_{t=M_g+1}^{M} \bigg((-1)^{m + l + \tau}  \nonumber \\
    &\quad \quad \quad \quad   \delta_{\hat{t}p} {\delta}_{p>M_g} + \matr{P}_{tp}  \bigg) \frac{j}{\sqrt{Z}} \tilde{\vect{\cal I}}_{t,\myparallel}^*(\vr) \,,  \label{eq:Hp_parallel_4} 
\end{align}
\end{subequations}
where $\hat{t} = (\tau, l, -m )$.

Comparing the expressions \eqref{eq:Ep_parallel_4}--\eqref{eq:Hp_parallel_4} to
\eqref{eq:Eo_freespace}--\eqref{eq:Ho_freespace} and using \eqref{eq:Smatrix_guided} shows that
\begin{align}
 \matr{S}_{tp} = \vec{\matr{I}}_{tp}  + \matr{P}_{t p}     
 \label{eq:SP_relation}
\end{align}
\end{enumerate}
where
$\vec{\matr{I}}$ is an (identity-like) real and symmetric matrix with entries
\begin{align}
    \vec{\matr{I}}_{tp} = \begin{cases} \delta_{tp}  & p \le M_g \\
    (-1)^{m+l+\tau} \delta_{\hat{t} p} & p > M_g\,.
    \end{cases}
    \label{eq:Ivec_def}
\end{align}

\subsection{Computation of the Frequency Derivative of the Scattering Matrix}

Knowledge of $\vect{\cal J}_p(\vr)$, $p=1,\hdots,M$ not only allows for the computation of $\matr{S}$, but also  $\matr{S}'$.  
Indeed, taking the frequency derivative of \eqref{eq:SP_relation} yields $\matr{S}_{tp}' = \matr{P}_{tp}'$, i.e.
\begin{align}
    \matr{S}_{tp}' &= -\frac{1}{2} \int_{d\Omega_{g,s}} \bigg [\vect{\cal E}_t^{\text{inc}}{'}(\vr) \cdot \vect{\cal J}_p(\vr) + \vect{\cal E}_t^{\text{inc}}(\vr) \cdot \vect{\cal J}_p'(\vr) \bigg]d\vr\,. \label{eq:dPmatrix_Continuous} 
\end{align}
The integrand in \eqref{eq:dPmatrix_Continuous} depends on $\vect{\cal J}_p'(\vr)$.
To eliminate $\vect{\cal J}_p'(\vr)$ from the picture, consider the integral over $d\Omega_{g,s}$ of the inner product of  $\vect{\cal J}_t(\vr)$ and the frequency derivative of \eqref{eq:MoM_IE_operator}.
The resulting expression, simplified by invoking the symmetry of the $\vect{\cal L}$ operator, reads
\begin{align}
&\int_{d\Omega_{g,s}} \vect{\cal E}_t^{\text{inc}}(\vr) \cdot \vect{\cal J}_p'(\vr) d\vr \label{eq:Jprime_expression2}
 \\
& =  \int_{d\Omega_{g,s}} \big\{ \vect{\cal J}_t(\vr)\cdot \vect{\cal E}_p^{\text{inc}}{'}(\vr)  + \vect{\cal J}_t(\vr) \cdot     \vect{\cal L}' \left[\vect{\cal J}_p\right](\vr)  \big \} d\vr\,. \nonumber
\end{align}
Substituting \eqref{eq:Jprime_expression2} into \eqref{eq:dPmatrix_Continuous} yields entries of the frequency derivative of the scattering matrix
\begin{align}
    \matr{S}_{tp}' &= -\frac{1}{2} \int_{d\Omega_{g,s}} \bigg\{ \vect{\cal E}_t^{\text{inc}}{'}(\vr) \cdot \vect{\cal J}_p(\vr)  + \vect{\cal E}_p^{\text{inc}}{'}(\vr) \cdot \vect{\cal J}_t(\vr)  \label{eq:dPmatrix_Continuous2} \\
    & \quad \quad \quad \quad \quad   +\vect{\cal J}_t(\vr) \cdot  \vect{\cal L}' \left[\vect{\cal J}_p\right](\vr)  \bigg \} d\vr \,.
    \nonumber
\end{align}

\subsection{Indirect Surface Integral Computation of the WS Time delay Matrix}

Using \eqref{eq:SP_relation} and \eqref{eq:dPmatrix_Continuous2} to populate the scattering matrix $\matr{S}$ and its derivative $\matr{S}'$ provides an alternative (indirect) method for computing $\matr{Q}$ via  \eqref{eq:WS_Main_Relation}. The direct and indirect approaches for computing $\matr{Q}$ produce equivalent though at face value quite different expressions for the renormalized energies and group delays of fields that interact with antennas and scatterers. Their equivalence in principle follows from the derivations in \cite{Smith_1960}, which used field-based methods to demonstrate WS relationship \eqref{eq:WS_Main_Relation} with $\matr{Q}$ expressed in terms of the energy-like overlap integrals \eqref{eq:Qeqn}; an alternative proof of the relationship leveraging the above-derived current-based expressions for $\matr{Q}$'s entries is presented below.  Note that the direct and indirect expressions for $\matr{Q}$ share an important property: they only require knowledge of currents densities, not their frequency derivatives. The scattering matrix-based method for computing $\matr{Q}$ thus provides an alternative approach for computing system energies.
Note that while the product of $j \matr{S}^\dag \matr{S}'$ stemming from \eqref{eq:SP_relation} and \eqref{eq:dPmatrix_Continuous2} is not manifestly self-adjoint, an alternative formula with this property can be obtained by using $\matr{Q} = \frac{j}{2} \left( \matr{S}^\dag \matr{S}' - \matr{S}'^\dag \matr{S} \right)$, which follows from $ \matr{S}^\dag \matr{S}' = - \matr{S}'^\dag \matr{S}$ as implied by the unitarity of $\matr{S}$.


\subsection{Method of Moments Implementation}

The above equations once again are easily implemented in the method of moments.  Indeed, substituting \eqref{eq:Jexpand_Basis1} into \eqref{eq:Smatrix_guided} and \eqref{eq:Ptp}, and using \eqref{eq:VMoM_Entries} yields
\begin{align}
    \matr{S} &=   \vec{\matr{I}} + \frac{1}{2} \matr{V}^T \matr{J} \,.
    \label{eq:SMatrix_MoM}
\end{align}
A discrete expression for $\matr{S}'$ similarly follows from \eqref{eq:dPmatrix_Continuous2}:
\begin{align}
    \matr{S}' =&  \frac{1}{2} \left(\matr{V}' \right)^T \matr{J} +  \frac{1}{2} \matr{J}^T \matr{V}' - \frac{1}{2} \matr{J}^T \matr{Z}' \matr{J}\,.
    \label{eq:Sprime_final}
\end{align}
Substituting ~\eqref{eq:SMatrix_MoM} and ~\eqref{eq:Sprime_final} into \eqref{eq:WS_Main_Relation} yields the following scattering matrix-based discrete expression for $\matr{Q}$
\begin{align}
    \matr{Q} = \frac{j}{2} \left(\vec{\matr{I}} + \frac{1}{2} \matr{V}^T \matr{J} \right)^\dag \left(  \left(\matr{V}' \right)^T \matr{J} +   \matr{J}^T \matr{V}' -  \matr{J}^T \matr{Z}' \matr{J}\right)\,.
    \label{eq:QIndirect_Final}
\end{align}

\subsection{Equivalence Between the Direct and Indirect Approaches to Compute $\matr{Q}$}
\label{eq:direct_indirect_equivalence}
This section presents an alternative proof of \eqref{eq:WS_Main_Relation} with $\matr{Q}$ expressed in terms of the energy-like overlap integrals \eqref{eq:Qeqn}, directly leveraging the current-based expressions for $\matr{Q}$ presented in Sections~\ref{sec:IE_Form} and \ref{sec:IE_Form_2}. To keep the notation compact, this exercise is performed starting from the method of moment expressions \eqref{eq:Q_MoM_Final} and \eqref{eq:QIndirect_Final}.

Using $\matr{V} \vec{\matr{I}} = \matr{V}^*$ (Identity A.6 in Appendix~\ref{App:IdentityA6}),  $\matr{V}' \vec{\matr{I}} = \matr{V}^*{'}$, and $\matr{J}' = \matr{Z}^{-1} \left(\matr{V}' - \matr{Z}' \matr{J} \right)$ (obtained by differentiating \eqref{eq:MoM2}) in \eqref{eq:QIndirect_Final} yields
\begin{align}
\matr{Q} =& \frac{j}{2}  \bigg[\matr{V}'^\dag \matr{J} - \matr{J}^\dag \matr{V}' + \matr{J}^\dag \matr{V}' + \matr{V}^\dag \matr{J}'   \nonumber \\
& \quad + \frac{1}{2}  \matr{J}^\dag \matr{V}^{*} \matr{V}'^T \matr{J} + \frac{1}{2}  \matr{J}^\dag \matr{V}^* \matr{V}^T \matr{J}'      \bigg] \label{eq:Qindirect_Simplify1}
\end{align}
Using $ \matr{J}^\dag \matr{V}' + \matr{V}^\dag \matr{J}' = \matr{J}^\dag \matr{Z}' \matr{J} + \matr{J}^\dag \left(\matr{Z} + \matr{Z}^*\right) \matr{J}' $ into \eqref{eq:Qindirect_Simplify1} yields
\begin{align}
    \matr{Q} =& \frac{j}{2}  \bigg[\matr{V}'^\dag \matr{J} - \matr{J}^\dag \matr{V}' + \matr{J}^\dag \matr{Z}' \matr{J}  + \frac{1}{2}  \matr{J}^\dag \matr{V}^* \matr{V}'^T \matr{J}       \nonumber \\
    & \quad + \matr{J}^\dag \underbrace{\left(\matr{Z} + \matr{Z}^* + \frac{1}{2} \matr{V}^* \matr{V}^T  \right)}_{= 0} \matr{J}'  \bigg] \label{eq:Qindirect_Simplify2}
\end{align}
where use was made of $\frac{1}{2} \matr{V}^* \matr{V}^T = -\left(\matr{Z} + \matr{Z}^* \right)$ (Identity A.7 in Appendix~\ref{App:IndentityA7}).
Finally, substituting \eqref{eq:Qindirect_Simplify2} into $
    \matr{Q} = \frac{1}{2} \left(\matr{Q} + \matr{Q}^\dag\right)
$ (because $\matr{Q}$ is self-adjoint) yields \eqref{eq:Q_MoM_Final}, thereby proving the equivalence of both approaches.

\section{Basis Transformations and Origin-Dependence of $\matr{Q}$}
\label{sec:WSModes}

This section comments on several attributes of the WS time delay matrix $\matr{Q}$ that derive from the above methods for evaluating its entries. These attributes complement the rich set of properties of $\matr{Q}$ and its eigenstates elucidated in \cite{TAP_1}.

\subsection{Unitary Transformations of Incoming Fields}

\vskip 6pt
\noindent
\textit{\ul{The methods for computing $\matr{Q}$ and $\matr{S}$ outlined in Sections {\ref{sec:IE_Form}} and {\ref{sec:IE_Form_2}} hold true regardless of the orthonormal basis that represents incoming fields on $d\Omega_{f}$ and $d\Omega_{g}$.}}
\vskip 6pt

Sections \ref{sec:computational_framework}--\ref{sec:IE_Form_2} focused on the characterization of $\matr{Q}$ and $\matr{S}$ using a basis of incoming guided waves and VSHs. 
The proposed schemes for computing $\matr{Q}$ and $\matr{S}$ however apply in any orthonormal basis.
Indeed, consider a set of incoming electric fields 
\begin{align}
    \vect{\widehat{\cal E}}_{t}^i(\vr) &= \sum_{t=1}^{M} \matr{W}_{tp}  \vect{\cal E}_{p}^i(\vr) \label{eq:UnitaryTransform_Einc}
\end{align}
where $\matr{W} = \begin{bmatrix} \matr{W}_{tp}; t,p = 1,\hdots, M \end{bmatrix}$ is a frequency-independent unitary matrix. Incoming magnetic fields, incident electric and magnetic fields, and current densities transform similarly. In the new basis, the moment method voltage matrix, its frequency derivative, and the current density matrix are $\widehat{\matr{V}} =  \matr{V} \matr{W}$, $\widehat{\matr{V}}' =  \matr{V'} \matr{W}$, and $\widehat{\matr{J}} =  \matr{J} \matr{W}$, while matrices $\matr{Q}$, $\matr{S}$, and  $\matr{S'}$ transform as $\widehat{\matr{Q}} = \matr{W}^\dag \matr{Q} \matr{W}$, $\widehat{\matr{S}} = \matr{W}^T \matr{S} \matr{W}$, and $\widehat{\matr{S}}' = \matr{W}^T \matr{S}' \matr{W}$, respectively. 
It is easily verified that expressions \eqref{eq:Qeqn}, \eqref{eq:Qeqn_decompose_1}-\eqref{eq:L_1prime}, \eqref{eq:Q_MoM_Final}, \eqref{eq:SP_relation}, \eqref{eq:dPmatrix_Continuous2}, and \eqref{eq:QIndirect_Final} for $\matr{Q}$ in terms incident fields, their frequency derivatives, and current densities continue to hold in the transformed system.

\subsection{Trace of $\matr{Q}$ and Wigner time delay}
\label{sec:traceQ}

The entries of $\matr{Q}$ and $\matr{S}$ depend on the basis used for expanding incoming waves as well as the position of the antennas and/or scatterers w.r.t. the spatial origin. \textit{\ul{The sum $\Tr \left( \matr{Q} \right) = \sum_{m=1}^M \matr{Q}_{mm}$ however is invariant under unitary transformations of the incoming fields and translations of the origin. In other words, the sum of the group delays experienced by all modes is a system constant, determined solely by its geometry.  Alternatively, the average group delay $\frac{1}{M} \Tr \left( \matr{Q}\right)$, also called the Wigner time delay, experienced by all modes is a system invariant.}}

The invariance of $\Tr \big( \matr{Q} \big)$ under unitary transformations $\matr{W}$ of the basis of incoming fields immediately follows from $\Tr \big(\widehat{\matr{Q}}\big) = \Tr \left( \matr{W}^\dag \matr{Q} \matr{W}\right) = \Tr \left( \matr{Q} \matr{W} \matr{W}^\dag\right) =  \Tr \left({\matr{Q}}\right)$ \cite{Horn_2017}.

The invariance of $\Tr \left( \matr{Q} \right)$ under translations of the origin $\vect{o} \rightarrow \vect{o}'$ is easily understood and demonstrated using a basis of incoming waves derived from the original guided waves and incident vector plane waves (VPWs)\footnote{The $M_f$ VSHs only approximately combine into VPWs in a ball of radius $a$; outside this ball, these fields become diverging beams.}
\begin{align}
    \widetilde{\vect{\cal E}}^{\text{inc}}_{p}(\vr) &\propto  e^{-j k \unit{k}(\theta_p,\phi_p) \cdot \unit{r}} \unit{p}_p\, \label{eq:Einc_pw}
\end{align}
where $p = M_g+1, \hdots, M = (\theta_p, \phi_p, \unit{p}_p)$, 
 $(\theta_p \in \{0, \pi \},\phi_p \in \{ 0, 2\pi \})$ are the azimuthal and polar angles characterizing the $p$-th plane wave's propagation direction $\unit{k}(\theta_p, \phi_p) = \cos \phi_p \sin \theta_p \unit{x} + \sin\phi_p \sin \theta_p \unit{y} + \cos \theta_p \unit{z}$, and $\unit{p}_p$ is the wave's polarization vector. In what follows, it is assumed that each $\unit{k}$ in the basis has a $-\unit{k}$ counterpart. Techniques for constructing the transformation matrices $\matr{W}$ that combine incident VSHs into approximate VPWs are detailed in~\cite{wittmann1988spherical, Hansen}.

 Let $\widetilde{\matr{Q}}$ and $\widetilde{\matr{Q}}_{\text{shift}}$ denote WS time delay matrices expressed in a basis of incoming waves derived from the original guided waves and incident VPWs defined w.r.t. origins $\vect{o}$ and $\vect{o}'$, respectively. To prove the invariance of the Wigner time delay w.r.t. shifts of the spatial origin, it is demonstrated below that $\Tr \big(\widetilde{\matr{Q}}\big) = \Tr \big(\widetilde{\matr{Q}}_{\text{shift}} \big)$. 
 The above discussion showed that this equality implies the shift-invariance of the Wigner time delay in any orthonormal basis.
 
 Intuitively, a shift of the origin $\vect{o} \rightarrow \vect{o}'$ does not change the diagonal elements of $\widetilde{\matr{Q}}$ corresponding to guided wave excitations as these move with the antennas and/or scatterers. The shift however does impose an extra group delay on a plane wave with propagation direction $\unit{k}$ that is proportional to $\unit{k}\cdot (\vect{o}-\vect{o}')$. This extra group delay however is the opposite of that experienced by a plane wave with propagation vector $-\unit{k}$. The sum of the changes in group delay for plane waves impinging on the antennas and/or scatterers from all direction therefore vanishes. 

To mathematically show that $\Tr \big(\widetilde{\matr{Q}}\big) = \Tr \big( \widetilde{\matr{Q}}_{\text{shift}}\big)$, 
let $\widetilde{\matr{V}}$ and $\widetilde{\matr{J}}$ denote the method of moments excitation and current matrices obtained in the VPW basis defined w.r.t. origin  $\vect{o}$. It follows from \eqref{eq:Q_MoM_Final} and the linearity of the trace operator that
\begin{align}
    \Tr \big( \widetilde{\matr{Q}} \big) 
    =&  \Re{-j \Tr \left( \widetilde{\matr{J}}^\dag \widetilde{\matr{V}}' \right)}  \nonumber \\
    &\quad +  \frac{j}{8} \Tr \bigg(\widetilde{\matr{J}}^\dag \Big(\widetilde{\matr{V}}^* \widetilde{\matr{V}}^T{'} -  \widetilde{\matr{V}}^*{'} \widetilde{\matr{V}}^T \Big) \widetilde{\matr{J}} \bigg )\,.\label{eq:Trace_Q}
\end{align}
When shifting the origin from $\vect{o}$ to $\vect{o}'$, incident fields from the guided ports remain fixed while those from $d\Omega_f$ undergo a phase shift $e^{-jk\unit{k}(\theta_p, \phi_p) \cdot \left(\vect{o}' - \vect{o}\right)}$. The method of moments excitation and current matrices in the shifted coordinate system therefore are
$\widetilde{\matr{V}}_{\text{shift}} = \widetilde{\matr{V}}\matr{D}$ and  
$\widetilde{\matr{J}}_{\text{shift}} = \widetilde{\matr{J}} \matr{D}$, where $\matr{D}$ is a diagonal matrix with entries $\matr{D}_{pp} = 1$ for $p \le M_g$ and $\matr{D}_{pp} =   e^{-jk\unit{k}(\theta_p, \phi_p) \cdot \left( \vect{o}' - \vect{o}\right)}$ otherwise.
Furthermore, the chain rule yields $\widetilde{\matr{V}}_{\text{shift}}' = \widetilde{\matr{V}}' \matr{D} + \widetilde{\matr{V}} \matr{F} \matr{D}$, where $\matr{F}$ is a diagonal matrix with entries $\matr{F}_{pp} = 0$ for $p\le M_g$ and $\matr{F}_{pp} = -j\sqrt{\mu \varepsilon} \unit{k}(\theta_p, \phi_p) \cdot \left(\vect{o}' - \vect{o}\right)$ otherwise.
Using the above expressions for $\widetilde{\matr{V}}_{\text{shift}}$,  $\widetilde{\matr{J}}_{\text{shift}}$, and $\widetilde{\matr{V}}_{\text{shift}}'$ in \eqref{eq:Q_MoM_Final} yields 
\begin{align}
    \Tr &\big( \widetilde{\matr{Q}}_{\text{shift}} \big) \nonumber \\
    =& \Re{-j \Tr \left(  \widetilde{\matr{J}}^\dag \widetilde{\matr{V}}'  +  \widetilde{\matr{J}}^\dag \widetilde{\matr{V}}\matr{F}  \right)} \nonumber \\
    &+ \frac{j}{8} \Tr \bigg( \widetilde{\matr{J}}^\dag \Big( \widetilde{\matr{V}}^*  \widetilde{\matr{V}}^T{'} + 2 \widetilde{\matr{V}}^*  \matr{F}^T \widetilde{\matr{V}}^T  - \widetilde{\matr{V}}^*{'}  \widetilde{\matr{V}}^T  \Big)\widetilde{\matr{J}}  \bigg)  \label{eq:Trace_Q_shift}
\end{align}
where the cyclic property of the trace operation was used to simplify the result.
Combining \eqref{eq:Trace_Q} and \eqref{eq:Trace_Q_shift} yields
\begin{align}
    \Tr \big(\widetilde{\matr{Q}}_{\text{shift}} - \widetilde{\matr{Q}} \big) 
    =&  \Re{-j \Tr \left(    \widetilde{\matr{J}}^\dag \widetilde{\matr{V}}\matr{F}  \right)} \nonumber \\
    &+ \frac{j}{4} \Tr \bigg(  \widetilde{\matr{V}}^T  \widetilde{\matr{J}}  \widetilde{\matr{J}}^\dag \widetilde{\matr{V}}^* \matr{F}  \bigg ) \,.  \label{eq:Trace_Qshift_Q_1}
\end{align}
Next, using $\widetilde{\matr{V}}^T  \widetilde{\matr{J}} = \widetilde{\matr{J}}^T \widetilde{\matr{V}} $ (because $\matr{S}$ in \eqref{eq:SMatrix_MoM} is symmetric) into \eqref{eq:Trace_Qshift_Q_1} along with Identity A.7 from Appendix~\ref{App:IndentityA7} yields  
\begin{align}
    \Tr \big(\widetilde{\matr{Q}}_{\text{shift}} - \widetilde{\matr{Q}} \big) 
     =&  \Re{-j \Tr \left(    \widetilde{\matr{J}}^\dag  \widetilde{\matr{V}} \matr{F}  \right)} \nonumber \\
      &+ \frac{j}{4} \Tr \bigg(  \widetilde{\matr{J}}^T \widetilde{\matr{V}}   \widetilde{\matr{V}}^\dag \widetilde{\matr{J}}^*  \matr{F}   \bigg ) \nonumber \\
      =& \Re{-2j \Tr \left(    \widetilde{\matr{J}}^\dag  \widetilde{\matr{V}} \matr{F}  \right)} \label{eq:Traceproof_1}
  \end{align}
Finally, substituting $\widetilde{\matr{J}}^\dag = \widetilde{\matr{V}}^\dag \left(\matr{Z}^{-1}\right)^*$ into \eqref{eq:Traceproof_1}  yields
\begin{align}
    \Tr\big(\widetilde{\matr{J}}^\dag  \widetilde{\matr{V}} \matr{F} \big) &= \Tr\big(\widetilde{\matr{V}}^\dag \left(\matr{Z}^{-1}\right)^* \widetilde{\matr{V}} \matr{F} \big) \nonumber \\
    &= 
    \sum_{m=M_g+1}^{M} \matr{F}_{mm} \sum_{t=1}^{N} \sum_{t'=1}^{N} \widetilde{\matr{V}}_{tm}^* \left(\matr{Z}^{-1}_{tt'}\right)^* \widetilde{\matr{V}}_{t'm} \nonumber \\ 
    &= 0\,. \label{eq:Trace_proof_final_result}
\end{align}
The last transition is due to the assumption that each VPW in the basis is accompanied by another one with the same polarization vector propagating in the opposite direction. These waves' method of moments voltage vectors are each others complex conjugate while their $\matr{F}$ entries sum to zero, ensuring that the sum in \eqref{eq:Trace_proof_final_result} vanishes.


\subsection{WS Modes}
\vskip 6pt
\noindent
\textit{\ul{Some Useful Properties of WS Modes (not discussed in~{\cite{TAP_1}})}} 
\vskip 6pt

Next, consider the unitary matrix $\widebar{\matr{W}}$ that simultaneously diagonalizes $\matr{Q}$ and $\matr{S'}$, while transforming $\matr{S}$ into the identity matrix. Owing to the fact that $\matr{Q}$ is self-adjoint, $\widebar{\matr{Q}} = \widebar{\matr{W}}^\dag \matr{Q} \widebar{\matr{W}}$ is a purely real diagonal matrix. Using this fact and $\widebar{\matr{S}} = \widebar{\matr{W}}^T \matr{S} \widebar{\matr{W}} = \matr{I}_M$ shows that   $\widebar{\matr{S}}' = \widebar{\matr{W}}^T \matr{S'} \widebar{\matr{W}}$ is a purely imaginary diagonal matrix \cite{TAP_1}. Fields described by $\widebar{\matr{W}}$'s column vectors are termed ``WS modes'', and experience well-defined group delays corresponding to the eigenvalues of $\matr{Q}$, viz. the diagonal elements of $\widebar{\matr{Q}}$. Computation of a system's WS time delay matrix using the methods of Sections \ref{sec:IE_Form} or \ref{sec:IE_Form_2} enables the construction of its WS modes and the associated interpretation/classification of the system's radiation and scattering properties \cite{TAP_1}.




Let $\{\widebar{\vect{\cal E}}_p^{i(o)}, \widebar{\vect{ \cal H}}_p^{i(o)} \}$  and 
$\{\widebar{\vect{\cal E}}_p, \widebar{\vect{ \cal H}}_p \}$ denote incoming (outgoing) and total electric and magnetic fields associated with the $p$-th WS mode excitation.
The current density induced on $d\Omega_{s}$ due to the $p$-th WS mode excitation is denoted by $\widebar{\vect{\cal J}}_p(\vr)$.
The following properties of WS mode fields and current densities are demonstrated below:
\begin{enumerate}
    \item $\widebar{\vect{\cal E}}_p$ and $\widebar{\vect{ \cal H}}_p$ are purely real and imaginary, respectively. 
    \item $\widebar{\vect{\cal J}}_p$ is purely imaginary.
    \item  $\widebar{\matr{Q}}^{\text{sca,sca}}_d$, the explicit origin-dependent component of $\widebar{\matr{Q}}^{\text{sca,sca}}$, vanishes identically, i.e. $\matr{\cal Q}^{\text{sca,sca}}_{d}\left(\widebar{\vect{\cal J}}_p, \widebar{\vect{\cal J}}_p \right) = 0$.
\end{enumerate}
To prove property \# 1, note that $\widebar{\matr{S}} = \matr{I}_M$ implies that the outgoing electric field is $\widebar{\vect{\cal E}}^o_p(\vr) = \widebar{\vect{\cal E}}^{i*}_p(\vr)$ while the total electric field is $\widebar{\vect{\cal E}}_p(\vr) = \widebar{\vect{\cal E}}_p^{i}(\vr) + \widebar{\vect{\cal E}}_p^{i*}(\vr)$.
Since $\widebar{\vect{\cal E}}_p^*(\vr) = \widebar{\vect{\cal E}}_p^{i*}(\vr) + \widebar{\vect{\cal E}}_p^{i}(\vr) = \widebar{\vect{\cal E}}_p(\vr)$, it follows that $\widebar{\vect{\cal E}}_p(\vr)$ is purely real. A similar argument for the magnetic field yields $\widebar{\vect{\cal H}}_p^*(\vr) = \widebar{\vect{\cal H}}_p^{i*}(\vr) - \widebar{\vect{\cal H}}_p^{i}(\vr) = -\widebar{\vect{\cal H}}_p(\vr)$. Therefore, $\widebar{\vect{\cal H}}_p(\vr)$ is purely imaginary.

To prove property \# 2, recall that $\widebar{\vect{\cal E}}_p(\vr)$ and $\widebar{\vect{\cal J}}_p(\vr)$ satisfy the vector wave equation for $\vr \in \Omega$:
\begin{align}
    \nabla \times \nabla \times \widebar{\vect{\cal E}}_p(\vr) - k^2 \widebar{\vect{\cal E}}_p(\vr) = - j \omega \mu \widebar{\vect{\cal J}}_p(\vr)\,.
    \label{eq:VectorWaveEqn}
\end{align}
Subtracting \eqref{eq:VectorWaveEqn} from its conjugate and using $\widebar{\vect{\cal E}}_p(\vr) = \widebar{\vect{\cal E}}_p^*(\vr)$
yields
$    -j \omega \mu \widebar{\vect{\cal J}}^*_p(\vr) - j \omega \mu \widebar{\vect{\cal J}}_p(\vr) = 0 \,.$
Since $\widebar{\vect{\cal J}}_p^*(\vr) = -\widebar{\vect{\cal J}}_p(\vr)$, it follows that $\widebar{\vect{\cal J}}_p(\vr)$ is purely imaginary.

To prove property \# 3, note that $\widebar{\matr{Q}}^{\text{sca,sca}}$ (see \eqref{eq:Qscasca_final}) is purely real because $\widebar{\matr{Q}}_{pp}$ and $\widebar{\matr{Q}}^{\text{sca,inc}}_{pp} + \widebar{\matr{Q}}^{\text{inc,sca}}_{pp}$ (see \eqref{eq:Qscainc_final} and \eqref{eq:Qincsca_final}) are purely real.
Furthermore, since $\widebar{\vect{\cal J}}_p$ is purely imaginary, $\widebar{\matr{Q}}_i^{\text{sca,sca}}(\widebar{\vect{\cal J}}_p, \widebar{\vect{\cal J}}_p)$ (see \eqref{eq:Operator_Qscasca_i_final})  is purely real. Since $\Re{\widebar{\matr{Q}}_{d}^{\text{inc,inc}}} = 0$ (see \eqref{eq:Operator_Qscasca_d_final}), it follows that $\widebar{\matr{Q}}_{d}^{\text{inc,inc}} = 0$.
 More generally, $\matr{Q}_{d,pp}^{\text{sca,sca}}$ vanishes whenever $\vect{\cal J}_{p}(\vr)$ has constant phase.

\vskip 6pt
\noindent
\textit{\ul{$\matr{S}'$ can be computed in terms of stored energies when the system is excited by WS modes}}
\vskip 6pt

The diagonalization of $\matr{Q}$ also facilitates the characterization of $\matr{S}'$ in terms of time delays (renormalized energies) associated with WS modes.
Indeed, using $\widebar{\matr{S}}=\matr{I}_M$ in WS relationship \eqref{eq:WS_Main_Relation} shows that $\widebar{\matr{S}}' = -j\widebar{\matr{Q}}$ , and 
expressing the frequency derivative of the scattering matrix as $\matr{S}' = \widebar{\matr{W}}^* \widebar{\matr{S}}' \widebar{\matr{W}}^\dag$ yields
\begin{subequations}
\begin{align}
    \matr{S}' &= -j \sum_{i=1}^{M} \left(\widebar{\matr{W}}_i \widebar{\matr{W}}_i^T\right)^* \widebar{\matr{Q}}_{ii} \label{eq:Sprime_Energy_a} \\
    &\approx -j \sum_{i=1, \abs{\widebar{\matr{Q}}_{ii}}>\bar{\varepsilon}}^{M} \left(\widebar{\matr{W}}_i \widebar{\matr{W}}_i^T\right)^* \widebar{\matr{Q}}_{ii} \,, 
    \label{eq:Sprime_Energy}
\end{align}
\end{subequations}
where $\widebar{\matr{W}}_i$ denotes the $i$-th column of $\widebar{\matr{W}}$. Equation \eqref{eq:Sprime_Energy} follows from \eqref{eq:Sprime_Energy_a} upon discarding WS modes that experience negligible group delays~\cite{TAP_1} (i.e. group delays in absolute value larger than $\bar{\varepsilon}$.) 
Equation \eqref{eq:Sprime_Energy} states that elements of $\matr{S}'$ consist of sums of renormalized stored energies / group delays weighed by the corresponding entries of the WS mode vectors $\widebar{\matr{W}}_i$, $i=1,\hdots, M$.

For one-port antennas, \eqref{eq:Sprime_Energy} may be used to relate the frequency derivative of antenna port's scattering coefficient $\matr{S}_{11}'$ to renormalized energy stored in the WS modes as
\begin{align}
    \matr{S}_{11}' \approx -j \sum_{i=1, \abs{\widebar{\matr{Q}}_{ii}} > \bar{\varepsilon}}^{M} \left( \widebar{\matr{W}}_{1i}^*\right)^2 \widebar{\matr{Q}}_{ii}\,.
    \label{eq:Sprime_Energy2}
\end{align}

\vskip 6pt
\noindent
\textit{\ul{Comparison of WS and Yaghjian-Best~{\cite{Best_2005}} methods for computing $\abs{\matr{S}'}$.}}
\vskip 6pt

It is instructive to compare \eqref{eq:Sprime_Energy2} to the celebrated approximation for the magnitude of the frequency derivative of the scattering parameter introduced by Yaghian and Best (YB) \cite{Best_2005}:

\begin{subequations}
\begin{align}
    \abs{\matr{S}_{11}'}_{\text{YB}} &\equiv \matr{Q}_{11} \label{eq:Sprime_YB_full}  \\
    &= \sum_{i=1}^M \abs{\widebar{\matr{W}}_{1i}}^2 \widebar{\matr{Q}}_{ii} \label{eq:Sprime_YB} \\
    &\approx \sum_{i=1, \abs{\widebar{\matr{Q}}_{ii}} > \bar{\varepsilon}}^{M} \abs{\widebar{\matr{W}}_{1i}}^2 \widebar{\matr{Q}}_{ii}\,. \label{eq:Sprime_YB2}
\end{align}
\end{subequations}
The second equality in \eqref{eq:Sprime_YB} follows from the expansion of $\matr{Q}_{11}$ in terms of entries of $\widebar{\matr{W}}$ and $\widebar{\matr{Q}}$. 

Comparing \eqref{eq:Sprime_Energy2} with  \eqref{eq:Sprime_YB2} implies that the Yaghian-Best formula yields an accurate estimate of $\abs{\matr{S}_{11}'}$ provided that the elements of $\widebar{\matr{W}}_{1i}$ with $\abs{\widebar{\matr{Q}}_{ii}} > \bar{\varepsilon}$ have constant phase (modulo $\pi$). This situation naturally arises in the quasi-static regime. It also holds true when the induced current density $\widebar{\vect{\cal J}}_1(\vr)$ due to the antenna port excitation has constant phase (modulo $\pi$). Indeed, it follows from $\vect{\cal J} = \widebar{\vect{\cal J}} \,\widebar{\matr{W}}^\dag$ and the fact that $\widebar{\vect{\cal J}}$ is imaginary (see property \#2 above) that $\widebar{\matr{W}}_{1i}$ has constant phase when $\vect{\cal J}_1$ has constant phase.  
Note that the reverse does not necessarily hold true as the YB formula remains accurate even when $\vect{\cal J}_1(\vr)$ contains contributions from WS mode currents $\widebar{\vect{\cal J}}_{i}$ modulated by ``out of phase" $\widebar{\matr{W}}_{1i}$'s for which $\abs{\widebar{\matr{Q}}_{ii}} < \bar{\varepsilon}$. 

Finally, note that when $\vect{\cal J}_1(\vr)$ has constant phase, $\abs{\matr{S}_{11}'}_{\text{YB}}$ can be computed with \eqref{eq:Sprime_YB2}. 
Alternatively, the YB estimate may be computed with \eqref{eq:Sprime_YB_full} using the methodology of \cite{VDB_2010}, i.e. with $\matr{Q}^{\text{sca,sca}}(\vect{\cal J}_1,\vect{\cal J}_1) = \matr{Q}^{\text{sca,sca}}_{i}(\vect{\cal J}_1,\vect{\cal J}_1)$. The origin-dependent term $\matr{Q}^{\text{sca,sca}}_d(\vect{\cal J}_1,\vect{\cal J}_1)$  can be safely ignored (see property \#3 above).

%% file: graphics/radiating_antenna_feed_model.tex
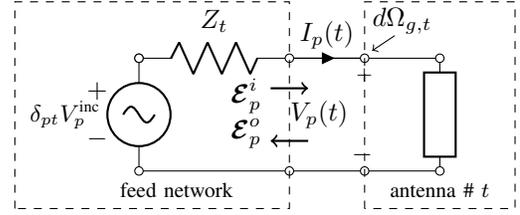
\begin{figure}[t]
\centering
\begin{circuitikz}[american]

\draw (0,0) to[vsourcesin, o-o, v^<={\small $\delta_{pt}V_{p}^{\text{inc}}$}] (0,1.5) to[R,l=$Z_t$, o-o] (2,1.5) to[short,i=$I_p(t)$, -o] (3,1.5) to[short] (4,1.5) to[generic, o-o] (4,0) to[short, -o] (3,0) to[short, -o] (2.0,0) to[short] (0,0);
\draw (3,1.25) to[open,v=$V_p(t)$] (3,0.25);

\draw[->, thick] (1.75,1.1) -- (2.25,1.1);
\node at (1.75,1) [left] {$\vect{\cal E}_p^i$};
\draw [->,thick] (2.25,0.4) -- (1.75,0.4);
\node at (1.75,0.5) [left] {$\vect{\cal E}_p^o$};

\draw[dashed]  (-1.65, -0.5) rectangle (2,2.25) ;
\draw[dashed] (3,-0.5) rectangle (5,2.25);
\node at (0.5,-0.25) {\footnotesize feed network};
\node at (4,-0.25) {\footnotesize antenna \# $t$};
\node at (3.5,2) {$d\Omega_{g,t}$};
\draw[->] (3.4,1.8) -- (3.1,1.6);
\end{circuitikz}
\caption{Antenna feed network in IE formulation.}
\label{fig:Antenna_feedmodel}
\end{figure}

%% file: results.tex
\section{Wigner-Smith Based Computation of Frequency Derivatives of Antenna Reflection Coefficient, Radiation Patterns, and Radar Cross Sections}
\label{sec:numerical_results}

\begin{figure*}[t]
\null \hfill \subfloat[  \label{fig:Dipole_antenna_error}]{\includegraphics[width=0.31\textwidth]{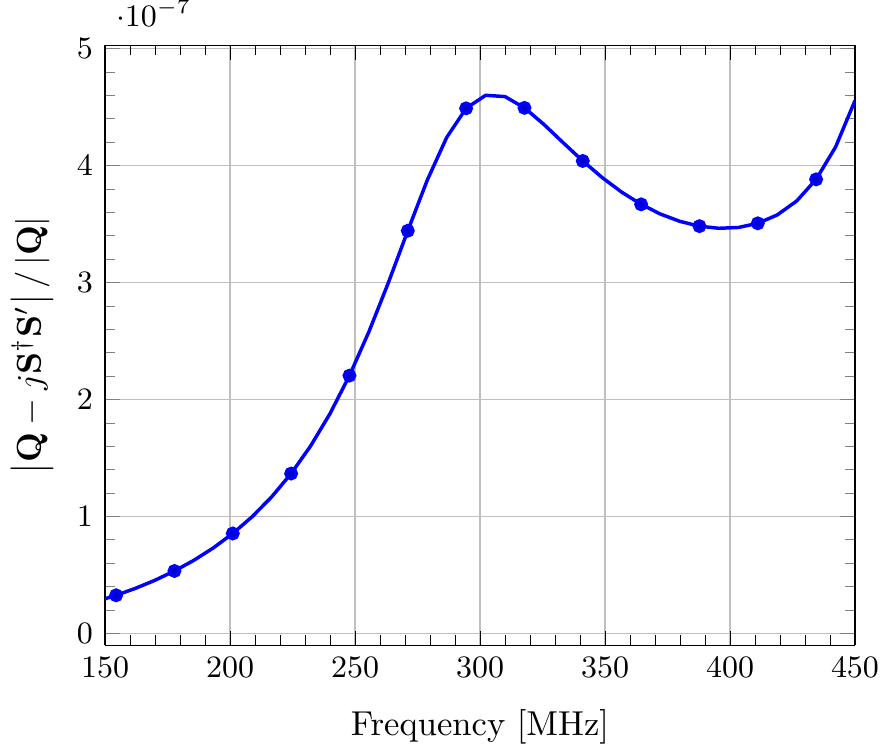}}
    \hfill
    \subfloat[ \label{fig:Dipole_antenna_S}]{\includegraphics[width=0.35\textwidth]{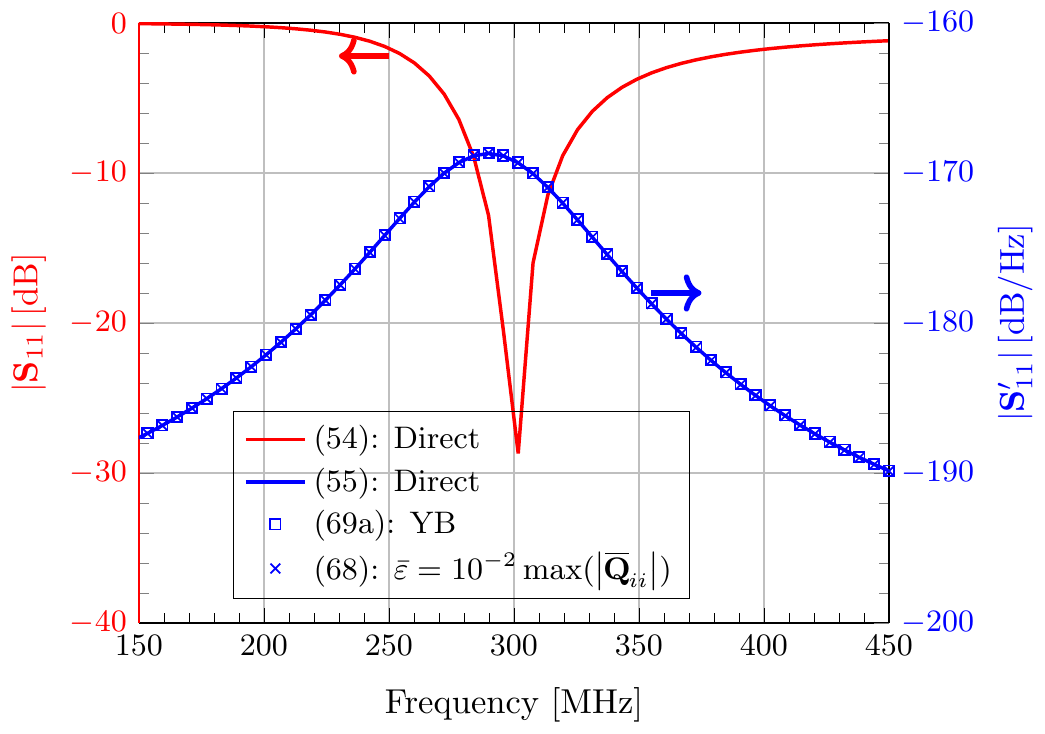}}   \hfill
\subfloat[ \label{fig:Dipole_convergence}]{\includegraphics[width=0.33\textwidth]{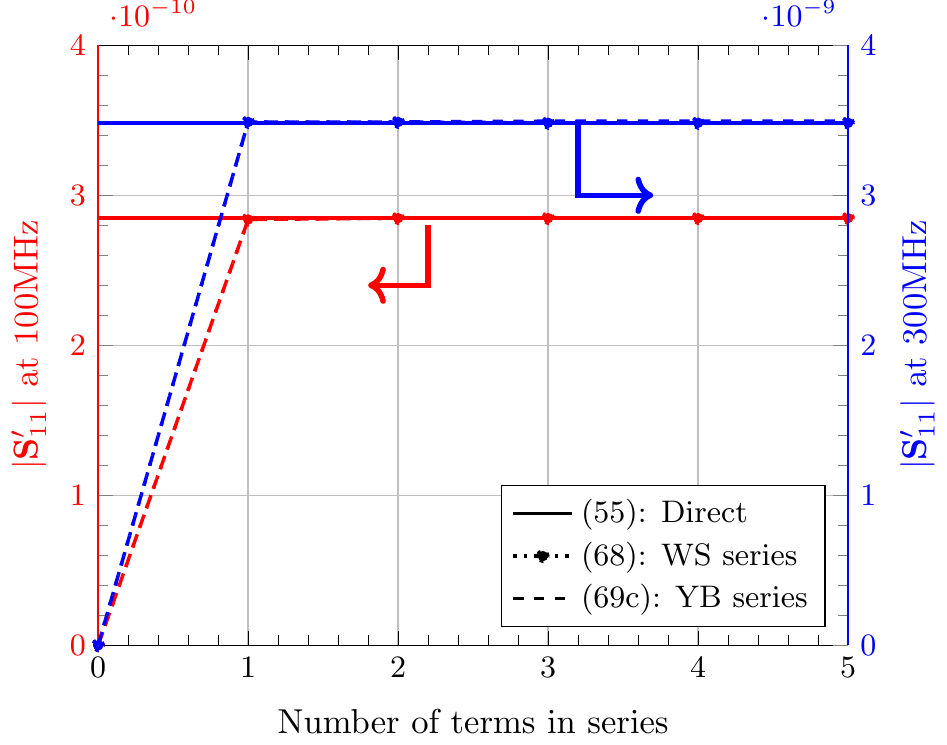}} 
  \hfill
    \null \\ 
    \null \hfill 
    \subfloat[ \label{fig:Dipole_antenna_Timedelays}]{\includegraphics[width=0.31\textwidth]{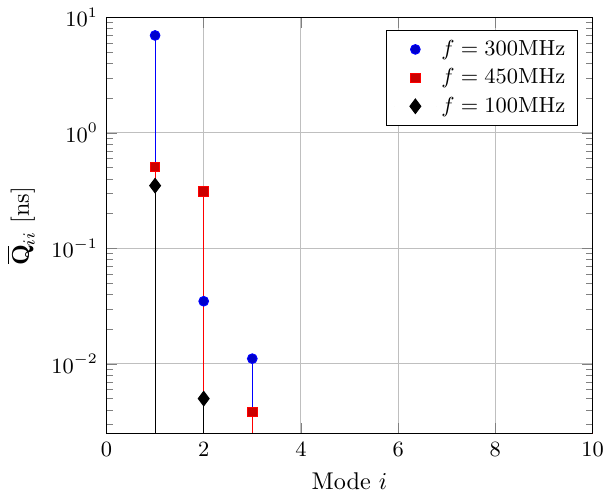}}  \hfill
    \subfloat[ \label{fig:Dipole_antenna_absW}]{\includegraphics[width=0.32\textwidth]{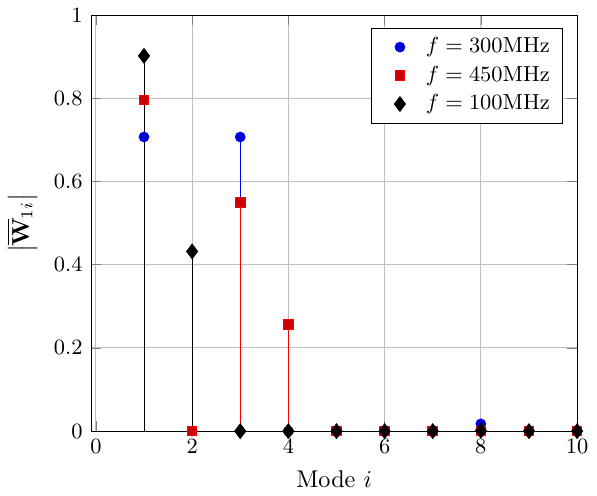}} 
    \hfill 
    \subfloat[ \label{fig:Dipole_antenna_angleW}]{\includegraphics[width=0.32\textwidth]{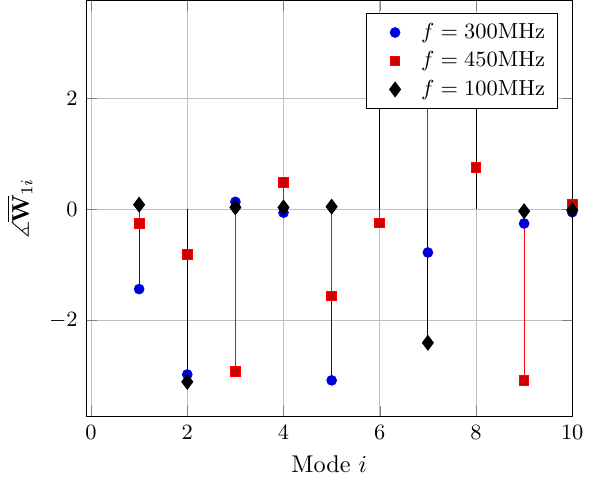}} 
    \hfill \null \\
    \null \hfill 
    \subfloat[ \label{fig:Dipole_antenna_J100}]{\includegraphics[width=0.33\textwidth]{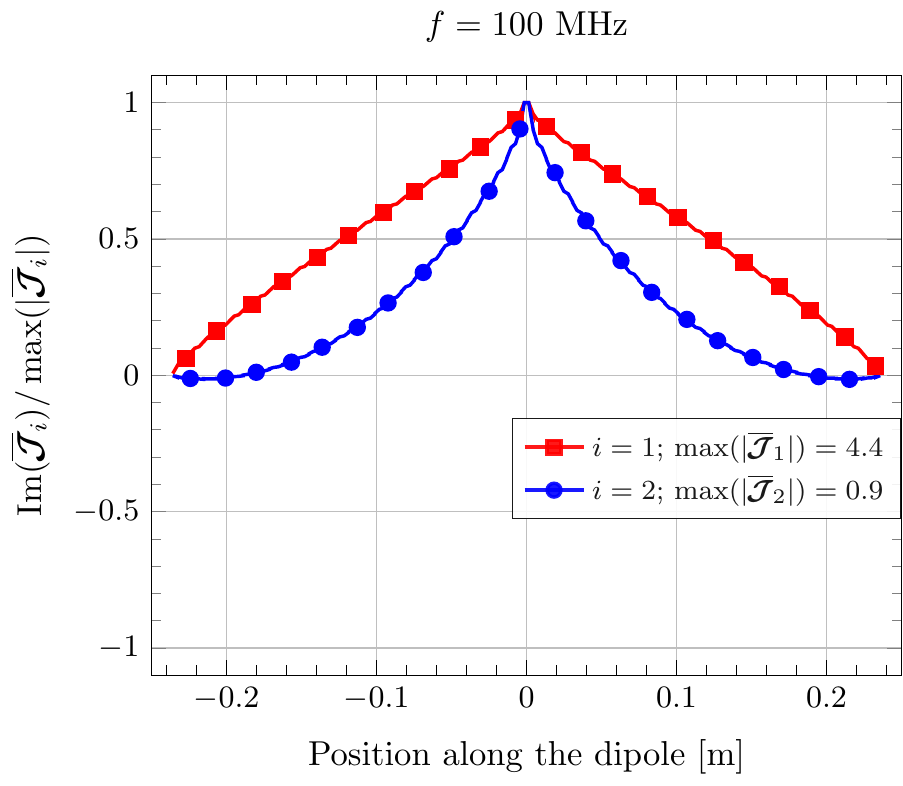}} 
    \hfill 
    \subfloat[ \label{fig:Dipole_antenna_J300}]{\includegraphics[width=0.33\textwidth]{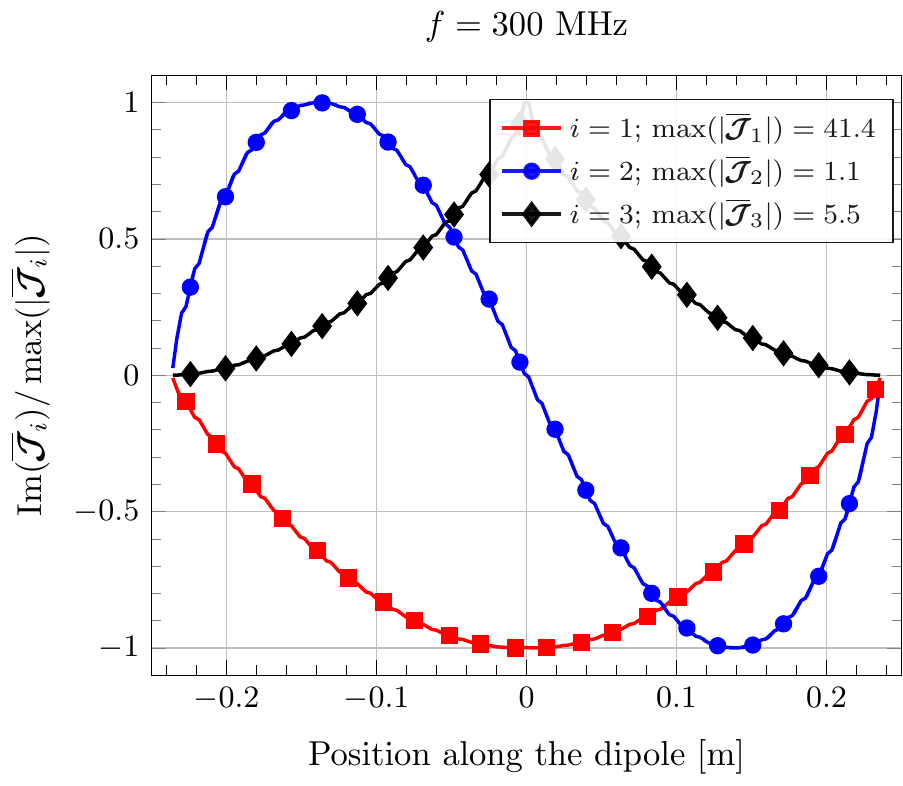}} 
    \hfill 
    \subfloat[ \label{fig:Dipole_antenna_J450}]{\includegraphics[width=0.32\textwidth]{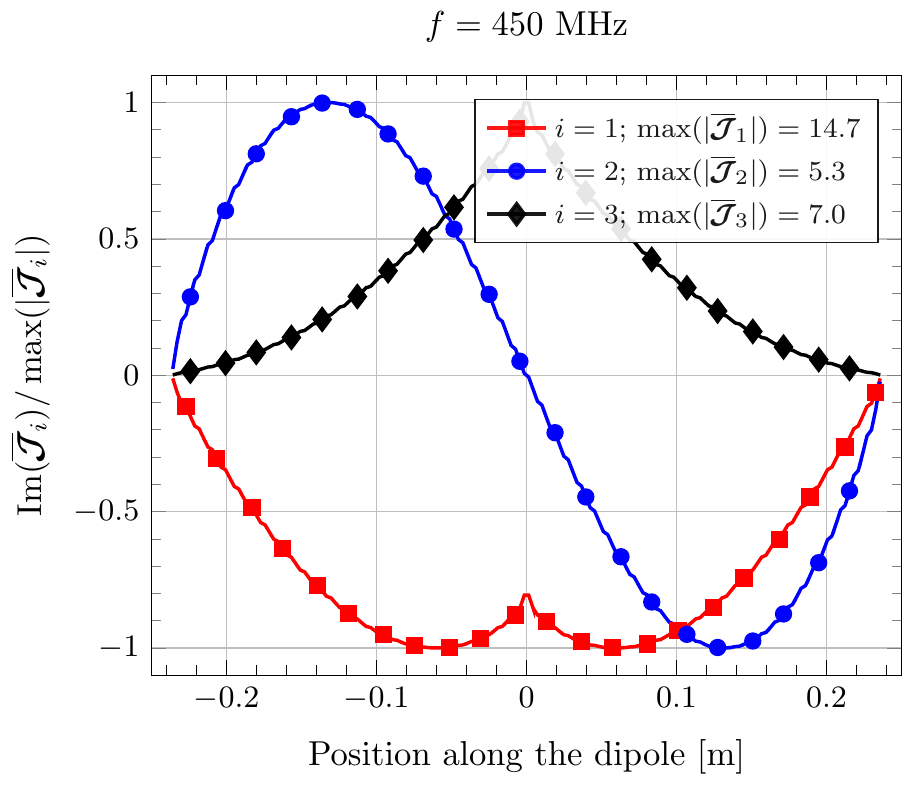}} 
    \hfill \null
\caption{Results for the center-fed strip dipole. (a) Norm of difference of $\matr{Q}$ computed with the direct and indirect approaches; (b) $\matr{S}_{11}$ and $\matr{S}_{11}'$; (c) Convergence of series for the YB estimate \eqref{eq:Sprime_YB2} and $\abs{\matr{S}_{11}'}$ \eqref{eq:Sprime_Energy2}; (d): Magnitude of time delay for select WS modes; (e): Magnitude of the $(1,i)$-th entry of $\widebar{\matr{W}}$ entry; (f) Phase of the $(1,i)$-th entry of $\widebar{\matr{W}}$; (g) WS mode currents at $f=100 \mathrm{MHz}$; (h) WS mode currents at $f = 300 \mathrm{MHz}$; (i) WS mode currents at $f=450 \mathrm{MHz}$.}
  \label{fig:Dipole_NewResults}
\end{figure*}

The methods of Sections \ref{sec:IE_Form} and \ref{sec:IE_Form_2} can be used to compute $\matr{Q}$ of arbitrary electromagnetic systems composed of PEC antennas and/or scatterers. Knowledge of $\matr{Q}$ in turn can be used to characterize average group delays or to compute the system's WS modes. These methods were applied in \cite[Sections IV.B--IV.C]{TAP_1}  to characterize the WS modes of a PEC strip, a cavity, and a simple antenna.

This section illustrates the evaluation of $\matr{S}'$, and along with it the frequency derivatives of other port observables, via \eqref{eq:Sprime_Energy}, i.e. as sums of renormalized energies of dominant WS modes weighed by the applicable eigenvector entries. 

The frequency derivative of an antenna ports' scattering matrix can be obtained by extracting the entries of the $M_g \times M_g$ block corresponding to the guided ports from $\matr{S}'$.
For one port antenna system, the resulting equation simplifies to \eqref{eq:Sprime_Energy2}.

To evaluate the frequency derivative of antenna patterns, note that they relate to their far-fields as
\begin{align}
    \vect{\cal F}_{\myparallel}(\theta,\phi) = e^{jkr} r  \vect{\cal E}^{\text{sca}}_{\myparallel}(r, \theta,\phi)\,.
    \label{eq:Fthetaphi}
\end{align} 
The derivative of $\vect{\cal F}_{\myparallel}(\theta,\phi)$ can be obtained by applying the chain rule to \eqref{eq:Fthetaphi}, making use of \eqref{eq:E_scatter} to compute $\vect{\cal E}_{\myparallel}^{\text{sca}}(r,\theta,\phi)$. 
The resulting expression requires $\matr{P}' = \matr{S}'$ which is computed using \eqref{eq:Sprime_Energy}.

Finally, to evaluate the frequency derivative of a scatterer's bistatic radar cross-section (RCS), note that it relates to its electric far-field and scattering pattern as
\begin{align} 
\text{RCS}(\theta,\phi) &= 
4\pi r^2 \vect{\cal E}^{\text{sca}}_{\myparallel}(r,\theta,\phi) \cdot \vect{\cal E}^{\text{sca}*}_{\myparallel}(r,\theta,\phi) \nonumber \\
&= 4 \pi \vect{\cal F}_{\myparallel}(\theta,\phi) \cdot \vect{\cal F}^*_{\myparallel}(\theta,\phi)\,. 
\label{eq:RCS_Eqn}
\end{align}
This relationship assumes that the scatterer is excited by a unit amplitude incident electric field.
The frequency derivative of the bistatic RCS can be computed by applying the chain rule to \eqref{eq:RCS_Eqn}, leveraging the above calculation for $\vect{\cal F}_{\myparallel}'(\theta,\phi)$.

\subsection{Dipole Antenna}
\label{sec:dipole_antenna}
First, consider a $z$-directed center-fed strip dipole of length $l=0.4746\mathrm{m}$ and width $w = 4\mathrm{mm}$. 
The spatial origin is located at the dipole center.
The antenna is fed by a $71 \Omega$ transmission line, which  matches its impedance at $f=300\mathrm{MHz}$.
The system is analyzed over the frequency range $150-450 \mathrm{MHz}$.
Matrices $\matr{S}$, $\matr{S}'$, and $\matr{Q}$ are constructed using $M_g = 1$ guided and $M_f = 126$ free-space ports.

Fig.~\ref{fig:Dipole_antenna_error} shows the relative norm of the difference between the $\matr{Q}$'s computed directly via \eqref{eq:Q_MoM_Final} and indirectly via \eqref{eq:QIndirect_Final}.  
The small difference between both results can be understood in light of the substitutions performed in Section \ref{eq:direct_indirect_equivalence} to show the equivalence between the direct and indirect method, some of which only hold true in the $M_f \rightarrow \infty$ limit.
The magnitude of the dipole reflection coefficient $\matr{S}_{11}$ and its frequency derivative $\matr{S}_{11}'$ are shown in Fig.~\ref{fig:Dipole_antenna_S}.
$\matr{S}_{11}$ was extracted from $\matr{S}$, which was computed using \eqref{eq:SMatrix_MoM}.
$\abs{\matr{S}_{11}'}$ was computed in three different ways: directly via \eqref{eq:Sprime_final}, using the energy-based formula \eqref{eq:Sprime_Energy2} with  $\bar{\varepsilon} = 10^{-2} \max \left(\abs{\widebar{\matr{Q}}_{ii}}\right)$, and via the YB estimator \eqref{eq:Sprime_YB_full}. For this dipole antenna, the results obtained using all three approaches match well over the entire frequency range.

To further understand why the YB estimate accurately computes $\abs{\matr{S}_{11}'}$ over the entire frequency range, $\matr{Q}$ is diagonalized and the resulting WS modes are sorted in descending order of their group delays' absolute value.
For this structure, all WS modes have non-negative time delays, i.e. $\widebar{\matr{Q}}_{ii} \ge 0$ for $i = 1,\hdots, M$.
Fig.~\ref{fig:Dipole_convergence} shows the convergence of the YB estimate \eqref{eq:Sprime_YB} and the energy-based series \eqref{eq:Sprime_Energy} for $\abs{\matr{S}_{11}'}$ vs. the number of modes retained in the series (by in effect varying $\bar{\varepsilon}$) at $f=100\mathrm{MHz}$
and $f=300\mathrm{MHz}$. Figs.~\ref{fig:Dipole_antenna_Timedelays}--\ref{fig:Dipole_antenna_angleW} show  $\widebar{\matr{Q}}_{ii}$ as well as the magnitude and phase of the corresponding $\widebar{\matr{W}}_{1i}$ at $f = 100\mathrm{MHz}$, $300\mathrm{MHz}$, and $450 \mathrm{MHz}$. Figs.~\ref{fig:Dipole_antenna_J100}--\ref{fig:Dipole_antenna_J450} show the current distribution on the dipole associated with the dominant WS modes at $f = 100\mathrm{MHz}$, $f = 300\mathrm{MHz}$, and $f = 450\mathrm{MHz}$. 
\begin{itemize}
\item At $100 \mathrm{MHz}$, only two WS modes have time delays greater than $\bar{\varepsilon} = 10^{-2} \max \left( \abs{\widebar{\matr{Q}}_{ii}} \right)$. Equation~\eqref{eq:Sprime_Energy2} therefore is accurate even if only two terms are retained in the sum. Moreover, the phases of $\widebar{\matr{W}}_{11}$ and $\widebar{\matr{W}}_{12}$ are equal, implying that \eqref{eq:Sprime_Energy2} and \eqref{eq:Sprime_YB} converge to the same value (Fig.~\ref{fig:Dipole_convergence} shows the convergence of the series for the YB estimate of  $\abs{\matr{S}_{11}'}$ at $f=100\mathrm{MHz}$). Using $\vect{\cal J} = \widebar{\vect{\cal J}} \,\widebar{\matr{W}}^\dag$, it follows that the total antenna current $\vect{\cal J}_1$ equals the sum of WS mode currents $\widebar{\vect{\cal J}}_{i}$ in Fig.~\ref{fig:Dipole_antenna_J100} scaled by $\widebar{\matr{W}}_{1i}^*$ in Figs.~\ref{fig:Dipole_antenna_absW}--\ref{fig:Dipole_antenna_angleW} for $i=1,2$.
Since $\big|\widebar{\vect{\cal J}}_2\big|$ is much smaller than $\big | \widebar{\vect{\cal J}}_1 \big|$,   so is WS mode \# 2's contribution to $\vect{\cal J}_1$. Finally, note that current $\vect{\cal J}_1$ has constant phase because phases of $\widebar{\matr{W}}_{11}$ and $\widebar{\matr{W}}_{12}$ are equal.
\item At $300 \mathrm{MHz}$ and $450 \mathrm{MHz}$, once again only two WS modes exhibit time delays greater than $\bar{\varepsilon} = 10^{-2} \max \left( \abs{\widebar{\matr{Q}}_{ii}} \right)$. 
Furthermore, as $\abs{\widebar{\matr{W}}_{12}} \approx 0$, WS mode \# 2 does not couple to the antenna port, i.e. it simply scatters off the dipole. The series in \eqref{eq:Sprime_Energy2} therefore can be truncated after the first WS mode and the YB estimator naturally yields the exact result (Fig.~\ref{fig:Dipole_convergence} shows the convergence of the series for the YB estimate of  $\abs{\matr{S}_{11}'}$ at $f=300\mathrm{MHz}$). 
At $f=300 \mathrm{MHz}$, ${\vect{\cal J}}_1$ is given by the sum of $\widebar{\vect{\cal J}}_i$ scaled by $\widebar{\matr{W}}_{1i}^*$ for $i = 1, 3$. Since,   $\big|\widebar{\vect{\cal J}}_1\big| $ is significantly larger than  $\big|\widebar{\vect{\cal J}}_3\big| $, $\vect{\cal J}_1$ has a near-constant phase.
At $f=450\mathrm{MHz}$,  $\vect{\cal J}_1$ is given by the sum of $\widebar{\vect{\cal J}}_i$ scaled by $\widebar{\matr{W}}_{1i}^*$ for $i = 1, 3,4$. Since   $\big|\widebar{\vect{\cal J}}_3\big| $ and $\big|\widebar{\vect{\cal J}}_4\big| $ are comparable to $\big|\widebar{\vect{\cal J}}_1\big| $, and the phases of $\widebar{\matr{W}}_{13}$ and $\widebar{\matr{W}}_{14}$ differ from those of $\widebar{\matr{W}}_{11}$, the phase of $\vect{\cal J}_1$ varies. This example demonstrates that while $\vect{\cal J}_1$ exhibiting constant phase guarantees accuracy of the YB estimator, it is not a  hard and fast  requirement.
\end{itemize}


\begin{figure*}[!t]
\null \hfill
    \subfloat[ \label{fig:yagi_antenna_S}]{\includegraphics[width=0.32\textwidth]{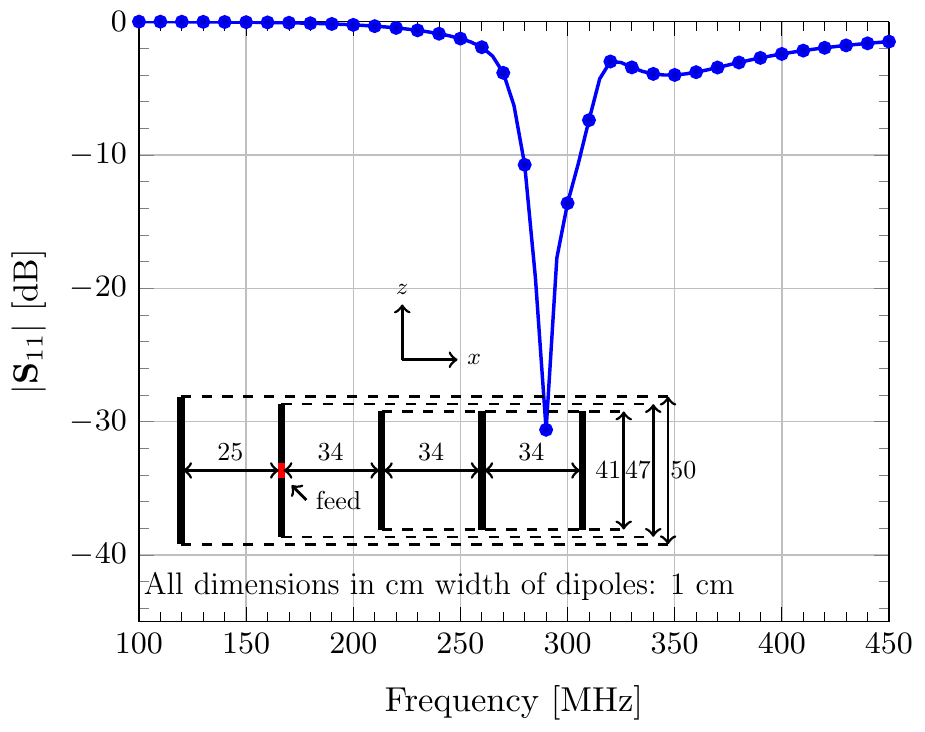}} 
    \hfill 
    \subfloat[ \label{fig:yagi_antenna_dS}]{\includegraphics[width=0.32\textwidth]{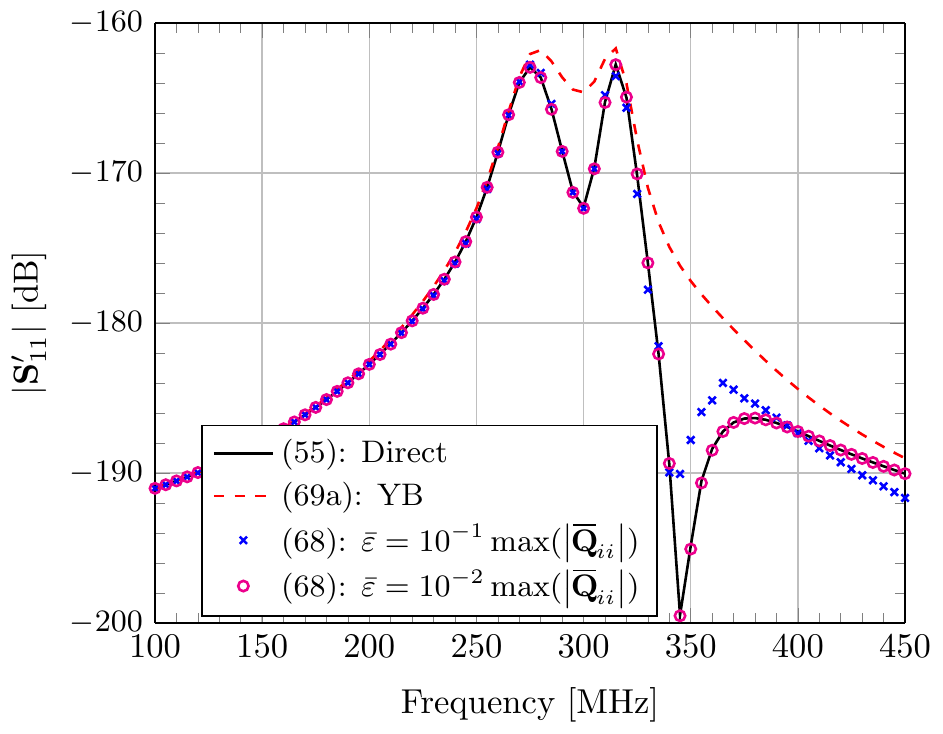}}  \hfill
    \subfloat[ \label{fig:yagi_origin_optimize}]{\includegraphics[width=0.31\textwidth]{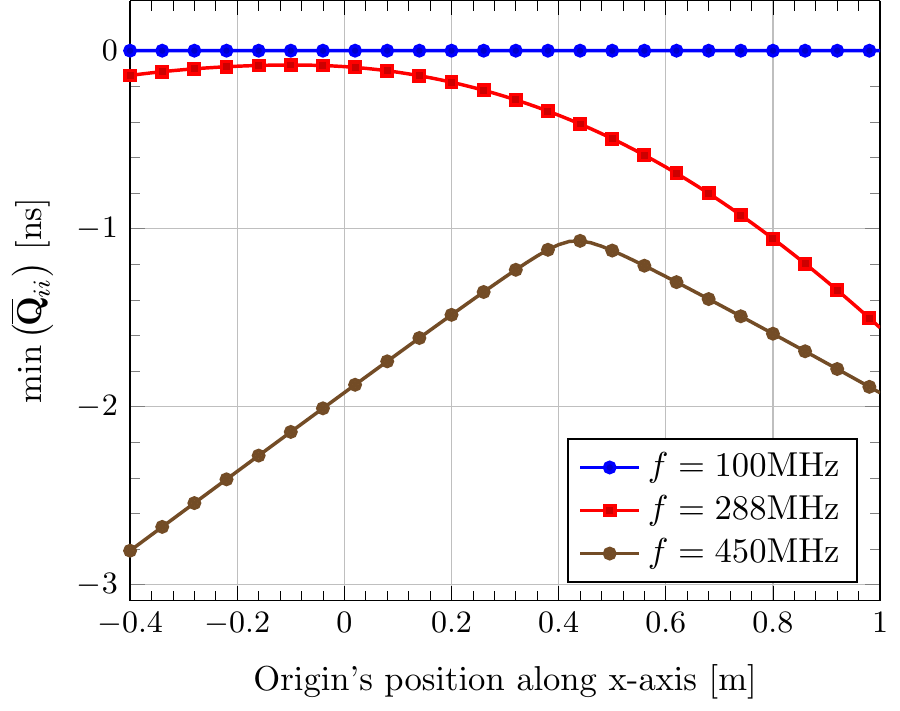}}
    \hfill \null \\
    \null \hfill 
    \subfloat[ \label{fig:yagi_antenna_ff}]{\includegraphics[width=0.31\textwidth]{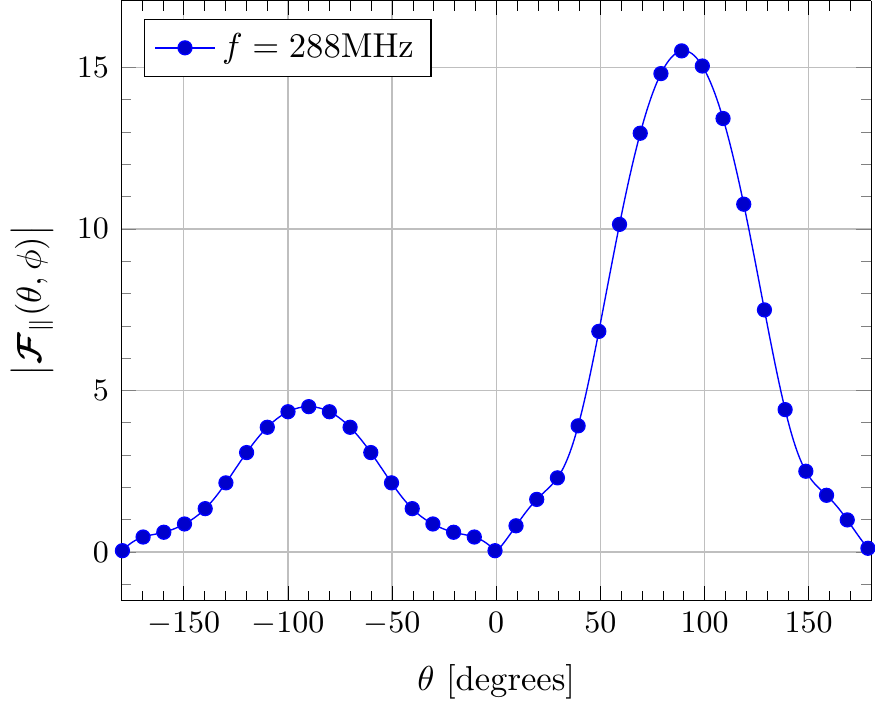}} \hfill
\subfloat[ \label{fig:yagi_antenna_dff}]{\includegraphics[width=0.32\textwidth]{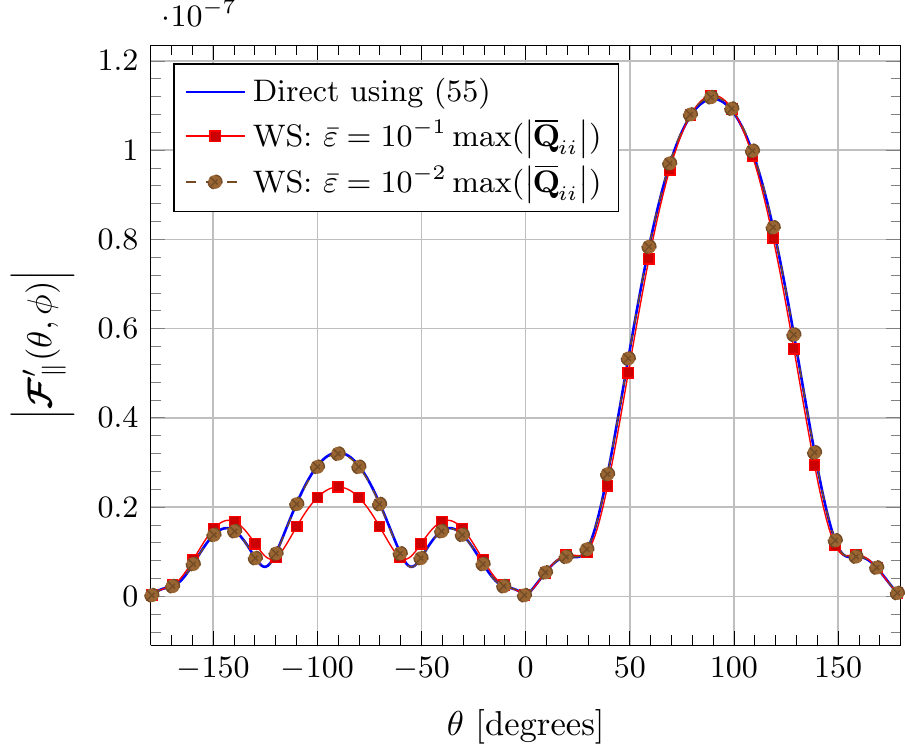}}
 \hfill \subfloat[\label{fig:yagi_antenna_series}]{\includegraphics[width=0.34\textwidth]{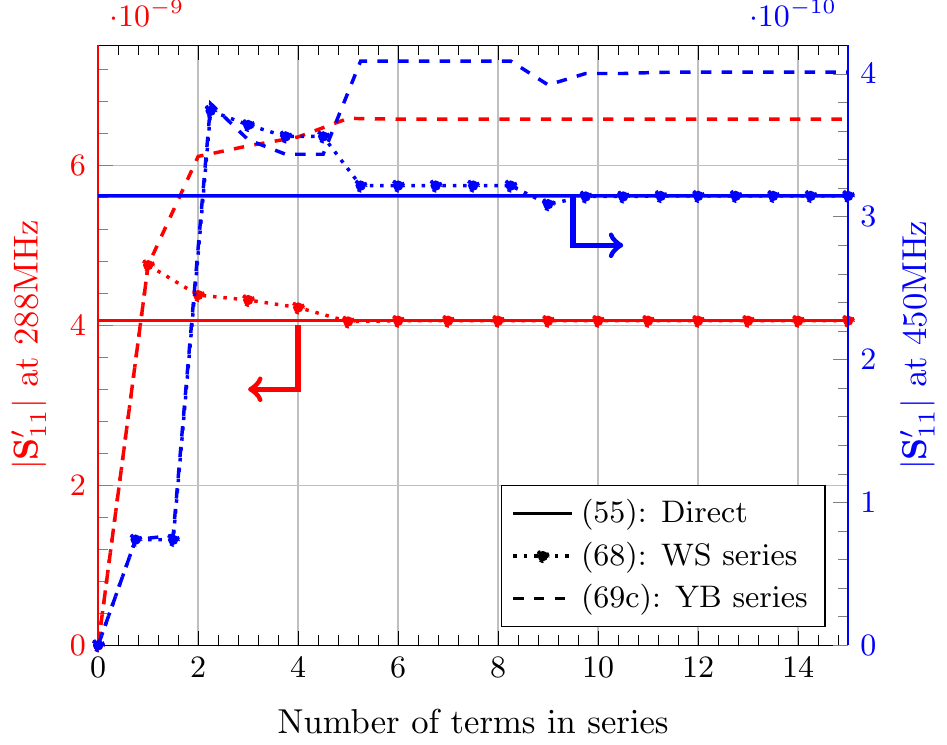}} \hfill 
\null \\
\null \hfill
\subfloat[ \label{fig:yagi_antenna_Timedelay}]{\includegraphics[width=0.33\textwidth]{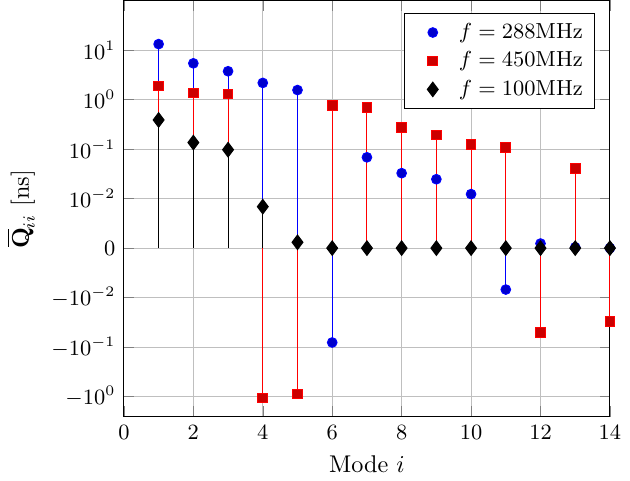}}  
\hfill 
\subfloat[ \label{fig:yagi_antenna_W}]{\includegraphics[width=0.33\textwidth]{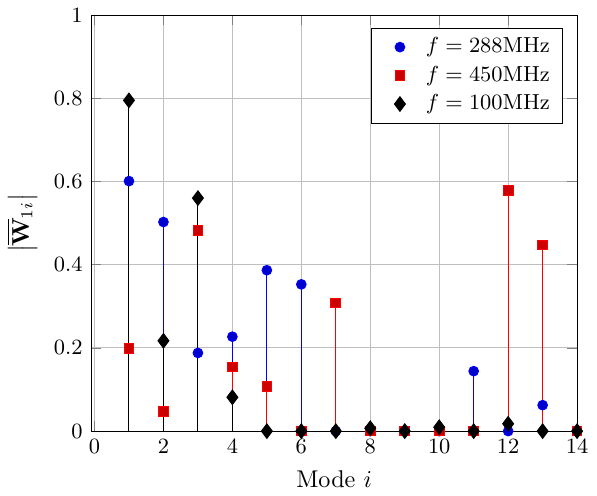}}
\hfill    
\subfloat[ \label{fig:yagi_antenna_Wangle}]{\includegraphics[width=0.33\textwidth]{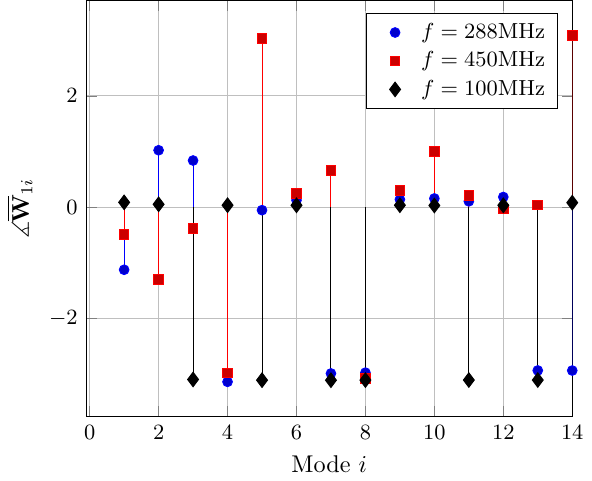}}
 \hfill \null
 \\
 \null \hfill   \subfloat[\label{fig:yagi_WS_J_1}]{\includegraphics[width=0.32\textwidth]{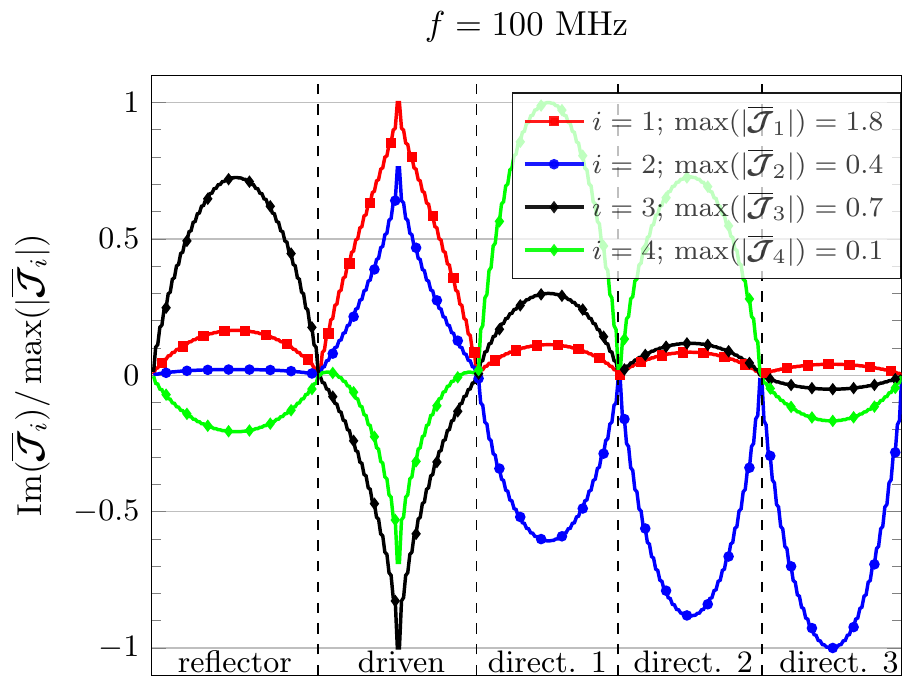}} \hfill 
\subfloat[ \label{fig:yagi_WS_J_2}]{\includegraphics[width=0.32\textwidth]{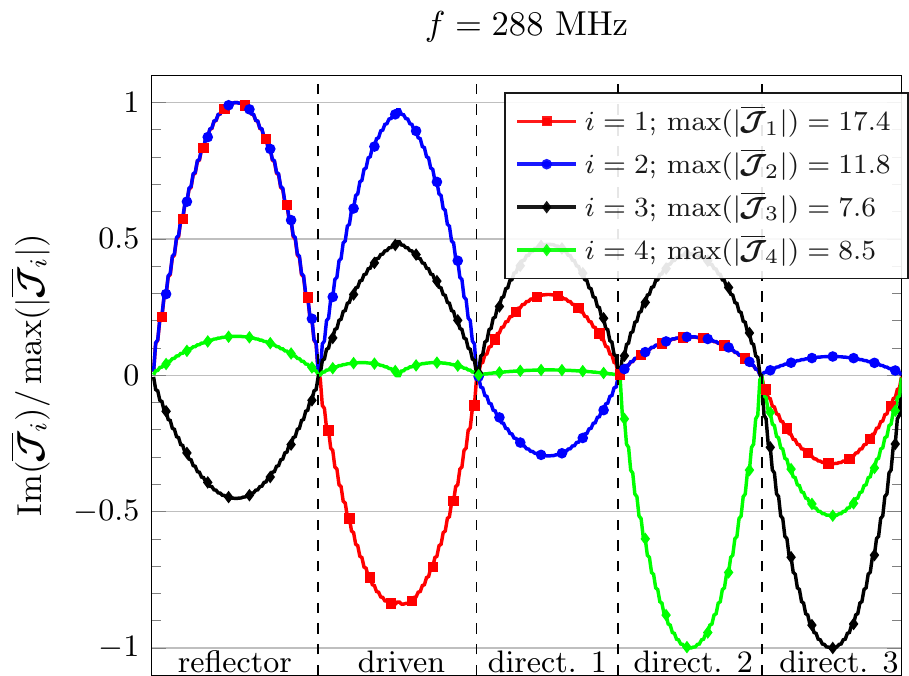}} \hfill
\subfloat[ \label{fig:yagi_WS_J_3}]{\includegraphics[width=0.32\textwidth]{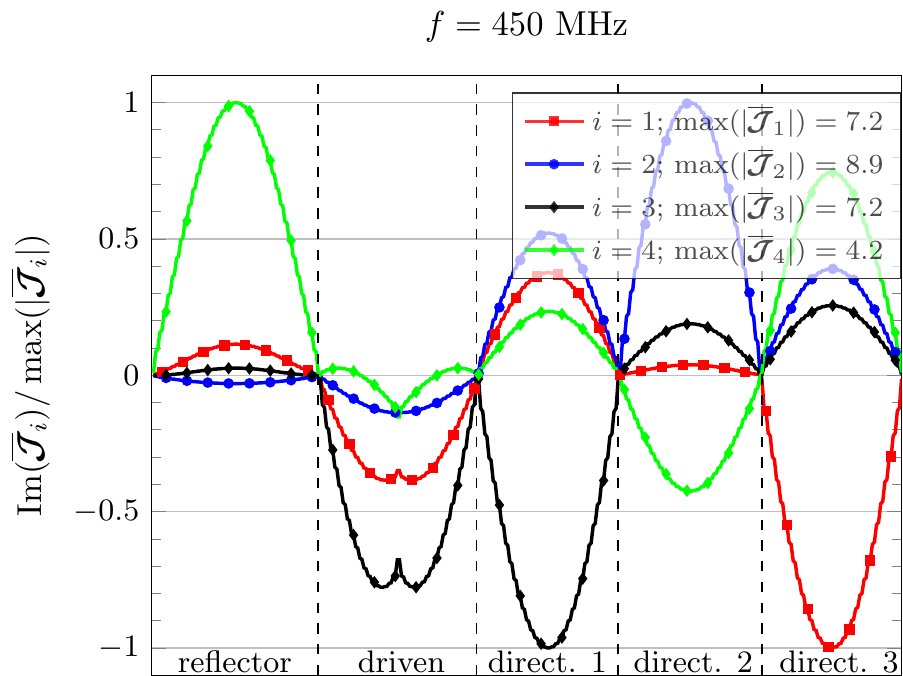}}
 \hfill \null 
    \caption{Results for the yagi antenna. (a) $\matr{S}_{11}$ (inset: geometry and dimensions); (b) $\matr{S}_{11}'$; (c) Smallest WS time delay vs. origin location along the $x$-direction; 
     (d) far-field pattern of the antenna in the $xz$ plane; (e) frequency derivative of the far-field pattern of the antenna in the $xz$ plane;
    (f) Convergence of \eqref{eq:Sprime_Energy2} and \eqref{eq:Sprime_YB} at $288\mathrm{MHz}$ and $450 \mathrm{MHz}$;  
    (g) Time delay of WS modes; (h) Magnitude of the $(1,i)$-th entry of $\widebar{\matr{W}}$; (i) Phase of the $(1,i)$-th entry of $\widebar{\matr{W}}$;  (j) Select WS mode currents at $f = 100 \mathrm{MHz}$; (k) Select WS mode currents at $f = 288\mathrm{MHz}$; (l) Select WS mode currents at $f = 450\mathrm{MHz}$.}
   \label{fig:Yagi_Results}
\end{figure*}

\subsection{Yagi Antenna}
\label{sec:yagi_example}
Second, consider a five-element Yagi antenna composed of one center-fed dipole, three directors, and one reflector (see the inset in Fig.~\ref{fig:yagi_antenna_S} for antenna dimensions). All elements are $z$-directed strips. The array extends along the $x$-direction.
The origin of the coordinate system initially is located at the center of the driven dipole.
The antenna is fed by a  $56 \Omega$  transmission line, which matches its impedance at $288 \mathrm{MHz}$. The system was analyzed over the frequency range of $100-450 \mathrm{MHz}$. Matrices $\matr{Q}$, $\matr{S}$, and $\matr{S}'$ were constructed using $M_g = 1$ and $M_f = 720$ ports.

The magnitude of the antenna port's reflection coefficient $\matr{S}_{11}$ and its frequency derivative $\matr{S}_{11}'$ are shown in Figs.~\ref{fig:yagi_antenna_S} and~\ref{fig:yagi_antenna_dS}.
$\matr{S}_{11}$ was extracted from $\matr{S}$, which was computed using \eqref{eq:SMatrix_MoM}.
$\abs{\matr{S}_{11}'}$ was computed in four different ways: directly via \eqref{eq:Sprime_final}, using the energy-based formula \eqref{eq:Sprime_Energy2} with  $\bar{\varepsilon} = 10^{-1} \max \left(\abs{\widebar{\matr{Q}}_{ii}}\right)$ and $\bar{\varepsilon} = 10^{-2} \max \left(\abs{\widebar{\matr{Q}}_{ii}}\right)$, and via the YB estimator \eqref{eq:Sprime_YB_full}. 
The results obtained from all three methods match up to $250 \mathrm{MHz}$; beyond this frequency the energy-based calculation with $\bar{\varepsilon} = 10^{-1} \max \left(\abs{\widebar{\matr{Q}}_{ii}}\right)$ and the YB estimate become inaccurate.  
 
Further comparison of the performance of the different methods for computing $\abs{\matr{S}_{11}'}$ requires reconsideration of the position of the spatial origin.  As discussed in Section \ref{sec:traceQ}, the choice of the origin affects the group delays / renormalized energies represented by the $\widebar{\matr{Q}}_{ii}$.  For a structure like the Yagi antenna that lacks symmetry, the question arises as to which choice of origin simplifies the interpretation of the WS mode picture.  
Here, a parametric sweep was performed to choose the origin along the $x$-axis that minimizes the absolute value of the smallest group delay; in practice this choice oftentimes results in the series in \eqref{eq:Sprime_Energy2} converging with the fewest possible terms.
Fig.~\ref{fig:yagi_origin_optimize} shows the smallest WS time delay as a function of origin's position ($x=0$ is the location of the driven strip dipole).  For this Yagi antenna, the smallest group delay is seen to be frequency dependent and always negative.  For $f=100\mathrm{MHz}$, $f=288\mathrm{MHz}$, and $f=450\mathrm{MHz}$, the ``optimal" origin is located at at $x=0 \mathrm{m}$, $x = -0.10 \mathrm{m}$, and $x = 0.44\mathrm{m}$, respectively. These origins are used in the WS decompositions discussed below.

The magnitude of the antenna's far-field in the $xz$ plane at $f=288 \mathrm{MHz}$ is shown in Fig.~\ref{fig:yagi_antenna_ff}.  
The frequency derivatives of the far-field computed
using $\matr{S}'$ obtained from \eqref{eq:Sprime_final} and   \eqref{eq:Sprime_Energy} with $\bar{\varepsilon} = 10^{-2} \max \left(\abs{\widebar{\matr{Q}}_{ii}}\right)$ match one another as shown in Fig.~\ref{fig:yagi_antenna_dff}.


To further explain of the different results obtained for $\abs{\matr{S}_{11}'}$, consider the group delays $\widebar{\matr{Q}}_{ii}$ and the corresponding entries of $\widebar{\matr{W}}_{i}$ at $f = 100\mathrm{MHz}$, $f = 300\mathrm{MHz}$, and $f = 450 \mathrm{MHz}$~(Figs.~\ref{fig:yagi_antenna_Timedelay}--\ref{fig:yagi_antenna_Wangle}). 
The currents along each element of the Yagi antenna due to select WS modes are shown in Fig.~\ref{fig:yagi_WS_J_1}--\ref{fig:yagi_WS_J_3}.
The following observations are in order:
\begin{itemize}
    \item At $100 \mathrm{MHz}$, only four modes have a non-negligible time delays when  $\bar{\varepsilon} = 10^{-2} \max \left(\abs{\widebar{\matr{Q}}_{ii}}\right)$. 
    Furthermore, the $\widebar{\matr{W}}_{1i}$ for $i = 1,\hdots, 4$ all have the same phase. 
    It immediately follows that the YB estimate agrees with the true $\abs{\matr{S}_{11}'}$. The current on the Yagi antenna $\vect{\cal J}_1$ can be computed as a superposition of WS mode currents $\widebar{\vect{\cal J}}_i$ for $i = 1,\hdots,4$ (Fig.~\ref{fig:yagi_WS_J_1}) scaled by the applicable $\widebar{\matr{W}}_{1i}^*$.  This current has constant phase due to the $\widebar{\matr{W}}_{1i}$ for $i=1,\hdots,4$ having constant phase.
    
    \item   At $288 \mathrm{MHz}$, five WS modes exhibit non-negligible group delays using $\bar{\varepsilon} = 10^{-2} \max \left(\abs{\widebar{\matr{Q}}_{ii}}\right)$. Therefore, the series in \eqref{eq:Sprime_Energy2} and \eqref{eq:Sprime_YB} converge using five terms (Fig.~\ref{fig:yagi_antenna_series}). However, since the phases of the corresponding $\widebar{\matr{W}}_{1i}$ vary, the YB estimate does not yield the true $\abs{\matr{S}_{11}'}$. 
    The antenna current $\vect{\cal J}_1$ is the superposition of $\widebar{\vect{\cal J}}_i$ (Fig.~\ref{fig:yagi_WS_J_2}) for $i=1,\hdots,6, 11,13$ scaled by the applicable $\widebar{\matr{W}}_{1i}$. Since the phases of the $\widebar{\matr{W}}_{1i}^*$ are not constant, the phase of $\vect{\cal J}_1$ also varies across the Yagi.
    \item Finally, at $450 \mathrm{MHz}$ there are $14$ modes that exhibit non-negligible time delays using $\bar{\varepsilon} = 10^{-2} \max \left(\abs{\widebar{\matr{Q}}_{ii}}\right)$. 
    Several of these modes have $\abs{\widebar{\matr{W}}_{1i}} \approx 0$ (e.g. 6, 8, and 9) and hence do not couple into the antenna port; their fields simply scatter off the antenna.
    The antenna current $\vect{\cal J}_1$ is obtained from the superposition of mode currents $\widebar{\vect{\cal J}}_i$ ( Fig.~\ref{fig:yagi_WS_J_3}), and has variable phase just like its scaling factors $\widebar{\matr{W}}_{1i}$.  
\end{itemize}

\begin{figure*}[t]
\null \hfill
    \subfloat[ \label{fig:Torus_fig}]{\input{graphics/Results/fig_Torus} }  \hfill
    \subfloat[ \label{fig:Torus_RCS} ]{\includegraphics[width=0.33\textwidth]{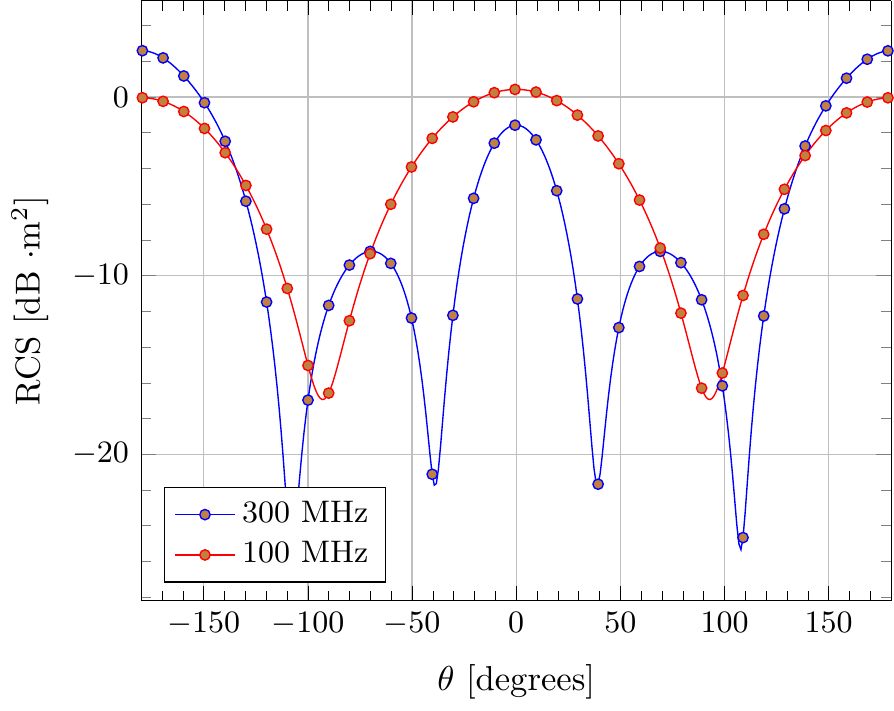}} 
    \hfill 
    \subfloat[ \label{fig:Torus_dRCS}]
    {\includegraphics[width=0.33\textwidth]{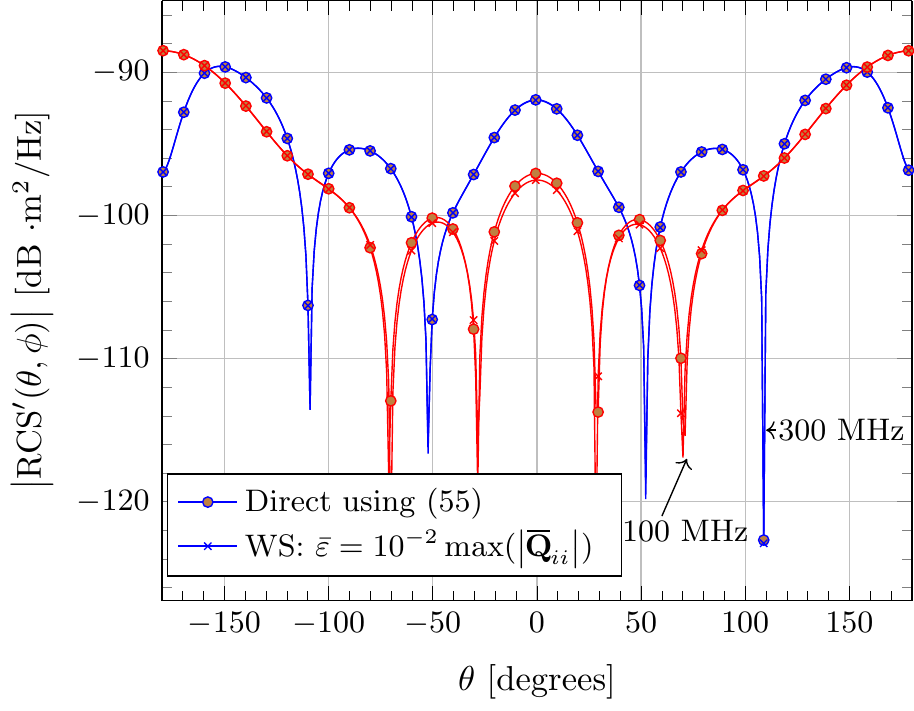}}
    \hfill
    \null 
    \caption{(a) PEC torus considered; (b) torus' bistatic RCS in the $xz$ plane; (c) frequency derivative of RCS in the $xz$ plane.}
   \label{fig:Torus_Results}
\end{figure*}
\subsection{Torus}

Finally, consider the PEC torus obtained by revolving a circle of radius $0.2 \mathrm{m}$ about the $z$-axis along a circle of radius $0.5 \mathrm{m}$ in the $xy$ plane (Fig.~\ref{fig:Torus_fig}). 
The torus is excited by an $x$-polarized incident plane-wave traveling in the $-\unit{z}$ direction.
The torus' bistatic RCS at $100 \mathrm{MHz}$ and $300 \mathrm{MHz}$ in the $xz$ plane is shown in Fig.~\ref{fig:Torus_RCS}.
The scattering matrix $\matr{S}$  and WS time delay matrix $\matr{Q}$ are computed using the integral formulations in Sec.~\ref{sec:IE_Form} and Eqns.~\eqref{eq:Qincinc_final}--\eqref{eq:Qscasca_final}  with $M_f = 126$.
Using $\matr{S}'$ obtained from \eqref{eq:Sprime_final} and from  \eqref{eq:Sprime_Energy} with $\bar{\varepsilon}  = 10^{-2} \max \left (\abs{\widebar{\matr{Q}}_{ii}} \right )$, the frequency derivative of the radar cross-section of the torus is computed at $100 \mathrm{MHz}$ and $300 \mathrm{MHz}$ (Fig.~\ref{fig:Torus_dRCS}).
It is observed that the frequency derivative of the RCS can be computed accurately at $100 \mathrm{MHz}$ and $300 \mathrm{MHz}$ using the proposed energy-based expression.


%% file: graphics/Results/fig_Torus.tex

\begin{tikzpicture}[thick,scale=0.8, every node/.style={scale=0.8}]
\centering
\node at (0,0) {\includegraphics[width=0.30\textwidth]{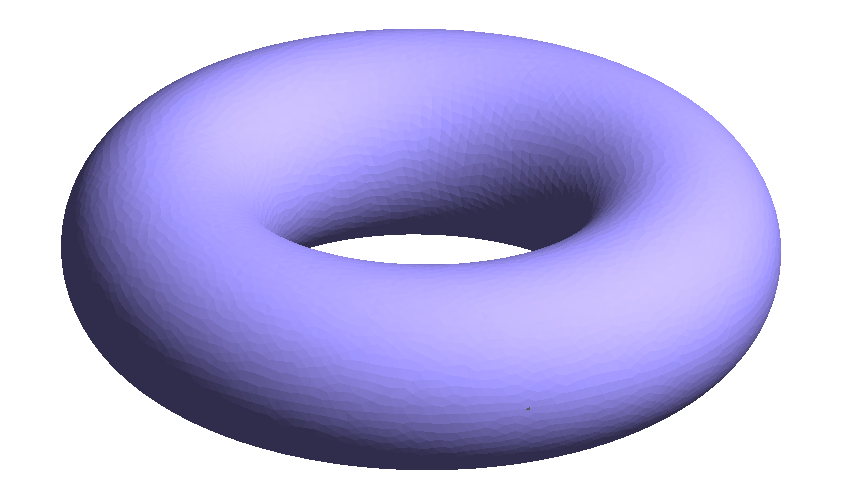}};

\begin{scope}[shift={(-2.75,-2)}]
\draw[->]  (0,0) -- (0,1);
\draw[->]  (0,0) -- (1,0);
\draw[->]  (0,0) -- (0.2,0.5);
\node at (1,0) [right] {\small $\unit{x}$};
\node at (0,1) [above] {\small $\unit{z}$};
\node at (0.2,0.5) [right] {\small $\unit{y}$};
\end{scope}

\begin{scope}[shift={(0,3)}]
\draw[->]  (0,0) -- (0,-0.5);
\draw[->]  (0,0) -- (0.5,0);
\node at (0,-0.5) [below] { $\unit{k}$};
\node at (0.5,0) [right] { $\vect{\cal E}^{\text{inc}}$};
\end{scope}

\draw[red,very thick, ->] (0,0) -- (1.7,0);
\node[black] at (1.0,-0.05)[below] {\large $0.5\mathrm{m}$};
\begin{scope}[yscale=0.5, shift={(0,0.3)}]
\draw[dashed] (0,0) circle (1.8);
\end{scope}

\begin{scope}[yscale=0.7, shift={(-1.7,0)}]
\draw[dashed] (0,0) circle (0.65);
\draw[red,thick, ->] (0,0) -- (-0.65,0);
\end{scope}
\node at (-2.3,0) [left] {\large $0.2\mathrm{m}$};

\end{tikzpicture}

%% file: Appendix.tex
\appendices

\section{Vector Spherical Harmonics and Properties}
\label{appendix:VSH_VSW}

\subsection{Vector Spherical Waves}
\label{appendix:VSW}

Electric fields in source-free shells in $\mathbb{R}^3$ can be expanded as
\begin{align}
    \vect{\cal E}(\vr) = \sum_{p} a_p \sqrt{Z} \vect{\cal B}_p(\vr)
    \label{eq:Appendix_Eexpand}
\end{align}
where the $\vect{\cal B}_p(\vr)$ are vector spherical wave (VSW) functions.  $\vect{\cal B}_p(\vr) = \vect{\cal I}_p(\vr), \vect{\cal O}_p(\vr),$  or $\vect{\cal W}_p(\vr)$ when representing incoming, outgoing, or standing waves, respectively.

The incoming VSW $\vect{\cal I}_p(\vr)$ with $p = (\tau, l,m)$ is defined as~\cite{colton2013integral, Hansen}
\begin{subequations}
\begin{align}
\vect{\cal I}_{1lm}(r,\theta,\phi)  &=  k j^{l+1} h_{l}^{(1)}(k r) \vect{\cal X}_{1 lm} (\theta,\phi) \label{eq:I_VSH_TE}
 \\
\vect{\cal I}_{2lm}(r,\theta,\phi)  &= \frac{j^{l}}{r}  \frac{\partial  k r h_{l}^{(1)}(k r)}{\partial (kr)}     \vect{\cal X}_{2 lm} (\theta,\phi)  \label{eq:I_VSH_TM} \\
&\quad + \sqrt{l(l+1)} \frac{j^{l}}{r}   h_{l}^{(1)}(k r) \vect{\cal X}_{3lm}(\theta,\phi)\,, \nonumber
\end{align}
\end{subequations}
where $\tau = \{1,2\} = \{\mathrm{TE}, \mathrm{TM}\}$ (to $r$), $l = \{ 1, \hdots, \infty \}$ and $m = \{-l,\hdots, l\}$ are modal indices, and $h_{l}^{(1)}(z)$ is the $l$-th order spherical Hankel function of the first kind~\cite{Abr64}.  
The vector spherical harmonics (VSH) $\vect{\cal X}_{\tau lm}(\theta,\phi)$ are
\begin{subequations}
\begin{align}
\vect{\cal X}_{1lm} (\theta, \phi) &= \frac{1}{\sqrt{l(l+1)}} \nabla \times \left( \vect{r} Y_{lm}(\theta,\phi)\right)  \\ 
\vect{\cal X}_{2lm}(\theta, \phi) &= \frac{1}{\sqrt{l(l+1)}} r \nabla Y_{lm}(\theta,\phi)\\
\vect{\cal X}_{3lm}(\theta,\phi) &= \unit{r} Y_{lm}(\theta,\phi)\,.
\end{align}
\end{subequations}
where the scalar spherical harmonic $Y_{lm}(\theta,\phi)$ is~\cite{kristensson, Hansen}
\begin{align}
Y_{lm}(\theta,\phi) = (-1)^{m} \sqrt{\frac{2l+1}{4 \pi} \frac{(l-m)!}{(l+m)!}} P_{l}^{m}(\cos \theta) e^{j m \phi} .
\label{eq:Ylm}
\end{align}
Here, $P_{l}^{m}(x)$ is the associated Legendre polynomial of degree $l$ and order $m$~\cite{Abr64}.

Outgoing electric fields are expanded in VSWs $\vect{\cal O}_p(\vr)$, obtained by swapping spherical Hankel functions of the first kind $h_l^{(1)}(kr)$ in \eqref{eq:I_VSH_TE}--\eqref{eq:I_VSH_TM} for their second kind counterparts $h_l^{(2)}(kr)$.

Standing electric fields are expanded in VSWs $\vect{\cal W}_p(\vr)$, obtained by swapping spherical Hankel functions of the first kind $h_l^{(1)}(kr)$ in \eqref{eq:I_VSH_TE}--\eqref{eq:I_VSH_TM} for twice the spherical Bessel function $j_l(kr)$.

Outgoing and standing VSWs can be expressed in terms of incoming VSWs as
\begin{subequations}
\begin{align}
    \vect{\cal O}_{\tau l m}(\vr) &= (-1)^{l+\tau + m} \vect{\cal I}_{\tau l (-m)}^*(\vr) \\
    \vect{\cal W}_{\tau l m}(\vr) &= \vect{\cal I}_{\tau l m} + \vect{\cal O}_{\tau l m} \\
    &= \vect{\cal I}_{\tau l m}(\vr) + (-1)^{l+\tau+m} \vect{\cal I}^*_{\tau l (-m)}(\vr)\,.
\end{align}
\end{subequations}
Finally, note that expansion \eqref{eq:Appendix_Eexpand} for the electric field implies an expansion of the magnetic field as
\begin{align}
    \vect{\cal H} (\vr) &= \frac{j}{k \sqrt{Z}} \sum_{p} a_p \nabla \times \vect{\cal B}_p \nonumber \\
    &= \frac{j}{\sqrt{Z}} \sum_{p} a_p \tilde{\vect{\cal B}}_p(\vr)
\end{align}
where $\tilde{\vect{\cal B}}_{\tau l m}(\vr) = j (-1)^{\tau+1}  \vect{\cal B}_{\bar{\tau} l m}(\vr)$, $\bar{1} = 2$, and $\bar{2} = 1$.

\subsection{Properties of Vector Spherical Harmonics}

The VSHs obey several key properties that are used throughout the paper.  Below, $p = (\tau,l,m)$ and $q = (\tau',l',m')$.
\begin{enumerate}
    \item Orthogonality:
\begin{align}
    \int_{0}^{2\pi} \int_{0}^{\pi} \vect{\cal X}_{p} (\theta,\phi) &\cdot \vect{\cal X}_{q}^*(\theta,\phi) \sin \theta d\theta d\phi \nonumber \\
    &= \delta_{\tau \tau'} \delta_{l l'} \delta_{m m'}\,. \label{eq:VSH_Orthogonality}
\end{align}
\item Conjugation:
\begin{align}
    \vect{\cal X}_{\tau l m}(\theta, \phi) &= (-1)^{m} \vect{\cal X}_{\tau l (-m)}^* (\theta, \phi) \,.
    \label{eq:VSH_Conjugation}
\end{align}
\item Cross-product:
\begin{align}
    \vect{\cal X}_{1lm}(\theta,\phi) \times \vect{\cal X}_{2lm}(\theta,\phi)  &=  \unit{r}\,. \label{eq:VSH_Crossproduct}
\end{align}
\end{enumerate}

\subsection{Vector Spherical Waves - Large Argument Approximation}

\label{appendix:large_VSW}

For large arguments, incoming, outgoing, and standing wave VSWs can be approximated as
\begin{subequations}
\begin{align}
    \vect{\cal I}_{\tau lm,\myparallel}(\vr)  &=  \frac{e^{jkr}}{r} \vect{\cal X}_{\tau lm}(\theta,\phi) \label{eq:VSH_I_TE_parallel}\\ 
    \vect{\cal O}_{\tau lm,\myparallel}(\vr)  &=   (-1)^{l+\tau} \frac{e^{-jkr}}{r} \vect{\cal X}_{\tau lm}(\theta,\phi) \label{eq:VSH_O_TE_parallel} \\
    \vect{\cal W}_{\tau lm,\myparallel}(\vr)  &=   \left[\frac{e^{jkr}}{r} + (-1)^{l+\tau} \frac{e^{-jkr}}{r} \right] \vect{\cal X}_{\tau lm} (\theta,\phi)\,. \label{eq:VSH_J_TE_parallel}
\end{align}
\end{subequations}
The frequency derivative of $\vect{\cal W}_{p,\myparallel}(\vr)$ is
\begin{subequations}
\begin{align}
    \vect{\cal W}_{\tau lm,\myparallel}'(\vr)  &= \frac{j k }{\omega}  \left[e^{jkr} - (-1)^{l+\tau} e^{-jkr} \right] \vect{\cal X}_{\tau lm} (\theta,\phi)\,. \label{eq:VSH_dJ_TE_parallel} 
\end{align}
\end{subequations}

\subsection{VSW Expansion of the $\vect{\cal L}$ Operator}

For observation points $\vr$ outside the sphere of radius $a$, the scattered field may be written in terms of VSWs \cite{kristensson}
\begin{subequations}
\begin{align}
\vect{\cal E}^{\text{sca}}(\vr) &=  \vect{\cal L} \left[\vect{\cal J}\right](\vr) \nonumber \\
&= - \frac{Z}{2} \sum_{t} \vect{\cal I}_t^*(\vr) \int_{d\Omega_s} \vect{\cal W}_t(\vrp) \cdot \vect{\cal J}_p(\vrp)d\vrp   \label{eq:VSW_expansion1} \\
&= - \frac{Z}{2} \sum_{t} \vect{\cal O}_t(\vr) \int_{d\Omega_s} \vect{\cal W}_t^*(\vrp) \cdot \vect{\cal J}_p(\vrp)d\vrp   \label{eq:VSW_expansion1b}
\end{align}
\end{subequations}
for $t = (\tau, l,m)$, $\tau = \{1,2\}$, $l = 1, \hdots, \infty$, $m = -l, \hdots, l$.
For observation points on $d\Omega_f$, \eqref{eq:VSW_expansion1} simplifies to
\begin{subequations}
\begin{align}
    \vect{\cal E}_\myparallel^{\text{sca}}(\vr) &=  \vect{\cal L}_\myparallel \left[\vect{\cal J}\right](\vr) \nonumber \\
&= - \frac{Z}{2} \sum_{t} \vect{\cal I}_{t,\myparallel}^*(\vr) \int_{d\Omega_s} \vect{\cal W}_t(\vrp) \cdot \vect{\cal J}_p(\vrp)d\vrp \,.  \label{eq:VSW_expansion2} \\
&= - \frac{Z}{2} \sum_{t} \vect{\cal O}_{t,\myparallel}(\vr) \int_{d\Omega_s} \vect{\cal W}_t^*(\vrp) \cdot \vect{\cal J}_p(\vrp)d\vrp \,.  \label{eq:VSW_expansion2b}
\end{align}
\end{subequations}

\section{Direct Computation of $\matr{Q}_{qp}^{\alpha,\beta}$ Integrals}
\label{App:QComputation}
This Appendix discusses how to evaluate integrals in \eqref{eq:Qqp_s}.

\subsection{Basic Identities}
\label{sec:Qcomp_BasisIdentities}
Expressions for $\matr{Q}_{qp}^{\alpha,
\beta}$ can be derived by manipulating Maxwell's equations and their frequency derivatives.
Consider two sets of electromagnetic fields in $\Omega$: $\{\vect{\cal E}_p^\beta(\vr), \vect{\cal H}_p^\beta(\vr) \}$ and $\{\vect{\cal E}_q^\alpha(\vr), \vect{\cal H}_q^\alpha(\vr) \}$.
The frequency derivative of Maxwell's equations for fields $\{\vect{\cal E}_p^\beta(\vr), \vect{\cal H}_p^\beta(\vr) \}$ reads
\begin{subequations}
\begin{align}
\nabla \times \vect{\cal H}_p^{\beta}{'}(\vr) &=  j \varepsilon \vect{\cal E}_p^{\beta}(\vr) + j\omega \varepsilon \vect{\cal E}_p^{\beta}{'}(\vr) + \vect{\cal J}_p^{\beta}{'}(\vr) \label{eq:Q_Comp_3} \\
\nabla \times \vect{\cal E}_p^{\beta}{'}(\vr) &= -j \mu \vect{\cal H}_p^{\beta}(\vr) - j\omega \mu \vect{\cal H}_p^{\beta}{'}(\vr)  \,. \label{eq:Q_Comp_4}
\end{align}
\end{subequations}
If $\beta = \text{inc}$, then $\vect{\cal J}_p^{\beta}(\vr) = 0$ since the incident field is source-free (within $\Omega$); otherwise, $\vect{\cal J}_p^{\beta}(\vr) = \vect{\cal J}_p(\vr)$. 
The conjugate of Maxwell's equations for fields $\{\vect{\cal E}_q^\alpha(\vr), \vect{\cal H}_q^\alpha(\vr) \}$ reads
\begin{subequations}
\begin{align}
\nabla \times \vect{\cal H}_q^{\alpha*}(\vr) &= -j\omega \varepsilon \vect{\cal E}_q^{\alpha*}(\vr) + \vect{\cal J}_q^{\alpha*}(\vr) \label{eq:Q_Comp_5} \\
\nabla \times \vect{\cal E}_q^{\alpha*}(\vr) &= j\omega \mu \vect{\cal H}_q^{\alpha*}(\vr) \,. \label{eq:Q_Comp_6}
\end{align}
\end{subequations}
Adding the dot-product of~\eqref{eq:Q_Comp_5} and $\frac{1}{2} \vect{\cal E}_p^{\beta}{'}(\vr)$ to the dot-product of~\eqref{eq:Q_Comp_3} and $\frac{1}{2}\vect{\cal E}_q^{\alpha*}$ yields
\begin{align}
\frac{1}{2}&\vect{\cal E}_p^{\beta}{'}(\vr) \cdot \nabla \times \vect{\cal H}_q^{\alpha*}(\vr) + \frac{1}{2} \vect{\cal E}_q^{\alpha*}(\vr) \cdot \nabla \times \vect{\cal H}_p^{\beta}{'}(\vr) \nonumber \\
&=   \frac{j}{2}\varepsilon \vect{\cal E}_q^{\alpha*} \cdot \vect{\cal E}_p^{\beta}(\vr)  + \frac{1}{2} \vect{\cal E}_q^{\alpha*}(\vr) \cdot \vect{\cal J}_p^{\beta}{'}(\vr) \nonumber \\
&\quad + \frac{1}{2} \vect{\cal E}_p^{\beta}{'}(\vr) \cdot \vect{\cal J}_q^{\alpha*}(\vr)\,. \label{eq:Q_Comp_7}
\end{align}
Similarly, adding the dot-product of \eqref{eq:Q_Comp_6} and $\frac{1}{2} \vect{\cal H}_p^{\beta}{'}(\vr)$ to the
dot-product of~\eqref{eq:Q_Comp_4} and $\frac{1}{2} \vect{\cal H}_q^{\alpha*}(\vr)$ yields
\begin{align}
\frac{1}{2} \vect{\cal H}_q^{\alpha*}(\vr) \cdot \nabla \times \vect{\cal E}_p^{\beta}{'}(\vr)   &+ \frac{1}{2} \vect{\cal H}_p^{\beta}{'}(\vr) \cdot \nabla \times \vect{\cal E}_q^{\alpha*}(\vr)  \nonumber \\
&=  - \frac{j}{2} \mu \vect{\cal H}_q^{\alpha*}(\vr) \cdot \vect{\cal H}_p^{\beta}(\vr) \label{eq:Q_Comp_8}\,.
\end{align}
Subtracting~\eqref{eq:Q_Comp_7} from~\eqref{eq:Q_Comp_8} produces
\begin{align}
\frac{j}{2} &\nabla \cdot \left(  \vect{\cal E}_p^{\beta}{'}(\vr) \times \vect{\cal H}_q^{\alpha*}(\vr) +  \vect{\cal E}_q^{\alpha*}(\vr)   \times \vect{\cal H}_p^{\beta}{'}(\vr)  \right)  \nonumber \\
&+ \frac{j}{2} \vect{\cal E}_q^{\alpha*}(\vr) \cdot \vect{\cal J}_p^{\beta}{'}(\vr) + \frac{j}{2} \vect{\cal E}_p^{\beta}{'}(\vr) \cdot \vect{\cal J}_q^{\alpha*}(\vr) \label{eq:Q_Comp_9}
 \\
&\quad  \quad \quad  = \frac{1}{2} \varepsilon \vect{\cal E}_q^{\alpha*}(\vr) \cdot \vect{\cal E}_p^{\beta}(\vr)  + \frac{1}{2} \mu \vect{\cal H}_q^{\alpha*}(\vr) \cdot \vect{\cal H}_p^{\beta}(\vr)\,. \nonumber
\end{align}
Finally, subtracting $\left[\frac{1}{2} \varepsilon \widehat{\vect{\cal E}}_q^{\alpha*}(\vr) \cdot \widehat{\vect{\cal E}}_p^{\beta}(\vr)  + \frac{1}{2} \mu \widehat{\vect{\cal H}}_q^{\alpha*}(\vr) \cdot \widehat{\vect{\cal H}}_p^{\beta}(\vr)\right]$ from both sides of \eqref{eq:Q_Comp_9}, integrating the resulting expression over the $\Omega$, and applying the divergence theorem yields
\begin{align}
\matr{Q}_{qp}^{\alpha, \beta} &= T_{qp,1}^{\alpha,\beta} + T_{qp,2}^{\alpha,\beta}  + T_{qp,S}^{\alpha,\beta}
- T_{qp,\myparallel}^{\alpha,\beta}
\label{eq:Qdecomp}
\end{align}
where
\begin{subequations}
\begin{align}
 T^{\alpha,\beta}_{qp,1} &= \frac{j}{2}  \int_{d\Omega_{g,s}}  \vect{\cal E}_q^{\alpha*}(\vr) \cdot \vect{\cal J}_p^\beta{'} (\vr) d\vr \label{eq:Qcomp_decomp_2} \\
  T^{\alpha,\beta}_{qp,2} &=  \frac{j}{2} \int_{d\Omega_{g,s}}  \vect{\cal E}_p^{\beta}{'}(\vr)  \cdot \vect{\cal J}_q^{\alpha*} (\vr)  d\vr  \label{eq:Qcomp_decomp_3}\\
  T^{\alpha,\beta}_{qp,\myparallel} &=   \frac{\varepsilon}{2}  \int_{\Omega} \widehat{\vect{\cal E}}_{q,\myparallel}^{\alpha*} (\vr) \cdot \widehat{\vect{\cal E}}_{p,\myparallel}^{\beta}(\vr) d\vr \nonumber \\
  &\quad \quad +   \frac{\mu}{2}  \int_{\Omega} \widehat{\vect{\cal H}}_{q,\myparallel}^{\alpha*}(\vr) \cdot \widehat{\vect{\cal H}}_{p,\myparallel}^{\beta}(\vr) d\vr \label{eq:Qcomp_decomp_parallel}\\
T_{qp,S}^{\alpha,\beta} &= \frac{j}{2} \int_{d\Omega_{f}} \unit{r}  \cdot \Big(  \vect{\cal E}_p^{\beta}{'} (\vr)  \times \vect{\cal H}_q^{\alpha*} (\vr) \nonumber \\
& \quad \quad \quad \quad  +  \vect{\cal E}_q^{\alpha*}(\vr)   \times \vect{\cal H}_p^{\beta}{'}(\vr)  \Big) d\vr \,. \label{eq:Qcomp_decomp_S}
\end{align}
\end{subequations}

\subsection{Computation of $\matr{Q}_{qp}^{\text{inc,inc}}$}
\label{sec:Qcomp_Qincinc}

Equation \eqref{eq:Qdecomp} implies that the $(q,p)$-th entry of $\matr{Q}^{\text{inc,inc}}$ is
\begin{align}
\matr{Q}_{qp}^{\text{inc,inc}} = T_{qp,S}^{\text{inc,inc}} - T_{qp,\myparallel}^{\text{inc,inc}}
\end{align}
because $T^{\text{inc,inc}}_{qp,1} = T^{\text{inc,inc}}_{qp,2} = 0$
since $\vect{\cal J}_p^i(\vr) = 0$ and $\vect{\cal J}_q^i(\vr) = 0$ (throughout $\Omega$).
Furthermore, $T_{qp,S}^{\text{inc,inc}} = T_{qp,\myparallel}^{\text{inc,inc}}$ as shown below, thus confirming \eqref{eq:Qincinc_final}.

\subsubsection{Computation of $T_{qp,S}^{\text{inc,inc}}$}

For $p \le M_g$, substituting the frequency derivative of \eqref{eq:Ei_antenna}--\eqref{eq:Hi_antenna} into \eqref{eq:Einc_antenna}--\eqref{eq:Hinc_antenna} yields $\vect{\cal E}_p^{\text{inc}}{'}(\vr) = \vect{\cal H}_p^{\text{inc}}{'}(\vr) = 0$. Therefore, $T_{qp,\myparallel}^{\text{inc,inc}} = 0$ when either $p\le M_g$ or $q \le M_g$.
For $p > M_g$ and $q > M_g$, use of the far-field approximations of \eqref{eq:Einc_freespace}--\eqref{eq:Hinc_freespace1} into \eqref{eq:Qcomp_decomp_S} yields
\begin{subequations}
\begin{align}
    T_{qp,S}^{\text{inc,inc}} &= \frac{j}{2} \int_{d\Omega_{f}} \unit{r}  \cdot \left[  \vect{\cal E}_p^{\text{inc}}{'}(\vr)  \times \vect{\cal H}_q^{\text{inc}*}(\vr) \right] d\vr \nonumber \\
    &\quad + \frac{j}{2} \int_{d\Omega_{f}} \unit{r}  \cdot  \left[\vect{\cal E}_q^{\text{inc}*} (\vr)    \times \vect{\cal H}_p^{\text{inc}}{'}(\vr)  \right] d\vr \nonumber \\
    &= \frac{j}{2} \int_{d\Omega_{f}} \unit{r} \cdot \bigg[\vect{\cal W}_{p,\myparallel}'(\vr)  \times (-1)^{\tau'}\vect{\cal W}_{\bar{q},\myparallel}^*(\vr) \nonumber \\
    & \quad \quad  \quad \quad \quad +\vect{\cal W}_{q,\myparallel}^*(\vr) \times (-1)^{\tau}\vect{\cal W}_{\bar{p},\myparallel}'(\vr)\bigg]  d\vr \\
    &=  2 R Z \varepsilon    \delta_{qp}\,. \label{eq:Ts_ii_1}
\end{align}
\end{subequations}
Here, $p = (\tau,l,m)$ and $q=(\tau', l', m')$;
 the final result \eqref{eq:Ts_ii_1} was obtained using Identity A.2 in Appendix~\ref{App:Identity_A2}.

\subsubsection{Computation of $T_{qp,\parallel}^{\text{inc,inc}}$}

For $p\le M_g$, $\widehat{\vect{\cal E}}^{\text{inc}}_{p,\myparallel}(\vr) = \widehat{\vect{\cal H}}^{\text{inc}}_{p,\myparallel}(\vr) = 0$
as is evident from the first terms in \eqref{eq:Ehat_1} and \eqref{eq:Hhat_1}, leading to $T_{qp,\myparallel}^{\text{inc,inc}} = 0$ when either $p\le M_g$ or $q \le M_g$. 
For $p> M_g$ and $q>M_g$, 
use of the far-field approximations of  \eqref{eq:Einc_freespace}--\eqref{eq:Hinc_freespace1}
into \eqref{eq:Qcomp_decomp_parallel} yields 
\begin{subequations}
\begin{align}
T^{\text{inc,inc}}_{qp,\myparallel} 
&= \frac{\varepsilon}{2}  \int_{\Omega} \widehat{\vect{\cal E}}_{q,\myparallel}^{\text{inc}*} (\vr) \cdot \widehat{\vect{\cal E}}_{p,\myparallel}^{\text{inc}}(\vr) d\vr \nonumber \\
  &\quad \quad +   \frac{\mu}{2}  \int_{\Omega} \widehat{\vect{\cal H}}_{q,\myparallel}^{\text{inc}*}(\vr) \cdot \widehat{\vect{\cal H}}_{p,\myparallel}^{\text{inc}}(\vr) d\vr \\
&= \frac{Z \varepsilon}{2}  \int_{\Omega} \big[ \vect{\cal W}_{p,\myparallel}(\vr) \cdot \vect{\cal W}_{q,\myparallel}^*(\vr) \nonumber \\
&\quad \quad \quad \quad  + (-1)^{\tau + \tau'} \vect{\cal W}_{\bar{p},\myparallel}(\vr) \cdot \vect{\cal W}_{\bar{q},\myparallel}^*(\vr)\big] d\vr \label{eq:Tinin_parallel_1} \\
&= 2 R Z \varepsilon  \delta_{qp} \label{eq:Tinin_parallel_2}
\end{align}
\end{subequations}
where the final result was obtained using Identity A.1 in Appendix~\ref{App:Identity_A1}.

\subsection{Computation of $\matr{Q}_{qp}^{\text{sca,inc}}$}

Equation \eqref{eq:Qdecomp} implies that the $(q,p)$-th entry of $\matr{Q}^{\text{sca,inc}}$ is
\begin{align}
    \matr{Q}_{qp}^{\text{sca,inc}} =   T^{\text{sca,inc}}_{qp,2} + T_{qp,S}^{\text{sca,inc}} - T_{qp,\myparallel}^{\text{sca,inc}} \label{eq:Qscin_decomp} 
\end{align}
because $T^{\text{sca,inc}}_{qp,1} = 0$ since $\vect{\cal J}_{p}(\vr) = 0$.
Furthermore, it can be shown that $T_{qp,S}^{\text{sca,inc}} - T_{qp,\myparallel}^{\text{sca,inc}} = 0$\,. Hence, the final result simplifies to key result \eqref{eq:Qscainc_final}, which is  non-zero only for $p>M_g$ because $\vect{\cal E}_p^{\text{inc}}{'}(\vr) = 0$ when $p \le M_g$.  

\subsubsection{Computation of $T_{qp,S}^{\text{sca,inc}}$}

For $p \le M_g$, $\vect{\cal E}_p^{\text{inc}}{'}(\vr) = \vect{\cal H}_p^{\text{inc}}{'}(\vr) = 0$ resulting in $T_{qp,S}^{\text{sca,inc}} = 0$.
For $p > M_g$, substituting the far-field approximation of \eqref{eq:Einc_freespace}--\eqref{eq:Hinc_freespace1} and the conjugate of
\eqref{eq:E_scatter}--\eqref{eq:H_scatter} with $p\rightarrow q$ into \eqref{eq:Qcomp_decomp_S} yields
\begin{align}
    T_{qp,S}^{\text{sca,inc}} &= \frac{j}{2} \int_{d\Omega_f} \unit{r}  \cdot \bigg[  \vect{\cal E}_p^{\text{inc}}{'}(\vr) \times \vect{\cal H}_q^{\text{sca}*}(\vr) \nonumber \\
    &\quad \quad \quad \quad \quad + \vect{\cal E}_q^{\text{sca}*}(\vr)   \times \vect{\cal H}_p^{\text{inc}}{'}(\vr)
    \bigg] d\vr \nonumber \\
    &= \frac{j}{2} \int_{d\Omega_f}  \sum_{t=M_g+1}^{M} \matr{P}_{tq}^* \unit{r}  \cdot \bigg[ \vect{\cal W}_{p,\myparallel}{'}(\vr) \times \vect{\cal I}_{\bar{t},\myparallel}(\vr)(-1)^{\tilde{\tau+1}} \nonumber \\
    & \quad \quad \quad  +   \vect{\cal W}_{\bar{p},\myparallel}'(\vr) \times \vect{\cal I}_{t,\myparallel}(\vr) (-1)^{\tau + 1} \bigg] d\vr  \\
    &= \varepsilon Z R \matr{P}_{p\hat{q}}^* (-1)^{\tau + l+m} \label{eq:TS_si_3}
\end{align}
where $t = (\tilde{\tau}, \tilde{l}, \tilde{m})$ and $\hat{q} = (\tau',l',-m')$; the final result \eqref{eq:TS_si_3} was obtained using Identity A.4 in Appendix~\ref{App:Identity_A4}.

\subsubsection{Computation of $T_{qp,\parallel}^{\text{sca,inc}}$}
For $p \le M_g$, $\widehat{\vect{\cal E}}_{p,\myparallel}^{\text{inc}}(\vr)=\widehat{\vect{\cal H}}_{p,\myparallel}^{\text{inc}}(\vr) = 0$ because the antenna ports $d\Omega_{g,p}$ are electrically small and hence produce negligible far-fields (the fields from a delta-gap or magnetic frill excitation are assumed entirely local). Therefore,  $T_{qp,\myparallel}^{\text{sca,inc}} = 0$ for $p \le M_g$. For $p > M_g$, 
$\widehat{\vect{\cal E}}_{p,\myparallel}^{\text{inc}}(\vr)$ and $\widehat{\vect{\cal H}}_{p,\myparallel}^{\text{inc}}(\vr)$ are given by the far-field approximation of \eqref{eq:Einc_freespace}--\eqref{eq:Hinc_freespace1}. Furthermore, $\widehat{\vect{\cal E}}_{q,\myparallel}^{\text{sca}}(\vr)$ and $\widehat{\vect{\cal H}}_{q,\myparallel}^{\text{sca}}(\vr)$ are given by the same expressions as ${\vect{\cal E}}_{q,\myparallel}^{\text{sca}}(\vr)$ and ${\vect{\cal H}}_{q,\myparallel}^{\text{sca}}(\vr)$ on $d\Omega_f$. Therefore, using the far-field approximation of \eqref{eq:E_scatter}--\eqref{eq:H_scatter} yields 
\begin{align}
    T_{qp,\myparallel}^{\text{sca,inc}} &= \frac{\varepsilon}{2}  \int_{\Omega} \widehat{\vect{\cal E}}_{q,\myparallel}^{\text{sca}*}(\vr) \cdot \widehat{\vect{\cal E}}_{p,\myparallel}^{\text{inc}}(\vr) d\vr \nonumber \\
    &\quad +   \frac{\mu}{2} \int_{\Omega} \widehat{\vect{\cal H}}_{q,\myparallel}^{\text{sca}*}(\vr) \cdot \widehat{\vect{\cal H}}_{p,\myparallel}^{\text{inc}}(\vr) d\vr \nonumber \\
    &=  \frac{Z \varepsilon}{2} \int_{\Omega} \sum_{t=M_g+1}^{M} \matr{P}_{tq}^* \big[\vect{\cal I}_{t,\myparallel}(\vr) \cdot \vect{\cal W}_{p,\myparallel}(\vr) + \nonumber \\
    &\quad \quad \quad  \vect{\cal I}_{\bar{t},\myparallel}(\vr) \cdot \vect{\cal W}_{\bar{p},\myparallel}(\vr) (-1)^{\tau + \tilde{\tau}+1} \big] d\vr \nonumber \\
    &= \varepsilon Z  R \matr{P}_{p\hat{q}}^* (-1)^{\tau + l + m}
\end{align}
where the final result was obtained using Identity A.3 in Appendix~\ref{App:Identity_A3}.

\subsection{Computation of $\matr{Q}_{qp}^{\text{sca,sca}}$}
\label{sec:Qcomp_Qscasca}

Equation \eqref{eq:Qdecomp} implies that the $(q,p)$-th entry of $\matr{Q}^{\text{sca,sca}}$ is
\begin{align}
\matr{Q}_{qp}^{\text{sca,sca}} = T^{\text{sca,sca}}_{qp,S} + T^{\text{sca,sca}}_{qp,1} + T^{\text{sca,sca}}_{qp,2} - T^{\text{sca,sca}}_{qp,\myparallel}\,. \label{eq:Qcomp_scsc_decomp}
\end{align}
All four terms on the RHS of \eqref{eq:Qcomp_scsc_decomp} are non-zero and dependent on $\vect{\cal J}_p(\vr)$ and $\vect{\cal J}_q(\vr)$.  Their computation is detailed below.  Their addition yields key result \eqref{eq:Qscasca_final}.

\subsubsection{Computation of $T^{\text{sca,sca}}_{qp,1}$}

Substituting \eqref{eq:Es_1} into \eqref{eq:Qcomp_decomp_2}
yields
\begin{align}
T^{\text{sca,sca}}_{qp,1} &= \frac{j}{2}  \int_{d\Omega_{g,s}}  \vect{\cal E}_q^{\text{sca}*}(\vr) \cdot \vect{\cal J}_p' (\vr) d\vr \nonumber \\
&= \frac{j}{2} \int_{d\Omega_{g,s}} \vect{\cal L}^*[\vect{\cal J}_q^*](\vr) \cdot \vect{\cal J}_p'(\vr) d\vr \label{eq:Tscsc_qp_1_a} \,.
\end{align}

\subsubsection{Computation of $T^{\text{sca,sca}}_{qp,2}$}

From  \eqref{eq:Qcomp_decomp_3} it follows that
\begin{align}
T^{\text{sca,sca}}_{qp,2}   
&= \frac{j}{2} \int_{d\Omega_{g,s}}  \vect{\cal E}_p^{\text{sca}}{'}(\vr)  \cdot \vect{\cal J}_q^* (\vr)  d\vr .
\label{eq:Tscsc_qp_2}
\end{align}
Using the chain rule, the frequency derivative of the scattered electric field  $\vect{\cal E}_p^{\text{sca}}{'}(\vr)$ \eqref{eq:Es_1} can be decomposed into two terms: one dependent on $\vect{\cal J}_{p}'(\vr)$ and the other on $G'(\vr,\vrp)$. Therefore, it follows that 
\begin{align}
T^{\text{sca,sca}}_{qp,2} &= T^{\text{sca,sca}}_{qp,2,J'} + T^{\text{sca,sca}}_{qp,2,G'}
\label{eq:Tscsc_qp_2b}
\end{align}
where  
\begin{align}
    T^{\text{sca,sca}}_{qp,2,J'} &= \frac{j}{2} \int_{d\Omega_{g,s}} \vect{\cal L} \left[ \vect{\cal J}_p' \right](\vr) \cdot \vect{\cal J}_q^*(\vr) d\vr \label{eq:Tscsc_2_Jprime}\\
    T^{\text{sca,sca}}_{qp,2,G'} &= \frac{j}{2} \int_{d\Omega_{g,s}} \vect{\cal L}'[\vect{\cal J}_p](\vr) \cdot \vect{\cal J}_q^*(\vr) d\vr.
\end{align}

\subsubsection{Evaluation of $T_{qp,S}^{\text{sca,sca}}$}

From \eqref{eq:Qcomp_decomp_S}, $T_{qp,S}^{\text{sca,sca}}$ is given by
\begin{align}
T_{qp,S}^{\text{sca,sca}} &= \frac{j}{2} \int_{d\Omega_{f}} \unit{r}  \cdot \Big(  \vect{\cal E}_{p,\myparallel}^{\text{sca}}{'} (\vr)  \times \vect{\cal H}_{q,\myparallel}^{\text{sca}*} (\vr) \nonumber \\
& \quad \quad \quad \quad  +  \vect{\cal E}_{q,\myparallel}^{\text{sca}*}(\vr)   \times \vect{\cal H}_{p,\myparallel}^{\text{sca}}{'}(\vr)  \Big) d\vr  \nonumber \\
&= \frac{j}{Z}  \int_{d\Omega_{f}}   \left(\vect{\cal E}_{p,\myparallel}^{\text{sca}}\right){'} \cdot \left(\vect{\cal E}_{q,\myparallel}^{\text{sca}*}\right)   d\vr
\label{eq:Tscsc_S_a}
\end{align}
where the last equality is due to  $-\unit{r} \times \vect{\cal H}_{q,\myparallel}^{\text{sca}*}(\vr) = \frac{1}{Z} \vect{\cal E}_{q,\myparallel}^{\text{sca}*}(\vr)$ 
and its frequency derivative.
Using the chain rule, the frequency derivative of the scattered electric field $\vect{\cal E}_{p,\myparallel}^{\text{sca}}(\vr)$ \eqref{eq:E_scatter} 
can be decomposed into two terms: one dependent on 
$\vect{\cal J}_p'(\vr)$ and the other on  $G_\infty'(\vrp,\vr)$. Using this decomposition, \eqref{eq:Tscsc_S_a} reads 
\begin{align}
    T_{qp,S}^{\text{sca,sca}} &=  T_{qp,S,J'}^{\text{sca,sca}} + T_{qp,S,G'}^{\text{sca,sca}}\,.
    \label{eq:Tscsc_S_decomp}
\end{align}
The first term on the RHS of \eqref{eq:Tscsc_S_decomp} reads 
\begin{subequations}
\begin{align}
    T_{qp,S,J'}^{\text{sca,sca}} &= \int_{d\Omega_{f}} \vect{\cal L}_{\myparallel}[\vect{\cal J}_p'](\vr) \cdot \vect{\cal L}_{\myparallel}^*[\vect{\cal J}_q^*](\vr) d\vr \nonumber \\  
&=  j  Z  \int_{ d\Omega_{g,s}} \int_{ d\Omega_{g,s}}  \bigg[ k^2 \vect{\cal J}_p'(\vrp) \cdot \vect{\cal J}_q^*(\vrq) -  \nonumber \\
& \quad  \nabla' \cdot \vect{\cal J}_p'(\vrp) \nabla'' \cdot \vect{\cal J}_q^*(\vrq) \bigg] \widetilde{G}_{1}(\vrp,\vrq) d\vrq d\vrp \label{eq:Tqp_S_Jprime1}
\end{align}
\end{subequations}
where $\vect{\cal L}_{\myparallel}[\vect{\cal J}](\vr)$ is the same as $\vect{\cal L}[\vect{\cal J}](\vr)$ (defined in \eqref{eq:LOperator}) with $G(\vr,\vrp) \rightarrow G_\infty(\vr,\vrp)$ and 
\begin{align}
    \widetilde{G}_{1}(\vrp,\vrq) &= \int_{d\Omega_f} G_{\infty}^*(\vr,\vrq) G_{\infty}(\vr,\vrp) d\vr \nonumber \\
    &= \frac{\sin(k D)}{4 \pi k D}\,. \label{eq:G1_eqn}
\end{align}
Using \eqref{eq:G1_eqn}, \eqref{eq:Tqp_S_Jprime1} may be written as
\begin{align}
    T_{qp,S,J'}^{\text{sca,sca}} &= -\frac{j}{2} \int_{d\Omega_{g,s}} \left[ \vect{\cal L}^*[\vect{\cal J}_q^*](\vr) + \vect{\cal L}[\vect{\cal J}_q^*](\vr) \right] \cdot \vect{\cal J}_p'(\vr) d\vr\,. \label{eq:Tqp_S_Jprime2}
\end{align}
The second term on the RHS of \eqref{eq:Tscsc_S_decomp} reads 
\begin{align}
    T_{qp,S,G'}^{\text{sca,sca}} &=  \frac{j}{Z} \int_{d\Omega_f} \vect{\cal L}_{\myparallel}'[\vect{\cal J}_p](\vr) \cdot \vect{\cal L}_{\myparallel}[\vect{\cal J}_q^*](\vr) d\vr
 \nonumber   \\
&= j  Z k \int_{d\Omega_{g,s}} \int_{ d\Omega_{g,s}} \big[ \vect{\cal J}_p(\vrp) \cdot \vect{\cal J}_q^*(\vrq) \big[ k'  \nonumber \\
&\quad \quad  \widetilde{G}_{1}(\vrp,\vrq)  + k \widetilde{G}_{2}(\vrp,\vrq)\big]  - \frac{1}{k}\nabla' \cdot \vect{\cal J}_p(\vrp)  \nonumber \\
& \quad \quad \quad \quad \quad  \nabla'' \cdot \vect{\cal J}_q^*(\vrq) \widetilde{G}_{2}(\vrp,\vrq) \big]  d\vrp d\vrq \,. \label{eq:Tscsc_S_2}
\end{align}
where $\vect{\cal L}_{\myparallel}'[\vect{\cal J}](\vr)$ is given by \eqref{eq:L_1prime} with $G(\vr,\vrp) \rightarrow G_\infty(\vr,\vrp)$
and
\begin{align}
\widetilde{G}_{2}(\vrp,\vrq) &= \int_{d\Omega_f} G_{\infty}^*(\vr,\vrq) G_{\infty}'(\vr,\vrp) d\vr \nonumber \\
&=-j\varepsilon Z R \widetilde{G}_1(\vrp,\vrq) - \vrp \cdot \unit{d}  \frac{\varepsilon Z }{4 \pi k^2 D^2} \nonumber \\
&\quad  \quad \quad \left[ k D \cos (k D) - \sin(k D) \right] 
\end{align}
using Identity A.5 in Appendix \ref{App:Identity_A5}.

\subsubsection{Computation of $T^{\text{sca,sca}}_{qp,\parallel}$}

From \eqref{eq:Qcomp_decomp_parallel}, $T_{qp,\myparallel}^{\text{sca,sca}}$ is given by
 \begin{align}
 T^{\text{sca,sca}}_{qp,\myparallel} &=   \frac{\varepsilon}{2}  \int_{\Omega} \widehat{\vect{\cal E}}_{q,\myparallel}^{\text{sca}*} (\vr) \cdot \widehat{\vect{\cal E}}_{p,\myparallel}^{\text{sca}}(\vr) d\vr \nonumber \\
  &\quad \quad +   \frac{\mu}{2}  \int_{\Omega} \widehat{\vect{\cal H}}_{q,\myparallel}^{\text{sca}*}(\vr) \cdot \widehat{\vect{\cal H}}_{p,\myparallel}^{\text{sca}}(\vr) d\vr \nonumber \\
  &= \varepsilon  \int_{\Omega} \widehat{\vect{\cal E}}_{q,\myparallel}^{\text{sca}*} (\vr) \cdot \widehat{\vect{\cal E}}_{p,\myparallel}^{\text{sca}}(\vr) d\vr 
  \label{eq:Tqp_sca_sca_parallel_a}
  \end{align}
  where the second equality follows from $\widehat{\vect{\cal H}}^{\text{sca}}_{p,\myparallel}(\vr) = \frac{1}{Z} \unit{r} \times \widehat{\vect{\cal E}}_{p,\myparallel}^{\text{sca}}(\vr)$.
  Substituting \eqref{eq:E_scatter} into \eqref{eq:Tqp_sca_sca_parallel_a} and simplifying the expression yields
\begin{align}
T^{\text{sca,sca}}_{qp,\myparallel} =& \frac{k Z R}{\omega} \Bigg[\int_{\Omega_s} \int_{\Omega_s}\bigg[ k^2  \vect{\cal J}_q^*(\vrq) \cdot \vect{\cal J}_p(\vrp) -     \label{eq:Tqp_sca_sca_Parallel_b} \\
&\,  \nabla'' \cdot \vect{\cal J}_q^*(\vrq) \nabla' \cdot \vect{\cal J}_p(\vrp) \bigg] \widetilde{G}_{1}(\vrp,\vrq)  d\vrq d\vrp\,. \nonumber
\end{align}

\subsubsection{Simplification}

Using \eqref{eq:Tscsc_qp_1_a}, \eqref{eq:Tscsc_2_Jprime}, and \eqref{eq:Tqp_S_Jprime2} it can be shown that $T_{qp,1}^{\text{sca,sca}} +  T_{qp,2,J'}^{\text{sca,sca}} + T_{qp,S,J'}^{\text{sca,sca}} = 0$.
Furthermore, the term proportional to $R$ in \eqref{eq:Tscsc_S_2} is identical to the term in $T_{qp,\myparallel}^{\text{sca,sca}}$ and therefore its contribution cancels when computing $\matr{Q}_{qp}^{\text{sca,sca}}$ using \eqref{eq:Qcomp_scsc_decomp}. 
Hence, the final result for $\matr{Q}_{qp}$ in \eqref{eq:Qscasca_final} follows from adding $T_{qp,2,G'}^{\text{sca,sca}}$ to the terms that are not proportional to $R$ in $T_{qp,S,G'}^{\text{sca,sca}}$.

Alternatively, the second term on the RHS of \eqref{eq:Tscsc_S_decomp} may be written using the dyadic VSW expansion of the Green's function \eqref{eq:VSW_expansion2} as
\begin{subequations}
\begin{align}
    T&_{qp,S,G'}^{\text{sca,sca}} \nonumber \\
    =&  \frac{jZ}{4} \sum_{t=1}^{\infty} \sum_{t' = 1}^\infty  \int_{d\Omega_{g,s}} \vect{\cal W}_t^*(\vrq) \cdot \vect{\cal J}_q^*(\vrq) d\vrq \bigg \{   \\
    & \int_{d\Omega_{g,s}} \vect{\cal W}_t'(\vrp) \cdot \vect{\cal J}_p(\vrp) d\vrp \int_{d\Omega_f} \vect{\cal I}_{t,\myparallel}^*(\vr) \cdot \vect{\cal I}_{t',\myparallel}(\vr) d\vr \nonumber \\
    &  + \int_{d\Omega_{g,s}} \vect{\cal W}_t(\vrp) \cdot \vect{\cal J}_p(\vrp) d\vrp \int_{d\Omega_f} \vect{\cal I}_{t,\myparallel}^*(\vr) \cdot \vect{\cal I}_{t',\myparallel}'(\vr) d\vr \bigg \} \nonumber \\
    =& \frac{jZ}{4} \sum_{t=1}^{\infty} \int_{d\Omega_{g,s}} \vect{\cal W}_t^*(\vrp) \cdot \vect{\cal J}_q^*(\vrp) d\vrp \bigg \{ \int_{d\Omega_{g,s}} \vect{\cal W}_t'(\vrp)   \nonumber \\
    &   \cdot \vect{\cal J}_p(\vrp) d\vrp - jkR \int_{d\Omega_{g,s}} \vect{\cal W}_t(\vrp) \cdot \vect{\cal J}_p(\vrp) d\vrp \bigg\} \,. \label{eq:Tscasca_S_G'_2}
\end{align}
\end{subequations}
Furthermore, the term that is proportional to $R$ in \eqref{eq:Tscasca_S_G'_2} cancels out with $T_{qp,\myparallel}^{\text{sca,sca}}$. Hence, the result in \eqref{eq:Tscasca_S_G'_2} and the self-adjoint property of $\matr{Q}^{\text{sca,sca}}$ was used to obtain \eqref{eq:Operator_Qscasca_d_final_2}.

\section{Various Identities Obtained Analytically}
\subsection{Identity A.1}
\label{App:Identity_A1}

This section evaluates
\begin{align}
    I_1(p,q) = I_{1,1}(p,q) + I_{1,2}(p,q)
\end{align}
where
\begin{subequations}
\begin{align}
    I_{1,1}(p,q) &= \int_{\Omega}\big[ \vect{\cal W}_{p,\myparallel}(\vr) \cdot \vect{\cal W}_{q,\myparallel}^*(\vr) d\vr \label{eq:I1_1-1a}\\
    I_{1,2}(p,q) &=  (-1)^{\tau + \tau'} \int_{\Omega} \big[ \vect{\cal W}_{\bar{p},\myparallel}(\vr) \cdot \vect{\cal W}_{\bar{q},\myparallel}^*(\vr) d\vr\, 
    \label{eq:I1_1-1b}
\end{align}
\end{subequations}
and $p=(\tau, l, m)$ and $q = ( \tau', l', m')$.
Let $p_{\mathrm{TE}} = (1,l,m)$ and $q_{\mathrm{TE}} = (1, l', m')$, substituting \eqref{eq:VSH_J_TE_parallel} into \eqref{eq:I1_1-1a}--\eqref{eq:I1_1-1b} yields
\begin{align}
  &I_{1,1} (p_{\mathrm{TE}},q_{\mathrm{TE}}) \nonumber \\
   &= \int_0^R \oint_{1}   \big[ 1 - (-1)^{l'} e^{2jkr} - (-1)^{l}e^{-2jkr}\nonumber \\
  &\quad + (-1)^{l+l'}\big]
  \vect{\cal X}_{p_{\mathrm{TE}}}(\theta,\phi) \cdot \vect{\cal X}_{q_{\mathrm{TE}}}^*(\theta,\phi) dS_1 dr  \nonumber \\
&= \delta_{ll'} \delta_{mm'}  \bigg[  2 R -  (-1)^{l} \frac{e^{2jk R}}{2jk} \left[1  -  e^{-4jkR}\right]\bigg] \label{eq:Identity_1_1}
\end{align}
and 
\begin{align}
  &I_{1,2}(p_{\mathrm{TE}},q_{\mathrm{TE}}) \nonumber \\
  &=   \int_0^R \oint_{1} \big[ 1 + (-1)^{l} e^{-2jkr} + (-1)^{l'}e^{2jkr} \nonumber \\
&\quad + (-1)^{l+l'} \big] \vect{\cal X}_{p_{\mathrm{TE}}}(\theta,\phi) \cdot \vect{\cal X}_{q_{\mathrm{TE}}}^*(\theta,\phi) dS_{1} dr  \nonumber \\
&= \delta_{ll'} \delta_{mm'}  \bigg[  2 R +  (-1)^{l} \frac{e^{2jk R}}{2jk} \left[1 -  e^{-4jkR} \right] \bigg] \label{eq:Identity_1_2}
\end{align}
where the inner integral is over the unit sphere, i.e. $\oint_{1} dS_1 = \int_{0}^{2\pi} \int_{0}^{\pi} \sin \theta d\theta d\phi$.
If $p_{\mathrm{TM}} = (2, l, m)$ and $q_{\mathrm{TM}} = (2, l', m')$, then $I_{1,1}(p_{\mathrm{TM}}, q_{\mathrm{TM}}) = I_{1,2}(p_{\mathrm{TE}},q_{\mathrm{TE}})$ and $I_{1,2}(p_{\mathrm{TM}},q_{\mathrm{TM}}) = I_{1,1}(p_{\mathrm{TE}},q_{\mathrm{TE}})$. Additionally, $I_{1,1}(p_{\mathrm{TE}}, q_{\mathrm{TM}}) = I_{1,2}(p_{\mathrm{TE}}, q_{\mathrm{TM}}) = I_{1,1}(p_{\mathrm{TM}}, q_{\mathrm{TE}}) = I_{1,2}(p_{\mathrm{TM}}, q_{\mathrm{TE}})  = 0$ due to the orthogonality property of the VSHs \eqref{eq:VSH_Orthogonality}.
It therefore follows from \eqref{eq:Identity_1_1} and \eqref{eq:Identity_1_2} that
\begin{align}
    I_1(p,q) = 4 R \delta_{p,q} \,.
\end{align}

\subsection{Identity A.2}
\label{App:Identity_A2}
This section evaluates 
\begin{align}
    I_2(p,q) = I_{2,1}(p,q) + I_{2,2}(p,q)
\end{align}
where
\begin{subequations}
\begin{align}
    I_{2,1}(p,q) &=  \int_{d\Omega_f} \unit{r} \cdot \big[\vect{\cal W}_{p,\myparallel}'(\vr)  \times (-1)^{\tau'}\vect{\cal W}_{\bar{q},\myparallel}^*(\vr)\big] d\vr \label{eq:I2_1-1a}\\
    I_{2,2}(p,q) &=    \int_{d\Omega_f} \unit{r} \cdot \big[ \vect{\cal W}_{q,\myparallel}^*(\vr) \times (-1)^{\tau}\vect{\cal W}_{\bar{p},\myparallel}'(\vr) \big] d\vr\,. 
    \label{eq:I2_1-1b}
\end{align}
\end{subequations}
Substituting \eqref{eq:VSH_J_TE_parallel}--\eqref{eq:VSH_dJ_TE_parallel} into \eqref{eq:I2_1-1a}--\eqref{eq:I2_1-1b} yields
 \begin{align}
 I_{2,1}&(p_{\mathrm{TE}},q_{\mathrm{TE}}) \nonumber \\
 =&  \frac{- j k}{\omega}  \oint_{1} \big[ 1 + (-1)^{l'} e^{2jkR}  + (-1)^{l}e^{-2jkR}  \nonumber \\
 &+ (-1)^{l+l'}\big] \unit{r} \cdot \vect{\cal X}_{p_{\mathrm{TE}}}(\theta,\phi)   \times \vect{\cal X}_{q_{\mathrm{TM}}}^*(\theta,\phi) R dS_1 \nonumber \\
 =& - j Z \varepsilon \delta_{ll'}\delta_{mm'}\left[ 2 + (-1)^{l} e^{2jkR} \left[1 + e^{-4jkR} \right] \right] \label{eq:Identity_2_1}
\end{align}
and
\begin{align}
 &I_{2,2}(p_{\mathrm{TE}},q_{\mathrm{TE}}) \nonumber \\
 =&  \frac{-j k}{\omega}  \oint_{1} \big[ 1 - (-1)^{l'} e^{2jkR}  - (-1)^{l}e^{-2jkR} \nonumber \\
 & + (-1)^{l+l'}\big] \unit{r} \cdot \vect{\cal X}_{q_{\mathrm{TE}}}(\theta,\phi) \times \vect{\cal X}_{p_{\mathrm{TM}}}^*(\theta,\phi) R dS_1 \nonumber \\ 
=& -j Z \varepsilon \delta_{ll'}\delta_{mm'}\left[ 2 - (-1)^{l} e^{2jkR}\left[1 + e^{-4jkR} \right] \right]\,. \label{eq:Identity_2_2}
\end{align}
A similar analysis yields
$I_{2,1}(p_{\mathrm{TM}}, q_{\mathrm{TM}}) = I_{2,2}(p_{\mathrm{TE}},q_{\mathrm{TE}})$ and $I_{2,2}(p_{\mathrm{TM}},q_{\mathrm{TM}}) = I_{2,1}(p_{\mathrm{TE}},q_{\mathrm{TE}})$. Additionally, $I_{2,1}(p_{\mathrm{TM}}, q_{\mathrm{TE}}) = I_{2,2}(p_{\mathrm{TM}}, q_{\mathrm{TE}}) = I_{2,1}(p_{\mathrm{TE}}, q_{\mathrm{TM}}) = I_{2,2}(p_{\mathrm{TE}}, q_{\mathrm{TM}})  = 0$ due to the cross-product property of the VSHs \eqref{eq:VSH_Crossproduct}.
It therefore follows from \eqref{eq:Identity_2_1} and \eqref{eq:Identity_2_2} that
\begin{align}
    I_2(p,q) = -j Z \varepsilon 4 R  \delta_{p,q} \,.
\end{align}

\subsection{Identity A.3}
\label{App:Identity_A3}
This section evaluates
\begin{align}
    I_3(p,q) = I_{3,1}(p,q) + I_{3,2}(p,q)
\end{align}
where
\begin{subequations}
\begin{align}
    I_{3,1}(p,q) &= \int_{\Omega} \vect{\cal I}_{q,\myparallel}(\vr) \cdot \vect{\cal W}_{p,\myparallel}(\vr) d\vr \label{eq:I3_1-1a} \\
    I_{3,2}(p,q) &=  (-1)^{\tau+\tau'+1}\int_{\Omega}  \vect{\cal I}_{\bar{q},\myparallel}(\vr) \cdot \vect{\cal W}_{\bar{p},\myparallel}(\vr) d\vr. 
    \label{eq:I3_1-1b}
\end{align}
\end{subequations}
Substituting \eqref{eq:VSH_O_TE_parallel} and \eqref{eq:VSH_J_TE_parallel} into \eqref{eq:I3_1-1a}--\eqref{eq:I3_1-1b} yields
\begin{align}
  I&_{3,1}(p_{\mathrm{TE}},q_{\mathrm{TE}})  \nonumber \\
  &= \int_0^R \oint_{1} \big[ e^{2jkr} + (-1)^{l+1}\big]
  \nonumber \\
& \quad \quad \quad  (-1)^{m'} \vect{\cal X}_{p_{\mathrm{TE}}}(\theta,\phi) \cdot \vect{\cal X}_{\hat{q}_{\mathrm{TE}}}^*(\theta,\phi) dS_1 dr \nonumber \\
&= \delta_{ll'} \delta_{m(-m')} (-1)^{m'}  \left[\frac{e^{2jk R} - 1}{2jk} + (-1)^{l+1} R  \right] \label{eq:Identity_3_1}
\end{align}
and
\begin{align}
  I_{3,2}&(p_{\mathrm{TE}},q_{\mathrm{TE}}) = -\int_0^R \oint_{1}   \big[ e^{2jkr} + (-1)^{l}\big] (-1)^{m'}
  \nonumber \\
& \quad  \vect{\cal X}_{p_{\mathrm{TM}}}(\theta,\phi) \cdot \vect{\cal X}_{\hat{q}_{\mathrm{TM}}}^*(\theta,\phi) dS_1 dr \nonumber \\
=& \delta_{ll'} \delta_{mm'}  (-1)^{m'} \left[   -\frac{e^{2jk R}    - 1}{2jk} + (-1)^{l+1} R\right]  \,. \label{eq:Identity_3_2}
\end{align}
Similar analysis yields
$I_{3,1}(p_{\mathrm{TM}}, q_{\mathrm{TM}}) = -I_{3,2}(p_{\mathrm{TE}},q_{\mathrm{TE}})$ and $I_{3,2}(p_{\mathrm{TM}},q_{\mathrm{TM}}) = -I_{3,1}(p_{\mathrm{TE}},q_{\mathrm{TE}})$. 
Additionally, $I_{3,1}(p_{\mathrm{TE}}, q_{\mathrm{TM}}) = I_{3,2}(p_{\mathrm{TE}}, q_{\mathrm{TM}}) = I_{3,1}(p_{\mathrm{TM}}, q_{\mathrm{TE}}) = I_{3,2}(p_{\mathrm{TE}}, q_{\mathrm{TE}})  = 0$ due to the orthogonality of the VSHs \eqref{eq:VSH_Crossproduct}.
It therefore follows from \eqref{eq:Identity_3_1} and \eqref{eq:Identity_3_2} that
\begin{align}
    I_3(p,q) = 2R (-1)^{\tau' + l' + m'}\delta_{p,\hat{q}} \,.
\end{align}

\subsection{Identity A.4}
\label{App:Identity_A4}

This section evaluates
\begin{align}
    I_4(p,q) = I_{4,1}(p,q) + I_{4,2}(p,q)
\end{align}
where
\begin{subequations}
\begin{align}
    I_{4,1}(p,q) &= (-1)^{\tau'+1}\int_{d\Omega_f} \unit{r}  \times \vect{\cal W}_{p,\myparallel}'(\vr) \cdot \vect{\cal I}_{\bar{q},\myparallel}(\vr) d\vr \label{eq:I4_1-1a} \\
    I_{4,2}(p,q) &= (-1)^{\tau+1}\int_{d\Omega_f}\unit{r}  \times \vect{\cal W}_{\bar{p},\myparallel}'(\vr) \cdot \vect{\cal I}_{{q},\myparallel}(\vr)   d\vr. 
    \label{eq:I4_1-1b}
\end{align}
\end{subequations}
Substituting \eqref{eq:VSH_O_TE_parallel} and \eqref{eq:VSH_dJ_TE_parallel} into \eqref{eq:I4_1-1a}--\eqref{eq:I4_1-1b} yields
\begin{align}
  I_{4,1}&(p_{\mathrm{TE}},q_{\mathrm{TE}}) \nonumber \\
  &= \frac{jk}{\omega}\oint_{1} \big[Re^{2jkR} - R(-1)^{l+1}\big] (-1)^{m'} \unit{r} \cdot
  \nonumber \\
&\quad \quad   \vect{\cal X}_{p_{\mathrm{TE}}}(\theta,\phi) \times \vect{\cal X}_{\hat{q}_{\mathrm{TM}}}^*(\theta,\phi) dS_1  \nonumber \\
&= j \varepsilon Z \delta_{ll'} \delta_{m(-m')}  (-1)^{m'} \big[  Re^{2jkR} + R (-1)^l\big]  \label{eq:Identity_4_1}
\end{align}
where $\hat{q}_{\mathrm{ TM}} = (\tau', l', -m') $ and
\begin{align}
  I&_{4,2}(p_{\mathrm{TE}},q_{\mathrm{TE}}) \nonumber \\
  &= j Z \varepsilon \oint_{1} (-1)^{m'} \big[  R e^{2jkR} - R (-1)^{l}\big] \unit{r} \cdot
  \nonumber \\
&\quad \quad   \vect{\cal X}_{p_{\mathrm{TM}}}(\theta,\phi) \times \vect{\cal X}_{\hat{q}_{\mathrm{TE}}}^*(\theta,\phi) dS_1  \nonumber \\
&= j Z \varepsilon \delta_{ll'} \delta_{m(-m')} (-1)^{m'} \big[-R e^{2jkR} + R(-1)^l \big]\,. \label{eq:Identity_4_2}
\end{align}
A similar analysis yields $I_{4,1}(p_{\mathrm{TM}}, q_{\mathrm{TM}}) = -I_{4,2}(p_{\mathrm{TE}},q_{\mathrm{TE}})$ and $I_{4,2}(p_{\mathrm{TM}},q_{\mathrm{TM}}) = -I_{4,1}(p_{\mathrm{TE}},q_{\mathrm{TE}})$. Additionally, $I_{4,1}(p_{\mathrm{TE}}, q_{\mathrm{TM}}) = I_{4,2}(p_{\mathrm{TE}}, q_{\mathrm{TM}}) = I_{4,1}(p_{\mathrm{TM}}, q_{\mathrm{TE}}) = I_{4,2}(p_{\mathrm{TM}}, q_{\mathrm{TE}})  = 0$ due to the cross-product property of the VSHs \eqref{eq:VSH_Crossproduct}.
It therefore follows from \eqref{eq:Identity_4_1} and \eqref{eq:Identity_4_2} that
\begin{align}
    I_4 = -2j\varepsilon Z R (-1)^{\tau'+l'+m'} \delta_{p,\hat{q}} \,.
\end{align}

\subsection{Identity A.5}
\label{sec:IdentityA5}
This section evaluates
\label{App:Identity_A5}
\begin{align}
\widetilde{G}_1(\vrp,\vrq) &= 
\int_{d\Omega_f} G_{\infty}^*(\vr,\vrq) G_{\infty}(\vr, \vrp) d\vr \label{eq:IdentityA5_1} \,.  
\end{align}
Substituting \eqref{eq:Ginf} into \eqref{eq:IdentityA5_1}
yields
\begin{align}
\widetilde{G}_1(\vrp,\vrq) &= \frac{1}{16\pi^2}  \oint_{1}e^{-jk \unit{r} \cdot \vr_{qp}} dS_{1}   \nonumber \\
&= \frac{1}{4\pi} \frac{\sin(k r_{qp})}{k r_{qp}}\,,
\end{align}
where the integral is evaluated using a coordinate transformation that aligns the $z'$ axis with the vector $\unit{d} = (\vrq - \vrp)/\abs{\vrq - \vrp}$.

\subsection{Identity A.6}
\label{App:IdentityA6}
To show $\matr{V} \vec{\matr{I}} = \matr{V}^*$, consider
\begin{align}
    (-1)^{-m + l+\tau} \vect{\cal W}_{\tau l (-m)}(\vr) &= (-1)^{-m+l+\tau} \big( \vect{\cal I}_{\tau l (-m)} (\vr) \nonumber \\
    &\quad \quad \quad + \vect{\cal O}_{\tau l (-m)}(\vr) \big) \nonumber \\
    &= \vect{\cal W}_{\tau l m}^*(\vr)\,. \label{eq:IdentityA_6}
\end{align}
Here, the last equality follows from using the conjugation property~\eqref{eq:VSH_Conjugation} of the VSWs.
Using \eqref{eq:IdentityA_6} with the definition of $\vec{\matr{I}}$ \eqref{eq:Ivec_def} yields $\matr{V} \vec{\matr{I}} = \matr{V}^*$ and differentiating this expression yields $\matr{V}' \vec{\matr{I}} = \matr{V}'^*$.

\subsection{Identity A.7}
\label{App:IndentityA7}
This section proves $-\frac{1}{2}\matr{V}^* \matr{V}^T = \left( \matr{Z} + \matr{Z}^* \right)$. Using the VSW expansion \eqref{eq:VSW_expansion1} and \eqref{eq:VSW_expansion1b}, the $(m,n)$-th entry of $\matr{Z}$ reads
\begin{subequations}
\begin{align}
    \matr{Z}_{mn} &= -\frac{Z}{2}\sum_{t=1}^{\infty}  \int_{d\Omega_{g,s}}   \vect{f}_m(\vr) \cdot \vect{\cal I}_t^*(\vr) d\vr \nonumber \\
    &\quad \int_{d\Omega_{g,s}} \vect{f}_n(\vrp) \cdot \vect{\cal W}_t(\vrp) d\vrp \label{eq:Zeqn_1}\\
    &= -\frac{Z}{2}\sum_{t=1}^{\infty}  \int_{d\Omega_{g,s}}   \vect{f}_m(\vr) \cdot \vect{\cal O}_t(\vr) d\vr \nonumber \\
    &\quad \int_{d\Omega_{g,s}} \vect{f}_n(\vrp) \cdot \vect{\cal W}_t^*(\vrp) d\vrp \label{eq:Zeqn_2}
\end{align}
\end{subequations}
From \eqref{eq:Zeqn_1} and \eqref{eq:Zeqn_2}, 
\begin{align}
    \matr{Z}_{mn} + \matr{Z}_{mn}^* &= -\frac{Z}{2}\sum_{t=M_g+1}^{M}  \int_{d\Omega_{g,s}}   \vect{f}_m(\vr) \cdot \vect{\cal W}_t^*(\vr) d\vr \nonumber \\
    &\quad \int_{d\Omega_{g,s}} \vect{f}_n(\vrp) \cdot \vect{\cal W}_t(\vrp) d\vrp
\end{align}
The above result can be compactly expressed as the $(m,n)$-th entry of $-\frac{1}{2} \matr{V}^* \matr{V}^T$.